\documentclass[aps,prd,onecolumn,preprintnumbers,groupedaddress,showpacs,nofootinbib,amssymb]{revtex4-2}
\usepackage{graphicx,color}
\usepackage{amsmath}
\usepackage{amssymb}
\usepackage{amsfonts}
\usepackage{bm}
\usepackage{cancel}
\usepackage{slashed}
\newcommand{\e}{\mathrm{e}}

\usepackage{graphicx,color}
\usepackage{amsmath}
\usepackage{amssymb}
\usepackage{amsfonts}
\usepackage{bm}
\usepackage{cancel}
\usepackage{slashed}
\usepackage{hyperref}

\usepackage{tikz}

\allowdisplaybreaks[4]
\begin{document}

\preprint{KEK-TH-2874, KEK-Cosmo-0434}

\title{ACT-compatible Inflation in Scalar Field Coupled $f\left(Q, \tilde R\right)$ Gravity}
\author{Shin'ichi~Nojiri,$^{1,2}$}
\email{nojiri@nagoya-u.jp}
\affiliation{$^{1)}$ Theory Center, High Energy Accelerator Research Organization (KEK), \\
Oho 1-1, Tsukuba, Ibaraki 305-0801, Japan} \affiliation{$^{2)}$
Kobayashi-Maskawa Institute for the Origin of Particles and the
Universe, Nagoya University, Nagoya 464-8602, Japan}
\author{S.~D.~Odintsov,$^{3,4,5}$}
\email{odintsov@ieec.cat} \affiliation{$^{3)}$ ICREA, Passeig Luis Companys, 23, 08010 Barcelona, Spain} \affiliation{$^{4)}$
Institute of Space Sciences (IEEC-CSIC) C. Can Magrans s/n, 08193
Barcelona, Spain} \affiliation{$^{5)}$ Institut d'Estudis
Espacials de Catalunya (IEEC), Edifici RDIT, Campus UPC, 08860
Castelldefels (Barcelona), Spain}
\author{V.~K.~Oikonomou,$^{6,7}$}
\email{v.k.oikonomou1979@gmail.com; voikonomou@gapps.auth.gr}
\affiliation{$^{6)}$Department of Physics, Aristotle University of Thessaloniki, Thessaloniki 54124, Greece}
\affiliation{$^{7)}$Center for Theoretical Physics, Khazar University, 41 Mehseti Str., Baku, AZ-1096, Azerbaijan}

\begin{abstract}
In this work, we construct models compatible with ACT constraints on inflation in the framework of $f\left(Q, \tilde R\right)$ gravity, with $f(R)$ gravity coupled to a scalar field.
The $f\left(Q, \tilde R\right)$ theory is equivalent to $f(Q, B)$ gravity models or $f(Q, C)$ gravity where $B$ or $C$ is the difference between $Q$ and the scalar curvature $\tilde R$ in Einstein's gravity, so, $B=Q-\tilde R$ or $C=\tilde R - Q$.
Using appropriate reconstruction techniques compatible with $f\left(Q, \tilde R\right)$ theory, we propose two functional behaviors of the Hubble rate $H$ that satisfy the constraints, and we construct models that realize these behaviors of the Hubble rate $H$ using $f\left(Q, \tilde R\right)$ gravity and $f(R)$ gravity coupled with a scalar field.
We also discuss the reheating stage after inflation.
Although the scenario of $f(R)$ gravity coupled with a scalar field is well-known, since the model can be rewritten in the form of Einstein's gravity coupled with two scalar fields, the non-trivial problem occurs in the $f\left(Q, \tilde R\right)$ gravity case, because the number of dynamical degrees of freedom in $f(Q)$ gravity is not settled.
\end{abstract}

\maketitle

\section{Introduction}\label{SecI}

Renewed constraints on the observational indices of inflation, like the spectral index of the primordial scalar perturbations $n_s$ and the tensor-to-scalar ratio $r$~\cite{AtacamaCosmologyTelescope:2025blo, AtacamaCosmologyTelescope:2025nti}, by the Atacama Cosmology Telescope (ACT) Data Release 6 (DR6)~\cite{AtacamaCosmologyTelescope:2025vnj, AtacamaCosmologyTelescope:2025nti, CosmoVerseNetwork:2025alb} have renewed activity on the research to verify the old models of the inflation and to propose new modified models which may be compatible with the newly obtained constraints~\cite{Yang:2026rzn, Gonuguntla:2026rkw, CastroJunior:2026gnm, Odintsov:2026dss, Yang:2026flt, Zambrano:2026fau, Latosh:2026ckf, Yogesh:2026esn, Ahmed:2026agd, Whittingham:2026cbo, Yuennan:2026fcn, Ahmed:2026msg, Peng:2026ofs, Nojiri:2026hij, Chattopadhyay:2026yam, Modak:2025grj, Thakur:2025uua, Wang:2025cpp, McDonough:2025lzo, Keskin:2025zqq, Chakraborty:2025wqn, Yuennan:2025mlg, Bezerra-Sobrinho:2025gfg, Fu:2025ciy, Afshar:2025ndm, Qiu:2025iqm, Yuennan:2025tyx, Pozdeeva:2025wsl, Hell:2025lbl, Zhu:2025twm, Ahmed:2025sfm,Yi:2025dms, Odintsov:2025eiv, Kallosh:2025rni, Gao:2025onc, Liu:2025qca, Yogesh:2025wak,Peng:2025bws, Yin:2025rrs, Byrnes:2025kit, Wolf:2025ecy, Aoki:2025wld, Gao:2025viy, Zahoor:2025nuq, Ferreira:2025lrd, Mohammadi:2025gbu, Choudhury:2025vso, Odintsov:2025wai, Q:2025ycf, Kouniatalis:2025orn, Hai:2025wvs, Dioguardi:2025vci, Yuennan:2025kde, Ajith:2025rvf, Kuralkar:2025hoz, Modak:2025bjv, Aoki:2025ywt, Ahghari:2025hfy, NooriGashti:2025gug, Deb:2025gtk, Ellis:2025zrf, Iacconi:2025odq,Wang:2025dbj, Asaka:2015vza, Oikonomou:2025htz, Choudhury:2025hnu, Singh:2025uyr, Kim:2025dyi, Garny:2026gcs, Odintsov:2026doe, DOnofrio:2025bol, Odintsov:2025jky, Oikonomou:2026qkj, Odintsov:2026cxz, Yang:2026rzn, Nojiri:2026ish}.
In this paper, we aim to construct models compatible with the constraints from the ACT data and others in the framework of $f\left(Q, \tilde R\right)$ gravity, with the $f(R)$ gravity being coupled with a scalar field.
Here $Q$ is the non-metricity scalar, and $\tilde R$ in $f\left(Q, \tilde R\right)$ and $R$ in $f(R)$ are scalar curvatures made from the Levi-Civita connection.
The gravity theory based on non-metricity is often called the symmetric teleparallel theory.
We should note that $f\left(Q, \tilde R\right)$ is equivalent to the gravity models called $f(Q, B)$ gravity in \cite{Capozziello:2023vne} or $f(Q, C)$ gravity in \cite{Gadbail:2023mvu}. Here $B$ or $C$ is the difference between $Q$ and the scalar curvature $\tilde R$ in Einstein's gravity, that is, $B=Q-\tilde R$ or $C=\tilde R - Q$.
The quantity $B$ or $C$ is a total derivative.
We should note that $f(Q, \tilde R)=f(Q, Q-B)=f(Q, Q+C)$, which indicates the equivalence between the $f\left(Q, \tilde R\right)$ gravity model and the $f(Q, B)$ or $f(Q, C)$ gravity models.
Also, even in the symmetric teleparallel theory, the conservation law is usually given in terms of the covariant derivative defined by the Levi-Civita connection.
The commutator of two covariant derivatives inevitably induces the curvature in Einstein's gravity, and therefore we can include the curvature in Einstein's
gravity in the symmetric teleparallel theory.
See also \cite{Nojiri:2024hau}.
We propose two behaviors of the Hubble rate $H$ that satisfy the constraints.
Next, we construct models that realize these behaviors of the Hubble rate $H$ using $f\left(Q, \tilde R\right)$ gravity and $f(R)$ gravity coupled with a scalar field.
Especially, we choose the Starobinsky model~\cite{Starobinsky:1980te} and its extensions in the $f(R)$ gravity sector.
We also discuss the reheating stage after inflation.
Although the scenario of $f(R)$ gravity coupled with a scalar field is well-known, because the model can be rewritten in the form of Einstein's gravity coupled with two scalar fields, the non-trivial problem occurs in the $f\left(Q, \tilde R\right)$ gravity because the number of the dynamical degrees of freedom in $f(Q)$ gravity is not settled~\cite{Hu:2022anq, DAmbrosio:2023asf, Heisenberg:2023lru, Hu:2023gui}.
The scalar mode in $f(R)$ gravity appears via the scale transformation, but the scalar field appearing via the same scale transformation in $f(Q)$ gravity is a ghost, but it does not propagate by the Hamilton constraint~\cite{Hu:2023gui}.
If we consider the perturbation from the flat background, although both the scalar mode and the massless spin-2 mode corresponding to the
graviton appear in $f(R)$ gravity, only the massless spin-2 mode appears in $f(Q)$ gravity \cite{Capozziello:2024vix}.
Therefore, it might be natural if a massless spin-2 graviton and a massive scalar mode emerge in $f\left(Q, \tilde R\right)$ gravity.
Based on these arguments, the reheating in $f\left(Q, \tilde R\right)$ gravity could not be so different from $f(R)$ gravity.
The $f(R)$ gravity coupled with a scalar field is described by Einstein gravity coupled with two scalar fields in the Einstein frame.
In models including two scalar fields, there is a scenario called curvaton~\cite{Enqvist:2001zp, Lyth:2001nq, Moroi:2001ct, Lyth:2002my, Sasaki:2006kq, Mazumdar:2010sa}.
In the curvaton model, one scalar field is an inflaton, which generates inflation, while the other scalar field is called the curvaton, which is light during inflation.
After inflation, the fluctuation of the curvaton is converted to the curvature fluctuation.
For our specific model of the $f(R)$ gravity coupled with a scalar field in this paper, we show the possibility of realizing this curvaton scenario by investigating the asymptotic behaviors of a tune-rate vector and the perturbation in the adiabatic approximation.
In our model, the additional scalar field plays the role of the inflaton, although the scalar mode in the $f(R)$ sector works as a curvaton.

\section{Brief Overview of $f\left(Q, \tilde R\right)$ Gravity}

In this section, we very briefly review the $f\left(Q, \tilde R\right)$ gravity, where $Q$ is the non-metricity scalar and $\tilde R$ is the scalar curvature constructed from the Levi-Civita connection.
The action of the $f\left(Q, \tilde R\right)$ gravity is given by,
\begin{align}
\label{Gnrl1QR}
S_\mathrm{QR} = \int d^4x \sqrt{-g} f\left(Q, \tilde R \right)\, .
\end{align}
Then, by varying the action with respect to the metric, one obtains,
\begin{align}
\label{fQG2}
0=&\, T_{\mu\nu} + \frac{1}{2} g_{\mu\nu} f
 - f_Q g^{\alpha\beta} g^{\gamma\rho} \left\{ -\frac{1}{4} \nabla_\mu g_{\alpha\gamma} \nabla_\nu g_{\beta\rho}
 - \frac{1}{2} \nabla_\alpha g_{\mu\gamma} \nabla_\beta g_{\nu\rho} \right. \nonumber \\
&\, + \frac{1}{2} \left( \nabla_\mu g_{\alpha\gamma} \nabla_\rho g_{\beta\nu}
+ \nabla_\nu g_{\alpha\gamma} \nabla_\rho g_{\beta\mu} \right)
+ \frac{1}{2} \nabla_\alpha g_{\mu\gamma} \nabla_\rho g_{\nu\beta}
+ \frac{1}{4} \nabla_\mu g_{\alpha\beta} \nabla_\nu g_{\gamma\rho}
+ \frac{1}{2} \nabla_\alpha g_{\mu\nu} \nabla_\beta g_{\gamma\rho} \nonumber \\
&\, \left. - \frac{1}{4} \left( \nabla_\mu g_{\alpha\beta} \nabla_\gamma g_{\nu\rho} + \nabla_\nu g_{\alpha\beta} \nabla_\gamma g_{\mu\rho} \right)
 - \frac{1}{2} \nabla_\alpha g_{\mu\nu} \nabla_\gamma g_{\beta\rho}
 - \frac{1} {4} \left(\nabla_\alpha g_{\gamma\rho} \nabla_\mu g_{\beta\nu} + \nabla_\alpha g_{\gamma\rho} \nabla_\nu g_{\beta\mu} \right)
\right\} \nonumber \\
&\, - \frac{g_{\mu\rho} g_{\nu\sigma}}{\sqrt{-g}} \partial_\alpha \left[ \sqrt{-g} f_Q \left\{ - \frac{1}{2} g^{\alpha\beta} g^{\gamma\rho} g^{\sigma\tau} \nabla_\beta g_{\gamma\tau}
+ \frac{1}{2} g^{\alpha\beta} g^{\gamma\rho} g^{\sigma\tau} \left( \nabla_\tau g_{\gamma\beta} + \nabla_\gamma g_{\tau\beta} \right) \right. \right. \nonumber \\
&\, \left. \left. + \frac{1}{2} g^{\alpha\beta} g^{\rho\sigma} g^{\gamma\tau} \nabla_\beta g_{\gamma\tau}
 - \frac{1}{2} g^{\alpha\beta} g^{\rho\sigma} g^{\gamma\tau} \nabla_\gamma g_{\beta\tau}
 - \frac{1}{4} \left( g^{\alpha\sigma} g^{\gamma\tau} g^{\beta\rho} + g^{\alpha\rho} g^{\gamma\tau} g^{\beta\sigma}
\right) \nabla_\beta g_{\gamma\tau}
\right\} \right] \nonumber \\
&\, - \frac{g_{\mu\rho} g_{\nu\sigma}}{\sqrt{-g}} {\Gamma^\rho}_{\alpha\eta} \left[ \sqrt{-g} f_Q \left\{ - \frac{1}{2} g^{\alpha\beta} g^{\gamma\eta} g^{\sigma\tau} \nabla_\beta g_{\gamma\tau}
+ \frac{1}{2} g^{\alpha\beta} g^{\gamma\eta} g^{\sigma\tau} \left( \nabla_\tau g_{\gamma\beta} + \nabla_\gamma g_{\tau\beta} \right) \right. \right. \nonumber \\
&\, \left. \left. + \frac{1}{2} g^{\alpha\beta} g^{\eta\sigma} g^{\gamma\tau} \nabla_\beta g_{\gamma\tau}
 - \frac{1}{2} g^{\alpha\beta} g^{\eta\sigma} g^{\gamma\tau} \nabla_\gamma g_{\beta\tau}
 - \frac{1}{4} \left( g^{\alpha\sigma} g^{\gamma\tau} g^{\beta\eta} + g^{\alpha\eta} g^{\gamma\tau} g^{\beta\sigma}
\right) \nabla_\beta g_{\gamma\tau}
\right\} \right] \nonumber \\
&\, - \frac{g_{\mu\rho} g_{\nu\sigma}}{\sqrt{-g}} {\Gamma^\sigma}_{\alpha\eta} \left[ \sqrt{-g} f_Q \left\{ - \frac{1}{2} g^{\alpha\beta} g^{\gamma\rho} g^{\eta\tau} \nabla_\beta g_{\gamma\tau}
+ \frac{1}{2} g^{\alpha\beta} g^{\gamma\rho} g^{\eta\tau} \left( \nabla_\tau g_{\gamma\beta} + \nabla_\gamma g_{\tau\beta} \right) \right. \right. \nonumber \\
&\, \left. \left. + \frac{1}{2} g^{\alpha\beta} g^{\rho\eta} g^{\gamma\tau} \nabla_\beta g_{\gamma\tau}
 - \frac{1}{2} g^{\alpha\beta} g^{\rho\eta} g^{\gamma\tau} \nabla_\gamma g_{\beta\tau}
 - \frac{1}{4} \left( g^{\alpha\eta} g^{\gamma\tau} g^{\beta\rho} + g^{\alpha\rho} g^{\gamma\tau} g^{\beta\eta}
\right) \nabla_\beta g_{\gamma\tau}
\right\} \right] \nonumber \\
&\,
 - 4 \tilde R_{\mu\nu} f_R - 4 g_{\mu\nu} \tilde\Box f_R + 4 \tilde\nabla_\mu \tilde\nabla_\nu f_R \, .
\end{align}
Here $f_Q \equiv \frac{\partial f }{\partial Q}$ and $f_R \equiv \frac{\partial f }{\partial \tilde R}$.
We should note that $\tilde\nabla_\mu$ is the covariant derivative given by the Levi-Civita connection and is different from $\nabla$.

We work in the spatially flat Friedmann-Lema\^{i}tre-Robertson-Walker (FLRW) spacetime whose metric is given,
\begin{align}
\label{FLRW}
ds^2 = - dt^2 + a(t)^2 \sum_{i=1,2,3} \left( dx^i \right)^2 \, .
\end{align}
In the spatially flat FLRW spacetime, we can choose the coincident gauge, where the connections vanish, and note that the Levi-Civita connection does not vanish.
Then, the covariant derivative $\nabla_\mu$ reduces to the partial derivative $\partial_\mu$.
In the spatially flat FLRW spacetime~\eqref{FLRW}, $Q$ and $\tilde R$ are given by,
\begin{align}
\label{QR}
Q=-6 H^2 \, , \quad \tilde R = 12 H^2 + 6\dot H\, .
\end{align}
In the following, we denote $\tilde R$ by simply $R$ as long as there is no confusion.

Then, the equations corresponding to the first and second Friedmann equations in Einstein's gravity are given by
\begin{align}
\label{G00QR}
0 =&\, - f - 12 H^2 f_Q + 6\left(H^2 + \dot H\right) f_R - 36 \left( 4H^2 \dot H + H \ddot H\right) f_{RR} + 2 \rho \, , \\
\label{GijQR}
0=&\, f + 4 a^{-3} \frac{d}{dt} \left( a^3 H f_Q \right) \nonumber \\
&\, - 2 \left(\dot H + 3H^2\right) f_R + 12 \left( 8H^2 \dot H + 4 {\dot H}^2 + 6 H \ddot H + \dddot H\right) f_{RR} + 72\left( 4H\dot H + \ddot H\right)^2 f_{RRR} + 2 p\, .
\end{align}
Also, the continuity equation holds for the matter fluids,
\begin{align}
\label{cons}
\dot\rho + 3 H \left( \rho + p \right) = 0 \, .
\end{align}
Because Eq.~\eqref{GijQR} can be obtained from \eqref{G00QR}, we neglect Eq.~\eqref{GijQR} hereafter.
For simplicity, we may consider the following model,
\begin{align}
\label{fQfR}
f\left(Q, R \right) = f_{(Q)}\left(Q \right) + f_{(R)}\left( R \right) \, .
\end{align}
Then Eq.~\eqref{G00QR} can be rewritten as follows,
\begin{align}
\label{G00QRfQfR}
0 =&\, - f_{(Q)} - 12 H^2 f_{(Q)}'
 - f_{(R)} + 6\left(H^2 + \dot H\right) f_{(R)}' - 36 \left( 4H^2 \dot H + H \ddot H\right) f_{(R)}'' + 2 \rho \, .
\end{align}
Here $f_{(Q)}' = \frac{df_{(Q)}}{dQ}$, $f_{(R)}' = \frac{df_{(R)}}{dR}$ and $f_{(R)}'' = \frac{d^2f_{(R)}}{dR^2}$.
Let us assume $H$ is given by a function of the cosmological time $t$ or $e$-folding number $N$,
\begin{align}
\label{HtHN}
H=H_{(t)}(t) \quad \mbox{or} \quad H=H_{(N)}(N)\, .
\end{align}
Here $H_{(t)}(t)$ and $H_{(N)}(N)$ are given functions.
The $e$-foldings number $N$ is defined by $a=\e^{N-N_0}$, where $N_0$ is the present value of the $e$-folding number.
By using \eqref{QR}, $t$ or $N$ are solved with respect to $Q$ as $t=t(Q)$ or $N=N(Q)$.
If we define $\rho_\mathrm{total}$ by,
\begin{align}
\label{rhototal}
\rho_\mathrm{total} \equiv \frac{1}{2} \left\{ - f_{(R)} + 6\left(H^2 + \dot H\right) f_{(R)}' - 36 \left( 4H^2 \dot H + H \ddot H\right) f_{(R)}'' \right\} + \rho \, ,
\end{align}
and the matter energy density $\rho$ is also given by a function of $t$ or $N$, $\rho_\mathrm{total}$ can be expressed as a function of $Q$, $\rho_\mathrm{total} = \rho_\mathrm{total} (Q)$.
Because $Q=-6H^2$, by using (\ref{G00QR}), we find
\begin{align}
\label{arbQtotal}
f_{(Q)}= \left( -Q \right)^\frac{1}{2} \int^Q dq \rho_\mathrm{total}(q) \left( - q \right)^{-\frac{3}{2}} \, .
\end{align}
By using \eqref{arbQtotal}, we can construct models to reproduce \eqref{HtHN}.
Since $\frac{d}{dt} = H \frac{d}{N}$ and $\frac{d^2}{dt^2} = H^2 \frac{d^2}{dN^2} + H H' \frac{d}{N}$ with $H'\equiv \frac{dH}{dN}$, Eq.~\eqref{rhototal} can be written as,
\begin{align}
\label{rhototalN}
\rho_\mathrm{total} = \frac{1}{2} \left\{ - f_{(R)} + 6\left(H^2 + H H'\right) f_{(R)}' - 36 \left( 4H^3 H' + H^2 {H'}^2 + H^3 H'' \right) f_{(R)}'' \right\} + \rho \, ,
\end{align}
which is used when we discuss the evolution of the universe in terms of the $e$-folding number $N$.

Then, for a given $f_{(R)}$ in \eqref{fQfR}, if the evolution of the Hubble rate $H$ in \eqref{HtHN} is also given, the corresponding $f_{(Q)}$ can be found by using \eqref{arbQtotal} and \eqref{rhototalN}.

\section{Inflationary Dynamics}

In terms of the cosmological time $t$ and the $e$-folding number $N$, we may define the slow-roll parameters $\epsilon$ and $\eta$ as follows,
\begin{align}
\label{slwrllprmtrs}
\epsilon = - \frac{\dot H}{H^2} = - \frac{H'}{H} \, , \quad \eta = \frac{\dot{\epsilon}}{\epsilon H} = \frac{\epsilon'}{\epsilon}\, .
\end{align}
We also define the end of inflation by $\epsilon=1$.
During inflation, the Hubble rate $H$ is almost constant, $\dot H\sim 0$.
The equation $\epsilon=1$ means that the time scale of the change of $H$, $\left| \frac{\dot H}{H} \right|^{-1} = - \left( \frac{\dot H}{H} \right)^{-1}$ is equal to the time scale of the expansion $H^{-1}$.
We observe the cosmic microwave radiation (CMB) with the wavenumber $k\sim 0.05\, \mathrm{Mpc}^{-1}$.
The emitted time $t_*$ and $e$-folding number $N_*$ are estimated to be
\begin{align}
\label{k}
k=a(t_*) H(t_*)=a(N_*) H(N_*)\, .
\end{align}
By using the estimated $t_*$ or $N_*$ and the slow-roll parameters $\epsilon$ and $\eta$ defined in \eqref{slwrllprmtrs} when $t=t_*$, the spectral index parameter $n_s$ and the tensor-to-scalar ratio parameter $r$ are given by,
\begin{align}
\label{prmtrs}
n_s - 1 = - 2 \epsilon(t_*) - \eta(t_*) = - 2 \epsilon(N_*) - \eta(N_*)\, , \quad r = 12 \epsilon (t_*) = 12 \epsilon (N_*) \, .
\end{align}
Only by the ACT observation~\cite{AtacamaCosmologyTelescope:2025nti}, we obtain the constraint on the value of $n_s$, as follows,
\begin{align}
\label{nsACTonly}
n_s = 0.974 \pm 0.009\, .
\end{align}
Because the low-$\ell$ B-mode polarisation of the CMB ($\ell\leq 100$) is not observed by the ACT, the constraint on the value of $r$ cannot be given only by the ACT observation.
We find the constraint on $r$ mainly from the BICEP/Keck observation.
By using the combined data of Planck, ACT, lensing, BAO, and BK18 (BICEP/Keck)~\cite{AtacamaCosmologyTelescope:2025nti}, we find the medians of $n_s$ and $r$ are given as follows,
\begin{align}
\label{nsr} n_s \sim 0.976 - 0.977\, , \quad r \sim 0.012 \ \mbox{or}\ r<0.036 \, .
\end{align}
We try to construct models satisfying the constraints in \eqref{nsr}.

\subsection{Model I}

First, we consider the following model, where the Hubble rate $H$ is given by,
\begin{align}
\label{mdl2}
H^2 = \frac{{H_0}^2}{\left(1 + A_0 \e^{4\alpha_0 \left( N - N_0 \right)}\right)^\frac{1}{\alpha_0}}\, .
\end{align}
Here, $H_0$, $A_0$, and $\alpha_0$ are positive constants, and $N_0$ is the present value of the $e$-folding number, as in \eqref{HtHN}. When $N\to -\infty$, $H^2 \to {H_0}^2$, that is, asymptotically de Sitter spacetime, which may correspond to
inflation.
On the other hand, when $N$ is large, if we choose $A_0 \e^{-4\alpha_0 N_0}$ to be large enough, or at least $A_0\gg 1$, $H^2$ behaves as $H^2 \propto \e^{-4N}$, which corresponds to the radiation-dominated era, which may be realized after inflation.

For the model~\eqref{mdl2}, the slow-roll parameters $\epsilon$ and $\eta$ in \eqref{slwrllprmtrs} have the following forms,
\begin{align}
\label{mdl2prmtrs}
\epsilon = \frac{2 A_0 \e^{4 \alpha_0 \left( N - N_0 \right)}}{1 + A_0 \e^{4 \alpha_0 \left( N - N_0 \right)}}\, , \quad
\eta = 4 \alpha_0 - \frac{4 \alpha_0 A_0 \e^{4 \left( N - N_0 \right)}}{1 + A_0 \e^{4 \left( N - N_0 \right)}}\, .
\end{align}
Then the end of the inflation defined by $\epsilon=1$ is given by,
\begin{align}
\label{einf}
N=N_\mathrm{end} \equiv N_0 + \frac{1}{4\alpha_0}\ln \left( A_0 \right)\, .
\end{align}
On the other hand, the $e$-foldings number $N_*$ when the CMB with the wavenumber $k\sim 0.05\, \mathrm{Mpc}^{-1}\sim 3\times 10^{-31}\, \mathrm{eV}$ was emitted is given by using \eqref{k},
\begin{align}
\label{CMBNmdl2}
k^{2\alpha_0} = \frac{{H_0}^{2\alpha_0} \e^{2\alpha_0\left(N_*-N_0\right)}}{1 + A_0 \e^{4 \alpha_0 \left( N_* - N_0 \right)}}\, ,
\quad
\e^{2\left(N_* -N_0\right)} = \frac{1}{2} \left( \frac{{H_0}^{2\alpha_0}}{k^{2\alpha_0} A_0} \pm \sqrt{ \frac{{H_0}^{4\alpha_0}}{k^{4\alpha_0} {A_0}^2} - \frac{4}{A_0}} \right) \, .
\end{align}
Because the solutions must be real numbers, the following condition should be satisfied,
\begin{align}
\label{resmdl2}
\frac{{H_0}^{4\alpha_0}}{4k^{4\alpha_0}} \geq A_0 \, .
\end{align}
Since $H_0\sim 10^{14}\, \mathrm{GeV}=10^{23}\, \mathrm{eV}$, the quantity $\frac{{H_0}^4}{4k^4}$ is very large, $\frac{{H_0}^4}{4k^4} \sim 10^{216}$ but if $\alpha_0$ is very small, $\frac{{H_0}^{4\alpha_0}}{4k^{4\alpha_0}} \sim \mathcal{O}(1)$.

If we assume $\frac{{H_0}^{4\alpha_0}}{4k^{4\alpha_0}} \gg A_0$ as a working hypothesis, we obtain,
\begin{align}
\label{CMBN2mdl2}
\e^{2\left(N_* -N_0\right)} \sim \frac{{H_0}^{2\alpha_0}}{k^{2\alpha_0} A_0} \quad \mbox{or} \quad \frac{k^{2\alpha_0}}{{H_0}^{2\alpha_0}} \, .
\end{align}
Because $N_* <N_0$, that is, $\e^{2 \left( N_* - N_0 \right)}<1$, the first solution is excluded.
If $\frac{{H_0}^4}{4k^4} \gg A_0$, we also find,
\begin{align}
\label{A0emdl2}
1 \gg A_0 \frac{k^{4\alpha_0}}{{H_0}^{4\alpha_0}} \sim A_0 \e^{4\left(N_* -N_0\right)} \, .
\end{align}
Then the slow-roll parameters in \eqref{mdl2prmtrs} are approximated to be,
\begin{align}
\label{mdl1prmtrsaprxmdl2}
\epsilon \sim 2 A_0 \e^{4 \left( N_* - N_0 \right)}\, , \quad
\eta \sim 4\alpha_0 - 4 A_0 \e^{4 \left( N_* - N_0 \right)}\, ,
\end{align}
which gives,
\begin{align}
\label{nswmdl2}
n_s - 1 \sim - 4A_0 \e^{4 \left( N_* - N_0 \right)} - 4\alpha_0 + 4 A_0 \e^{4 \left( N_* - N_0 \right)} = - 4\alpha_0 \, , \quad
r = 24 A_0 \e^{4 \left( N_* - N_0 \right)}\, .
\end{align}
If $n_s=0.976$ and $r=0.012$, we find,
\begin{align}
\label{cnstrntrns}
\alpha_0=0.006\, , \quad A_0 \e^{4 \left( N_* - N_0 \right)} = 0.0005\, .
\end{align}
Therefore, the constraints in \eqref{nsr} can be satisfied.
We should note that Eq.~\eqref{cnstrntrns} is consistent with the assumptions \eqref{A0emdl2} and that $\alpha_0$ is very small.

\subsection{Model II}

We now consider the following model,
\begin{align}
\label{mdl5}
H^2 = \frac{{H_0}^2}{1 + \alpha_0 \e^{\alpha_1 N} + \beta_0 \e^{4N}}\, .
\end{align}
Here $\alpha_0$, $\alpha_1$, and $\beta_0$ are positive constants and we choose $0<\alpha_1 \ll 4$.
When $N$ is negative and $\left| N \right| \gg 1$, $H^2$ goes to a constant $H^2 \to {H_0}^2 \left( 1 - \alpha_0 \e^{\alpha_1 N} \right)$, which can generate inflationary dynamics.
The term $- \alpha_0 \e^{\alpha_1 N}$ plays the role of the slow-roll.
On the other hand, when $N$ is positive and $N\gg 1$, we find $H^2 \sim \frac{{H_0}^2}{\beta_0}\e^{-4N}$, which corresponds to the radiation-dominated Universe after the inflationary era.
Now the slow roll parameters \eqref{slwrllprmtrs} are given by,
\begin{align}
\label{slrllprmtrmdl5}
\epsilon = \frac{1}{2} \frac{\alpha_0 \alpha_1 \e^{\alpha_1 N} + 4 \beta_0 \e^{4N}} {1 + \alpha_0 \e^{\alpha_1 N} + \beta_0 \e^{4N}}\, , \quad
\eta = \frac{\alpha_0 {\alpha_1}^2 \e^{\alpha_1 N} + 16 \beta_0 \e^{4N}}{\alpha_0 \alpha_1 \e^{\alpha_1 N} + 4 \beta_0 \e^{4N}}
 - \frac{\alpha_0 \alpha_1 \e^{\alpha_1 N} + 4 \beta \e^{4N}} {1 + \alpha_0 \e^{\alpha_1 N} + \beta_0 \e^{4N}}\, .
\end{align}
Thus, the end of inflation occurs at $\epsilon\left( N_\mathrm{end} \right)=1$, which is,
\begin{align}
\label{endmdl5}
0 = 2 + \left( 2 - \alpha_1 \right) \alpha_0 \e^{\alpha_1 N_\mathrm{end}} - 2 \beta_0 \e^{4N_\mathrm{end}} \, .
\end{align}
Because if $N_\mathrm{end}\to -\infty$, the right hand side of Eq.~\eqref{endmdl5} goes to $2>0$ and if $N_\mathrm{end}\to +\infty$, the right hand side of Eq.~\eqref{endmdl5} behaves as $- 2 \beta_0 \e^{4N_\mathrm{end}}<0$.
Therefore, Eq.~\eqref{endmdl5} always has a solution, although it is difficult to express $N_\mathrm{end}$ by elementary functions.

The emitted $e$-folding number $N_*$ of the observed CMB radiation with the wave number $k\sim 0.05\, \mathrm{Mpc}^{-1}\sim 3\times 10^{-31}\, \mathrm{eV}$ given by \eqref{k} can be estimated to be,
\begin{align}
\label{kmdl2}
k = H_0 \e^{N_* - N_0} \, .
\end{align}
Here $N_0$ is the present value of the $e$-foldings number and we write $a(N)$ as $a(N)=\e^{N - N_0}$.
The parameter $H_0$ could be estimated to be $H_0\sim 10^{14}\, \mathrm{GeV}=10^{23}\, \mathrm{eV}$.
% {\color{blue}Because the CMB could be emitted during inflation, it could be natural to assume $N_*$ is negative and $\left| N_* \right| \gg 1$.????????}
Because the primordial curvature perturbations that seeded the CMB
anisotropies were generated during inflation, $N_*$ is negative
and $\left| N_* \right| \gg 1$. Then the slow-roll parameters in
\eqref{slrllprmtrmdl5} are estimated to be,
\begin{align}
\label{slrllprmtrmdl5_2}
\epsilon \sim \frac{1}{2}\alpha_0 \alpha_1 \e^{\alpha_1 N_*} \, , \quad
\eta \sim \alpha_1 - \alpha_0 \alpha_1 \e^{\alpha_1 N_*} \sim \alpha_1 \, .
\end{align}
Then the spectral index parameter $n_s$ and the tensor-to-scalar ratio parameter $r$ \eqref{prmtrs} are given by,
\begin{align}
\label{prmtrsmdl5}
n_s - 1 = - 2 \epsilon(N_*) - \eta(N_*) \sim - \alpha_0 \alpha_1 \e^{\alpha_1 N_*} - \alpha_1 \sim - \alpha_1 \, , \quad
r = 12 \epsilon (N_*) \sim 6 \alpha_0 \alpha_1 \e^{\alpha_1 N_*} \, ,
\end{align}
which gives,
\begin{align}
\label{estmtn}
\alpha_1 \sim 0.024 \, , 6 \alpha_0 \alpha_1 \e^{\alpha_1 N_*}<0.038 \, .
\end{align}
The value of $\alpha_1$ is consistent with the assumption $0<\alpha_1 \ll 4$. Hence, we obtain a model consistent with the observations by the ACT, etc.

\section{Models of $f\left(Q, \tilde R\right)$ Gravity Describing Inflation}\label{SecIV}

\subsection{Inflation Models of $f\left(Q, \tilde R\right)$ Gravity}

We now construct a model whose solution is given by \eqref{mdl2}, that is, Model I.
Because $Q=-6H^2$, Eq.~\eqref{mdl2} tells,
\begin{align}
\label{sol}
A_0 \e^{4\alpha_0 \left( N - N_0 \right)} = - 1 + \left( - \frac{6{H_0}^2}{Q} \right)^{\alpha_0}\, .
\end{align}
Then we find,
\begin{align}
\label{R}
R = 12 {H_0}^2 \left( - \frac{Q}{6{H_0}^2} \right)^{\alpha_0+1} \, ,
\end{align}
and,
\begin{align}
\label{rhototalN2}
\rho_\mathrm{total} =&\, \frac{1}{2} \left[ - f_{(R)} + 6{H_0}^2\left\{- \left( - \frac{Q}{6{H_0}^2} \right) + 2\left( - \frac{Q}{6{H_0}^2} \right)^{\alpha_0+1} \right\} f_{(R)}'
 - \left( 6{H_0}^2 \right)^2 \left\{ 8\left( \alpha_0 + 1 \right) \left( - \frac{Q}{6{H_0}^2} \right)^{2\alpha_0+2} \right. \right. \nonumber \\
&\, \left. \left. \left. \qquad \qquad - 8\left(\alpha_0 + 1 \right) \left( - \frac{Q}{6{H_0}^2} \right)^{\alpha_0+2}
+ 4 \left( - \frac{Q}{6{H_0}^2} \right) \right\} f_{(R)}'' \right] \right|_{R = 12 {H_0}^2 \left( - \frac{Q}{6{H_0}^2} \right)^{\alpha_0+1}}
+ \rho \, .
\end{align}
Then by using \eqref{arbQtotal}, we can find $f_{(Q)}$ corresponding to \eqref{mdl2}. For the derivations of Eqs.~\eqref{R} and \eqref{rhototalN2}, see Appendix~\ref{SecIVA}.

For the Model II in \eqref{mdl5}, we find,
\begin{align}
\label{mdl5Q}
Q = - \frac{6{H_0}^2}{1 + \alpha_0 \e^{\alpha_1 N} + \beta_0 \e^{4N}}\, ,
\end{align}
but it is difficult to explicitly solve Eq.~\eqref{mdl5Q} with respect to $N$ as a function of $Q$ as in \eqref{sol}.
Therefore, we cannot write down the explicit form of $f_{(Q)}$.

In order to specify the model, we may assume $f_{(R)}$,
\begin{align}
\label{f(R)}
f_{(R)} = f_m R^m\, .
\end{align}
Where $f_m$ is a constant. If we consider the $f(R)$ gravity theory where the action is only given by \eqref{f(R)}, the model generates a spacetime whose evolution is identical to that of the Universe in Einstein's gravity with a perfect fluid whose equation of state parameter $w$ is given by \cite{Nojiri:2003ft},
\begin{align}
\label{PQV3}
w = - 1 - \frac{2(m-2)}{3(m-1)(2m-1)}\, .
\end{align}
Especially, when $m=\frac{5}{4}$, we find $w=\frac{1}{3}$, that is, the radiation.
We neglect the contribution from matter, $\rho=0$.
Then, by using Eq.~\eqref{arbQtotal}, we find,
\begin{align}
\label{arbQtotalmdl2}
f_{(Q)}
=&\, \frac{f_m}{2} \left(12 {H_0}^2\right)^m \left( -\frac{Q}{6{H_0}^2} \right)^{m\left(\alpha_0+1\right)} \left[
\frac{2\alpha_0 + 1}{2\left\{\left(m-1\right)\left(\alpha_0+1\right) + \frac{1}{2}\right\}} \left( -\frac{Q}{6{H_0}^2} \right)^{-\alpha_0}
 - \frac{2 \left( \alpha_0 + 1\right)}{m \left(\alpha_0+1\right) - \frac{1}{2}} \right. \nonumber \\
&\, \left. \qquad \qquad - \frac{m \left( m - 1 \right)}{\left(m-2\right)\left(\alpha_0+1\right) - \frac{1}{2}} \left( -\frac{Q}{6{H_0}^2} \right)^{-2\left(\alpha_0+1\right)} \right]
+ C \left( -\frac{Q}{6{H_0}^2} \right)^\frac{1}{2} \, .
\end{align}
Here $C$ is the constant of integration. The derivation of \eqref{arbQtotalmdl2} is given in Eq.~\eqref{arbQtotalmdl2Ap} in the Appendix.

Then we obtain a model consistent with the ACT observations, as in \eqref{nsr},
\begin{align}
\label{rmfqmdl1}
f(Q,R) =&\, \frac{f_m}{2} \left(12 {H_0}^2\right)^m \left( -\frac{Q}{6{H_0}^2} \right)^{m\left(\alpha_0+1\right)} \left[
\frac{2\alpha_0 + 1}{2\left\{\left(m-1\right)\left(\alpha_0+1\right) + \frac{1}{2}\right\}} \left( -\frac{Q}{6{H_0}^2} \right)^{-\alpha_0}
 - \frac{2 \left( \alpha_0 + 1\right)}{m \left(\alpha_0+1\right) - \frac{1}{2}} \right. \nonumber \\
&\, \left. \qquad \qquad - \frac{m \left( m - 1 \right)}{\left(m-2\right)\left(\alpha_0+1\right) - \frac{1}{2}} \left( -\frac{Q}{6{H_0}^2} \right)^{-2\left(\alpha_0+1\right)} \right]
+ C \left( -\frac{Q}{6{H_0}^2} \right)^\frac{1}{2} + f_m R_m \, , \nonumber \\
&\, \alpha_0=0.006\, , \quad A_0 \e^{4 \left( N - N_0 \right)} = 0.0005\, ,
\end{align}
which reproduces the Hubble rate $H$ of Model I \eqref{mdl2}.
The $f_{(Q)}$ corresponding to Model II \eqref{mdl5} cannot be given by elementary functions.

\subsection{Non-metricity Corrections to the Starobinsky model and its Extensions}

The action of the Starobinsky model~\cite{Starobinsky:1980te} is given by,
\begin{align}
\label{Strbnskymdl}
S_\mathrm{Starobinsky} = \frac{1}{2\kappa^2} \int d^4 x \sqrt{-g} \left( R + \alpha R^2 \right) \, .
\end{align}
with a constant $\alpha$. In order to avoid the tachyon mode, we assume $\alpha$ is positive.
The Starobinsky model predicts,
\begin{align}
\label{Strbnskynsr}
n_s \sim 0.964\, , \quad r\sim 0.004 \, ,
\end{align}
which are consistent with Planck 2018 results~\cite{Planck:2018vyg}, but there is a $2\sigma$ tension with the combined data of Planck, ACT, lensing, BAO, BK18 (BICEP/Keck)~\cite{AtacamaCosmologyTelescope:2025nti} in \eqref{nsr}.
There also appear constraints on the equation of state parameter $w$ and the temperature $T$ during the reheating \cite{Drees:2025ngb}.

We should note that the tension is improved by adding the corrections from $f(Q)$ gravity as in \eqref{arbQtotalmdl2},
\begin{align}
\label{arbQtotalmdl2strbnsky}
f_{(Q)} = \frac{1}{4\kappa^2 } &\, \left( f_{(1)}(Q) + \alpha f_{(2)}(Q) \right) \nonumber \\
= \frac{1}{4\kappa^2} &\, \left[
 - \left(12 {H_0}^2\right) \left( -\frac{Q}{6{H_0}^2} \right) \right. \nonumber \\
&\, \left. + \alpha \left(12 {H_0}^2\right)^2 \left( -\frac{Q}{6{H_0}^2} \right)^{2\left(\alpha_0+1\right)} \left\{
\frac{4\alpha_0 + 3}{2 \alpha_0 + \frac{5}{2}} \left( -\frac{Q}{6{H_0}^2} \right)^{-\alpha_0}
 - \frac{2 \left( \alpha_0 + 1\right)}{2\alpha_0 + \frac{3}{2}} + \frac{4}{3} \left( -\frac{Q}{6{H_0}^2} \right)^{-2\alpha_0} \right\} \right] \nonumber \\
&\, + C \left( -\frac{Q}{6{H_0}^2} \right)^\frac{1}{2}\, .
\end{align}
Therefore, we obtain a model consistent with the ACT observations.
We may consider the model as an extension of the Starobinsky model, where the power $R^m$ of the scalar curvature $R$ in \eqref{f(R)} is added to the Einstein-Hilbert action, which is linear in $R$.
The action of the Starobinsky model~\cite{Starobinsky:1980te} is given by,
\begin{align}
\label{Rmmdl}
S_{R+R^m} = \frac{1}{2\kappa^2} \int d^4 x \sqrt{-g} \left( R + f_m R^m \right) \, .
\end{align}
In the general $f(R)$ gravity theory, if the algebraic equation,
\begin{align}
\label{adS}
0= - R f'(R) + 2 f(R)\, ,
\end{align}
has a positive solution, the de Sitter spacetime is an exact solution of the $f(R)$ gravity.
For the model~\eqref{Rmmdl}, Eq.~\eqref{adS} has a form,
\begin{align}
\label{adS2}
0 = - R - m f_m R^m + 2 R + 2 m f_m R^m
= R - \left( m - 2 \right) f_m R^m \, .
\end{align}
Therefore, if $\left( m - 2 \right) f_m>0$, there is a solution describing the de Sitter spacetime,
\begin{align}
\label{adS3}
R = R_\mathrm{dS} \equiv \left\{ \left( m - 2 \right) f_m \right\}^{- \frac{1}{m-1}}\, .
\end{align}
The de Sitter spacetime may correspond to inflation in the early Universe.
The de Sitter solution is stable if,
\begin{align}
\label{adS4}
 - R_\mathrm{dS} + \frac{f'\left( R_\mathrm{dS} \right)}{f''\left( R_\mathrm{dS} \right)}> 0\, .
\end{align}
On the other hand, the de Sitter solution is unstable if,
\begin{align}
\label{adS5}
 - R_\mathrm{dS} + \frac{f'\left( R_\mathrm{dS} \right)}{f''\left( R_\mathrm{dS} \right)}< 0\, .
\end{align}
If the de Sitter space is stable, inflation does not end, and therefore the model is not realistic for describing the inflationary regime.
However, it might be realistic to describe the late-time accelerating expansion. In order to describe realistic inflation, the de Sitter spacetime must be unstable as in \eqref{adS5}.
For the model~\eqref{Rmmdl}, we find,
\begin{align}
\label{adS6}
 - R_\mathrm{dS} + \frac{f'\left( R_\mathrm{dS} \right)}{f''\left( R_\mathrm{dS} \right)}
=&\, - R_\mathrm{dS} + \frac{1 + m f_m {R_\mathrm{dS}}^{m-1}}{m \left( m - 1 \right) {R_\mathrm{dS}}^{m-2}} \nonumber \\
=&\, \left\{ \left( m - 2 \right) f_m \right\}^{- \frac{1}{m-1}} + \frac{1}{m \left( m - 1 \right)} \left\{ \left( m - 2 \right) f_m \right\}^{\frac{m-2}{m-1}}
+ \frac{1}{m-1} \left\{ \left( m - 2 \right) f_m \right\}^{- \frac{1}{m-1}} \nonumber \\
=&\, \frac{\left\{ \left( m - 2 \right) f_m \right\}^{- \frac{1}{m-1}}}{m\left( m-1 \right)} \left\{ m^2 + \left( m - 2 \right) f_m \right\}
\, .
\end{align}
Since we require $\left( m - 2 \right) f_m>0$, if $0<m<1$, the model~\eqref{Rmmdl} may describe the de Sitter spacetime, which may correspond to inflation in the early Universe.

Even if we do not consider the de Sitter spacetime, there is a possibility that the model~\eqref{Rmmdl} could describe inflation.
As is well known, $f(R)$ gravity can be rewritten in the scalar-tensor form \cite{Maeda:1988ab, Nojiri:2003ft} with the scalar potential.
In the scalar-tensor form, the action is given by the Einstein-Hilbert action with a minimally coupled scalar field with a potential. If the scalar potential satisfies the slow-roll conditions, the model may describe inflation.
In the model~\eqref{Rmmdl}, however, it is difficult to describe realistic inflation~\cite{Nojiri:2007cq, Odintsov:2023weg}.
If $m>2$, the potential becomes steep and $\epsilon$ and $r$ tend to become larger, and it becomes difficult to satisfy the slow-roll conditions.
Therefore, it becomes difficult to fit the model to the observational data.
On the other hand, if $m<2$, the potential becomes flat, and it becomes difficult to realize the end of inflation and reheating.

It is also possible to construct realistic models by adding contributions from $f(Q)$ gravity to the action~\eqref{Rmmdl},
\begin{align}
\label{RmmdlfQ}
\frac{1}{2\kappa^2} \left( R + f_m R^m \right) \Rightarrow &\, \frac{1}{2\kappa^2} \left( R + f_m R^m \right) + f(Q)\, , \nonumber \\
f_{(Q)} = \frac{1}{4\kappa^2 } &\, \left( f_{(1)}(Q) + f_m f_{(m)}(Q) \right) \nonumber \\
= \frac{1}{4\kappa^2} &\, \left[
 - \left(12 {H_0}^2\right) \left( -\frac{Q}{6{H_0}^2} \right) \right. \nonumber \\
&\, + f_m \left(12 {H_0}^2\right)^m \left( -\frac{Q}{6{H_0}^2} \right)^{m\left(\alpha_0+1\right)} \left\{
\frac{2\left(\alpha_0 + 1\right)m (m-1) -1}{2\left\{\left(m-1\right)\left(\alpha_0+1\right) + \frac{1}{2}\right\}} \left( -\frac{Q}{6{H_0}^2} \right)^{-\alpha_0} \right. \nonumber \\
&\, \left. \left. - \frac{4 \left( \alpha_0 + 1\right)m(m-1)}{m \left(\alpha_0+1\right) - \frac{1}{2}}
 - \frac{m \left( m - 1 \right)}{\left(m-2\right)\left(\alpha_0+1\right) + \frac{3}{2}} \left( -\frac{Q}{6{H_0}^2} \right)^{-2\alpha_0} \right\} \right]
+ C \left( -\frac{Q}{6{H_0}^2} \right)^\frac{1}{2} \, .
\end{align}
Hence, models consistent with the observations by ACT can be obtained.
We should note that the case of Einstein's gravity, that is, the case $m=0$, is special. In this case, we find
\begin{align}
\label{RmmdlfQEinstein}
f_{(Q)} = \frac{Q}{2\kappa^2} + C \left( -\frac{Q}{6{H_0}^2} \right)^\frac{1}{2} \, .
\end{align}
Because $Q=-R + \mbox{total derivative}$, the first term exactly cancels the Hilbert-Einstein term, and there only remains the last term $C \left( -\frac{Q}{6{H_0}^2} \right)^\frac{1}{2}$, which does not contribute to the Friedmann equations.
Because the action of the gravity sector substantially vanishes, any evolution of the Universe is a solution if there is no matter.

Then we obtain a model consistent with the ACT observations, as in \eqref{nsr},
\begin{align}
\label{rmfqmdl1RmR}
f(Q, R) =&\,
\frac{1}{4\kappa^2} \left[
 - \left(12 {H_0}^2\right) \left( -\frac{Q}{6{H_0}^2} \right)
+ f_m \left(12 {H_0}^2\right)^m \left( -\frac{Q}{6{H_0}^2} \right)^{m\left(\alpha_0+1\right)} \left\{
\frac{2\left(\alpha_0 + 1\right)m (m-1) -1}{2\left\{\left(m-1\right)\left(\alpha_0+1\right) + \frac{1}{2}\right\}} \left( -\frac{Q}{6{H_0}^2} \right)^{-\alpha_0} \right. \right. \nonumber \\
&\, \left. \left. - \frac{4 \left( \alpha_0 + 1\right)m(m-1)}{m \left(\alpha_0+1\right) - \frac{1}{2}}
 - \frac{m \left( m - 1 \right)}{\left(m-2\right)\left(\alpha_0+1\right) + \frac{3}{2}} \left( -\frac{Q}{6{H_0}^2} \right)^{-2\alpha_0} \right\} \right]
+ C \left( -\frac{Q}{6{H_0}^2} \right)^\frac{1}{2} + \frac{1}{2\kappa^2} \left( R + f_m R^m \right) \, , \nonumber \\
&\, \alpha_0=0.006\, , \quad A_0 \e^{4 \left( N - N_0 \right)} = 0.0005\, ,
\end{align}
which reproduces the Hubble rate $H$ of Model I \eqref{mdl2}.
When $m=2$, the model~\eqref{rmfqmdl1RmR} can be regarded as a correction to the Starobinsky model~\eqref{Strbnskymdl}.
The $f_{(Q)}$ corresponding to Model II \eqref{mdl5} cannot be given by elementary functions again.

\subsection{The Models with the Decreasing Contribution from non-metricity in the Late-time Universe}

We may also consider models where the $f(Q)$ part of the gravity only plays a role in the inflationary epoch, and the $f(Q)$ part can be neglected in the present universe.
This can be done by modifying $f(Q)$ in \eqref{arbQtotalmdl2}, for example, as
\begin{align}
\label{late}
f_{(Q)} = f_m f_{(m)}(Q) \Rightarrow \tilde f_{(Q)} \equiv \frac{1}{2}f_m f_{(m)}(Q) \left\{ 1 + \tanh \left( \left( -\frac{Q}{6{H_1}^2} \right) - \left( -\frac{Q}{6{H_1}^2} \right)^{-1} \right) \right\}\, .
\end{align}
Here $H_1$ is a positive constant and we choose $H_1\ll H_0$ but $H_1$ is much larger than the present value of $H$, that is, the Hubble constant, $H_1 \gg 70\,\mathrm{km\,s}^{-1}\mathrm{Mpc}^{-1} \sim 10^{-33}\,\mathrm{eV}$.
Then, when $H\gg H_1$ or $-Q = 6H^2 \gg 6{H_1}^2$ as in the epoch of the inflation, we find $\tanh \left( \left( -\frac{Q}{6{H_1}^2} \right) - \left( -\frac{Q}{6{H_1}^2} \right)^{-1} \right) \sim \tanh \left( \frac{H^2}{{H_1}^2} \right) \to 1 - 2\e^{- \frac{H^2}{{H_1}^2}} + \mathcal{O}\left( \e^{-2 \frac{H^2}{{H_1}^2}} \right)$, and therefore $\tilde f_{(Q)} \to f_m f_{(m)}(Q)$.
Then the model describes inflation consistent with observations from ACT.
On the other hand, when $H\ll H_1$ or $-Q = 6H^2 \ll 6{H_1}^2$ as in the present Universe, since $\tanh \left( \left( -\frac{Q}{6{H_1}^2} \right) - \left( -\frac{Q}{6{H_1}^2} \right)^{-1} \right) \sim - \tanh \left( \frac{{H_1}^2}{H^2} \right) \to - 1 + 2\e^{- \frac{{H_1}^2}{H^2}} + \mathcal{O}\left( \e^{-2 \frac{{H_1}^2}{H^2}} \right)$, $\tilde f_{(Q)}$ decreases very rapidly as $\tilde f_{(Q)}\to f_m f_{(m)}(Q)\e^{- \frac{{H_1}^2}{H^2}} \to 0$.

Then, in the early Universe, the $f(Q)$ term generates inflation combined with the $f(R)$ term, and realizes its end.
The spectral index parameter $n_s$ and the tensor-to-scalar ratio parameter $r$ are consistent with the ACT observations.
In the late Universe, the $f(Q)$ term decreases, and the $f(R)$ term may realize the dark energy epoch and the phantom crossing indicated by the Dark Energy Spectroscopic Instrument (DESI) observations~\cite{DESI:2024mwx, DESI:2025zgx}.

%%%%%%%%%%%%%%%%%%%%%%%%%%%%%%%%%%%%%wednesday evening

\subsection{The Reheating Era}

In the case of $f(R)$ gravity, the scalar mode contained within the theory acts as the inflaton and induces reheating.
On the other hand, in the case of $f(Q)$ gravity, as shown in \cite{Hu:2023gui}, the conformal mode of the metric tensor corresponding to the scalar mode of $f(R)$ gravity becomes a ghost, but does not propagate due to the Hamiltonian constraint.
Since the non-metricity scalar $Q$ also contains a metric tensor, it is not clear whether an Einstein frame corresponding to such a system can be considered.
Consequently, in theories where the action describes both $f(R)$ and $f(Q)$, it is unclear whether an Einstein frame exists, and it is not clear how phenomena such as reheating should be handled.

In fact, in the case of $f(Q)$ gravity, the physical degrees of freedom are also unclear~\cite{Hu:2022anq, DAmbrosio:2023asf, Heisenberg:2023lru, Hu:2023gui}. Despite long-standing discussion, the problem has not been completely solved.
However, a perturbative calculation of the $f(Q)$ gravitational theory in flat spacetime is available in \cite{Capozziello:2024vix}; this paper shows that the only propagating mode is the ordinary massless spin-2 graviton.
This is consistent with the finding, as shown in \cite{Hu:2023gui}, that the scalar mode, which is a conformal mode, does not propagate.
Therefore, if we add the $f(R)$ gravity term, which has flat spacetime as a solution, to this theory and consider the perturbation, the action of the massless spin-2 graviton, combined with that arising from $f(Q)$, merely changes the gravitational constant; however, it appears that a scalar mode also emerges from the $f(R)$ term.
If this argument is correct, it seems likely that a massless spin-2 graviton and a massive scalar mode will emerge as perturbative degrees of freedom.

During reheating following inflation, the expansion around flat spacetime is not a good approximation, so non-linear degrees of freedom may also contribute to reheating; however, if a massive scalar mode exists, it will likely drive reheating in the same way
as a standard inflaton.

We can rewrite $f(R)$ gravity in the scalar-tensor form, as follows.
By introducing the auxiliary field $A$, we rewrite the action of the $f(R)$ gravity,
\begin{align}
\label{JGRG7}
S_{f(R)}= \frac{1}{2\kappa^2}\int d^4 x \sqrt{-g} f(R) \, ,
\end{align}
in the following form,
\begin{align}
\label{JGRG21}
S=\frac{1}{2\kappa^2}\int d^4 x \sqrt{-g} \left\{f'(A)\left(R-A\right) + f(A)\right\}\, .
\end{align}
Compared with \eqref{fQfR}, $f_{(R)}\left( R \right) = \frac{f(R)}{2\kappa^2}$\, .

The variation of the action \eqref{JGRG21} with respect to $A$ tells $A=R$.
By substituting $A=R$ into the action \eqref{JGRG21}, the action in \eqref{JGRG7} is reproduced.
On the other hand, by rescaling the metric tensor in the following way,
\begin{align}
\label{JGRG22}
g_{\mu\nu}\to \e^\sigma g_{\mu\nu}\, ,\quad \sigma = -\ln f'(A)\, ,
\end{align}
we obtain the Einstein frame action as follows,
\begin{align}
\label{JGRG23}
S_E =&\, \frac{1}{2\kappa^2}\int d^4 x \sqrt{-g} \left( R - \frac{3}{2}g^{\rho\sigma}
\partial_\rho \sigma \partial_\sigma \sigma - V(\sigma)\right) \, ,\nonumber \\
V(\sigma) =&\, \e^\sigma g\left(\e^{-\sigma}\right)
 - \e^{2\sigma} f\left(g\left(\e^{-\sigma}\right)\right) = \frac{A}{f'(A)} - \frac{f(A)}{f'(A)^2}\, .
\end{align}
Here $g\left(\e^{-\sigma}\right)$ is given by solving the equation $\sigma =- \ln f'(A)$ as $A=g\left(\e^{-\sigma}\right)$.
Due to the scale transformation \eqref{JGRG22}, a coupling of the scalar field $\sigma$ with usual matter arises.
This coupling generates the decay of the coherent state of $\sigma$ into the matter.

We just assume that the reheating could be described by the $f(R)$ sector as an approximation.
For the model \eqref{Rmmdl}, we find,
\begin{align}
\label{msigma}
\e^{-\sigma} = f'(A) = 1 + m f_m A^{m-1}\quad \mbox{or} \quad A = g(\e^{-\sigma}) = \left( \frac{\e^{-\sigma} -1}{m f_m} \right)^\frac{1}{m-1}\, .
\end{align}
Therefore, we obtain the following potential,
\begin{align}
\label{mpotential}
V(\sigma) =\left(m-1\right) f_m \e^{2\sigma} \left( \frac{\e^{-\sigma} -1}{m f_m} \right)^\frac{m}{m-1} \, .
\end{align}
When $m\neq 2$, there is a branch cut at $\e^{-\sigma}=1$ and we require$\frac{\e^{-\sigma} -1}{m f_m}>0$, which corresponds to $A^{m-1}>0$ as we find from Eq.~\eqref{msigma}.
For the derivation of \eqref{mpotential}, see \eqref{mpotentialAp}

Because,
\begin{align}
\label{mpotentialprime}
V'(\sigma)
=&\, \left\{m -2 - 2\left(m-1\right) \e^{\sigma} \right\} \frac{\e^\sigma}{m} \left( \frac{\e^{-\sigma} -1}{m f_m} \right)^\frac{1}{m-1} \, .
\end{align}
Therefore $V'(\sigma)$ vanishes when,
\begin{align}
\label{zeros}
\e^\sigma = \frac{m-2}{2\left( m - 1 \right)} \, , \quad 1\, .
\end{align}
For the derivation of \eqref{mpotentialprime}, see \eqref{mpotentialprimeAp}.
If $mf_m>0$, because we require $\e^{-\sigma}>1$ or $\e^\sigma<1$, which require $m>2$ and $f_m>0$ or $m<0$ and $f_m<0$.
On the other hand, if $mf_m<0$, the requirement $\e^{-\sigma}<1$ or $\e^\sigma>1$ tells $0<m<1$ and $f_m<0$.
We should note that if $1<m<2$, $\frac{m-2}{2\left( m - 1 \right)}$ is negative, which could conflict with $\e^\sigma>0$.
If $\e^\sigma<0$, $f'(A)$ is also negative, but when $f'(A)<0$, anti-gravity appears, and the graviton becomes a ghost, rendering the model physically inconsistent.

When $\e^\sigma = \frac{m-2}{2\left( m - 1 \right)}$, we find,
\begin{align}
\label{mpotentialprimeprime}
V''(\sigma)
= - \frac{2\left(m-1\right)}{m} \e^{2\sigma} \left( \frac{\e^{-\sigma} -1}{m f_m} \right)^\frac{1}{m-1} \, .
\end{align}
We should note that if $m>2$ or $m<1$, $V''(\sigma)$ is negative when $\e^\sigma = \frac{m-2}{2\left( m - 1 \right)}$ and the model becomes tachyonic, and therefore the oscillation required for the reheating does not occur.
If $0<m<1$, $V''(\sigma)$ is positive when $\e^\sigma = \frac{m-2}{2\left( m - 1 \right)}$ and therefore the reheating might occur by the oscillation of the scalar mode $\sigma$.

Even if $m<0$ or $m>1$ $\left( m\neq 2\right)$, the reheating might occur by quantum effects in the curved spacetime~\cite{Parker:1968mv, Parker:1969au, Parker:1971pt, Birrell:1982ix, Buchbinder:1992gdx}, that is, the particle creation in the expanding universe.
Furthermore, even if the scalar mode $\sigma$ does not oscillate, if the value of the scalar mode changes very rapidly, particle creation might occur~\cite{Traschen:1990sw, Felder:1998vq}.

When $m=2$, which corresponds to the Starobinsky model, the potential is given by
\begin{align}
\label{StrbnskyV}
V(\sigma) = \alpha \e^{2\sigma} \left( \frac{\e^{-\sigma} -1}{2 \alpha} \right)^2 \, , \quad f_2 = \alpha \, .
\end{align}
The potential has a minimum at $\e^\sigma=1$. As is well known, reheating could occur, and the particles could be created.

Although the $f(Q)$ sector plays a role in improving the model against the constraints given by the ACT, its role in reheating is
not clear.
The scalar mode, which generates the reheating, could appear from the $f(R)$ sector; we cannot say anything about the non-perturbative degrees of freedom in the $f(Q)$ sector, which may also give any contributions to the reheating.
Even for the scalar mode coming from the $f(R)$ sector, the structure of the kinetic term could be changed by the contribution from the $f(Q)$ sector, which might be clarified after the number and the structure of the dynamical freedoms in the $f(Q)$ gravity, although the coupling of $f(R)$ gravity to the $f(Q)$ gravity could make the Hamiltonian structure much more complicated.

\section{$F(R)$ gravity coupled with a scalar field}\label{SecV}

In the previous section, we observed that the Starobinsky model~\eqref{Strbnskymdl} and other $f(R)$ gravity theories, as in \eqref{Rmmdl}, are improved to satisfy the constraints from the ACT observations by adding $f(Q)$ gravity as a correction.
In this section, we consider a similar scenario by using a scalar field $\phi$.

We consider the following Lagrangian,
\begin{align}
\label{fRsclr}
S_{F(R)-\mathrm{scalar}} = \frac{1}{2\kappa^2} \int d^4 x \sqrt{-g} \left\{ f(R) - \frac{1}{2}\omega(\phi)\partial_\mu \phi \partial^\mu\phi - U\left( \phi \right)\right\}\, .
\end{align}
Then the equations corresponding to the Friedmann equations have the following forms,
\begin{align}
\label{fRsclrFeq1}
0 =&\, - f + 6\left(H^2 + \dot H\right) f_R - 36 \left( 4H^2 \dot H + H \ddot H\right) f_{RR} + \frac{1}{2}\omega(\phi){\dot \phi}^2 + U\left( \phi \right)\, , \\
\label{fRsclrFeq1B}
0=&\, f - 2 \left(\dot H + 3H^2\right) f_R + 12 \left( 8H^2 \dot H + 4 {\dot H}^2 + 6 H \ddot H + \dddot H\right) f_{RR} + 72\left( 4H\dot H + \ddot H\right)^2 f_{RRR}
+ \frac{1}{2}\omega(\phi){\dot \phi}^2 - U\left( \phi \right)\, .
\end{align}
Here we neglect the contributions from matter.
Eqs.~\eqref{fRsclrFeq1} and \eqref{fRsclrFeq1B} can be rewritten in the form
\begin{align}
\label{fRsclromega}
\omega(\phi){\dot \phi}^2 =&\, - 4 \dot H f_R + 12 \left( 4 H^2 \dot H - 4 {\dot H}^2 - 3 H \ddot H - \dddot H\right) f_{RR} - 72\left( 4H\dot H + \ddot H\right)^2 f_{RRR} \, , \\
\label{fRsclrV}
U\left( \phi \right) =&\, f + 2 \left( \dot H + 3 H^2\right) f_R + 6 \left( 20 H^2 \dot H + 4 {\dot H}^2 + 9 H \ddot H + \dddot H\right) f_{RR} + 36 \left( 4H\dot H + \ddot H\right)^2 f_{RRR} \, .
\end{align}
Further we rewrite Eqs.~\eqref{fRsclromega} and \eqref{fRsclrV} by using the $e$-folding number $N$ as follows,
\begin{align}
\label{fRsclromegaB}
\omega(\phi)H^2 {\phi'}^2
=&\, - 4 H H' f_R + 12 \left( 4 H^3 H' - 7 H^2 {H'}^2 - 3 H^3 H'' - H {H'}^3 - 4 H^2 H' H'' - H^3 H''' \right) f_{RR} \nonumber \\
&\, - 72 \left( 4H^2 H' + H {H'}^2 + H^2 H'' \right)^2 f_{RRR} \, , \\
\label{fRsclrVB}
U\left( \phi \right)
=&\, f + 2 \left( H H' + 3 H^2\right) f_R
+ 6 \left( 20 H^3 H' + 13 H^2 {H'}^2 + 4 H^2 H' H'' + H^3 H''' \right) f_{RR} \nonumber \\
&\, + 36 \left( 4H^2 H' + H {H'}^2 + H^2 H'' \right)^2 f_{RRR} \, .
\end{align}
Then if we consider a model, where $\omega(\phi)$ and $U\left( \phi \right)$ are given by a function $\Phi(\phi)$ as follows,
\begin{align}
\label{fRsclromegaPhi}
\omega(\phi)
=&\, - \frac{4 \Phi'(\phi)}{\Phi(\phi)} f_R \left( R \to 12 \Phi(\phi)^2 + 6\Phi(\phi)\Phi'(\phi) \right) \nonumber \\
&\, + 12 \left( 4 \Phi(\phi) \Phi'(\phi) - 7 \Phi'(\phi)^2 - 3 \Phi(\phi) \Phi''(\phi) - \frac{\Phi'(\phi)^3}{\Phi(\phi)} - 4 \Phi'(\phi) \Phi''(\phi) - \Phi(\phi) \Phi'''(\phi) \right) \nonumber \\
&\, \qquad \times f_{RR} \left( R \to 12 \Phi(\phi)^2 + 6\Phi(\phi)\Phi'(\phi) \right) \nonumber \\
&\, - 72 \left( 4 \Phi'(\phi) + \frac{\Phi'(\phi)^2}{\Phi(\phi)} + \Phi''(\phi) \right)^2 f_{RRR} \left( R \to 12 \Phi(\phi)^2 + 6\Phi(\phi)\Phi'(\phi) \right) \, , \\
\label{fRsclrVPhi}
U\left( \phi \right)
=&\, f \left( R \to 12 \Phi(\phi)^2 + 6\Phi(\phi)\Phi'(\phi) \right) + 2 \left( \Phi(\phi) \Phi'(\phi) + 3 \Phi(\phi)^2\right) f_R \left( R \to 12 \Phi(\phi)^2 + 6\Phi(\phi)\Phi'(\phi) \right) \nonumber \\
&\, + 6 \left( 20 \Phi(\phi)^3 \Phi'(\phi) + 13 \Phi(\phi)^2 \Phi'(\phi)^2 + 4 \Phi(\phi)^2 \Phi'(\phi) \Phi''(\phi) + \Phi(\phi)^3 \Phi'''(\phi) \right) \nonumber \\
&\, \qquad \times f_{RR} \left( R \to 12 \Phi(\phi)^2 + 6\Phi(\phi)\Phi'(\phi) \right) \nonumber \\
&\, + 36 \left( 4\Phi(\phi)^2 \Phi'(\phi) + \Phi(\phi) \Phi'(\phi)^2 + \Phi(\phi)^2 \Phi''(\phi) \right)^2 f_{RRR} \left( R \to 12 \Phi(\phi)^2 + 6\Phi(\phi)\Phi'(\phi) \right) \, ,
\end{align}
a solution is given by,
\begin{align}
\label{solB}
H=\Phi(N)\, , \quad \phi = N \, .
\end{align}
Especially if we choose,
\begin{align}
\label{mdl2sclr}
\Phi(\phi) = \frac{H_0}{\left(1 + A_0 \e^{4\alpha_0 \left( \phi - N_0 \right)}\right)^\frac{1}{2\alpha_0}}\, , \quad
\alpha_0=0.006\, , \quad A_0 \e^{4 \left( N_* - N_0 \right)} = 0.0005\, ,
\end{align}
we obtain the Hubble rate $H$ of Model I \eqref{mdl2} with $\phi=N$, which is consistent with the ACT observations.
We may also consider $f(R)$ as the Starobinsky model~\eqref{Strbnskymdl} or the model in \eqref{Rmmdl}.
The explicit forms of $\omega(\phi)$ and $U(\phi)$ are given by,
\begin{align}
\label{omgmdl1}
\omega\left( \phi \right)
=&\, \frac{8 A_0 \e^{4 \alpha_0 \left( \phi - N_0 \right)}}{1 + A_0 \e^{4 \alpha_0 \left( \phi - N_0 \right)}}
f_R \left( \frac{12{H_0}^2}{\left(1 + A_0 \e^{4\alpha_0 \left( \phi - N_0 \right)}\right)^{\frac{1}{\alpha_0}+1}} \right)
+ \frac{12{H_0}^2}{\left(1 + A_0 \e^{4\alpha_0 \left( \phi - N_0 \right)}\right)^\frac{1}{\alpha_0}}
\left\{ - \frac{8 A_0 \e^{4 \alpha_0 \left( \phi - N_0 \right)}}{1 + A_0 \e^{4 \alpha_0 \left( \phi - N_0 \right)}} \right. \nonumber \\
&\, + \frac{\left( 24\alpha_0 + 32{\alpha_0}^2 \right) A_0 \e^{4 \alpha_0 \left( \phi - N_0 \right)}}{\left( 1 + A_0 \e^{4 \alpha_0 \left( \phi - N_0 \right)} \right)^2}
 - \frac{\left(40- 32 \alpha_0 \right) {A_0}^2 \e^{8 \alpha_0 \left( \phi - N_0 \right)}}{\left( 1 + A_0 \e^{4 \alpha_0 \left( \phi - N_0 \right)} \right)^2} \nonumber \\
&\, \left. - \frac{\left( 64 {\alpha_0}^2 + 80 \alpha_0 \right){A_0}^2 \e^{8 \alpha_0 \left( \phi - N_0 \right)}}{\left( 1 + A_0 \e^{4 \alpha_0 \left( \phi - N_0 \right)} \right)^3}
+ \frac{\left(32 \alpha_0 + 48 \right) {A_0}^3 \e^{12 \alpha_0 \left( \phi - N_0 \right)}}{\left( 1 + A_0 \e^{4 \alpha_0 \left( \phi - N_0 \right)} \right)^3}
\right\} f_{RR} \left( \frac{12{H_0}^2}{\left(1 + A_0 \e^{4\alpha_0 \left( \phi - N_0 \right)}\right)^{\frac{1}{\alpha_0}+1}} \right) \nonumber \\
&\, - \frac{72{H_0}^2}{\left(1 + A_0 \e^{4\alpha_0 \left( \phi - N_0 \right)}\right)^\frac{1}{\alpha_0}} \left[
 - \frac{2 A_0 \e^{4 \alpha_0 \left( \phi - N_0 \right)}}{1 + A_0 \e^{4 \alpha_0 \left( \phi - N_0 \right)}}
 - \frac{8\alpha_0 A_0 \e^{4 \alpha_0 \left( \phi - N_0 \right)}}{\left( 1 + A_0 \e^{4 \alpha_0 \left( \phi - N_0 \right)} \right)^2} \right. \nonumber \\
&\, \left. + \frac{8 {A_0}^2 \e^{8 \alpha_0 \left( \phi - N_0 \right)}}{\left( 1 + A_0 \e^{4 \alpha_0 \left( \phi - N_0 \right)} \right)^2} \right]^2
f_{RRR} \left( \frac{12{H_0}^2}{\left(1 + A_0 \e^{4\alpha_0 \left( \phi - N_0 \right)}\right)^{\frac{1}{\alpha_0}+1}} \right) \, , \\
\label{Vmdl1}
U\left( \phi \right)
=&\, f \left( \frac{12{H_0}^2}{\left(1 + A_0 \e^{4\alpha_0 \left( \phi - N_0 \right)}\right)^{\frac{1}{\alpha_0}+1}} \right)
+ \frac{2{H_0}^2 \left( 3 + A_0 \e^{4\alpha_0 \left( \phi - N_0 \right)}\right)}{\left(1 + A_0 \e^{4\alpha_0 \left( \phi - N_0 \right)}\right)^{\frac{1}{\alpha_0} +1}}
f_R \left( \frac{12{H_0}^2}{\left(1 + A_0 \e^{4\alpha_0 \left( \phi - N_0 \right)}\right)^{\frac{1}{\alpha_0}+1}} \right) \nonumber \\
&\, + \frac{6{H_0}^4}{\left(1 + A_0 \e^{4\alpha_0 \left( \phi - N_0 \right)}\right)^\frac{2}{\alpha_0}}
\left[ - \frac{40 A_0 \e^{4 \alpha_0 \left( \phi - N_0 \right)}}{1 + A_0 \e^{4 \alpha_0 \left( \phi - N_0 \right)}}
 - \frac{32{\alpha_0}^2 A_0 \e^{4 \alpha_0 \left( \phi - N_0 \right)}}{\left( 1 + A_0 \e^{4 \alpha_0 \left( \phi - N_0 \right)} \right)^2}
+ \frac{\left(32 \alpha_0 + 52 \right){A_0}^2 \e^{8 \alpha_0 \left( \phi - N_0 \right)}}{\left( 1 + A_0 \e^{4 \alpha_0 \left( \phi - N_0 \right)} \right)^2}
\right. \nonumber \\
&\, \left.
+ \frac{\left( 64 {\alpha_0}^2 + 80 \alpha_0 \right){A_0}^2 \e^{8 \alpha_0 \left( \phi - N_0 \right)}}{\left( 1 + A_0 \e^{4 \alpha_0 \left( \phi - N_0 \right)} \right)^3}
 - \frac{\left(32 \alpha_0 + 40 \right) {A_0}^3 \e^{12 \alpha_0 \left( \phi - N_0 \right)}}{\left( 1 + A_0 \e^{4 \alpha_0 \left( \phi - N_0 \right)} \right)^3} \right]
f_{RR} \left( \frac{12{H_0}^2}{\left(1 + A_0 \e^{4\alpha_0 \left( \phi - N_0 \right)}\right)^{\frac{1}{\alpha_0}+1}} \right) \nonumber \\
&\, + \frac{{H_0}^6}{\left(1 + A_0 \e^{4\alpha_0 \left( \phi - N_0 \right)}\right)^\frac{3}{\alpha_0}}
\left[ - \frac{8 A_0 \e^{4 \alpha_0 \left( \phi - N_0 \right)}}{1 + A_0 \e^{4 \alpha_0 \left( \phi - N_0 \right)}}
- \frac{8\alpha_0 A_0 \e^{4 \alpha_0 \left( \phi - N_0 \right)}}{\left( 1 + A_0 \e^{4 \alpha_0 \left( \phi - N_0 \right)} \right)^2} \right. \nonumber \\
&\, \left. + \frac{8 {A_0}^2 \e^{8 \alpha_0 \left( \phi - N_0 \right)}}{\left( 1 + A_0 \e^{4 \alpha_0 \left( \phi - N_0 \right)} \right)^2}
\right]^2
f_{RRR} \left( \frac{12{H_0}^2}{\left(1 + A_0 \e^{4\alpha_0 \left( \phi - N_0 \right)}\right)^{\frac{1}{\alpha_0}+1}} \right) \, .
\end{align}
The derivations of Eqs.~\eqref{omgmdl1} and {Vmdl1} are given in \eqref{omgmdl1Ap} and {Vmdl1Ap}.

Even for Model II in \eqref{mdl5}, if we choose,
\begin{align}
\label{mdl5Phi}
\Phi = \frac{H_0}{\sqrt{1 + \alpha_0 \e^{\alpha_1 N} + \beta_0 \e^{4N}}}\, , \quad
\alpha_1 \sim 0.024 \, , \quad 6 \alpha_0 \alpha_1 \e^{\alpha_1 N_*}<0.038 \, ,
\end{align}
the Hubble rate in \eqref{mdl5} with $\phi=N$ is reproduced, and we obtain another model consistent with the ACT observations.
The explicit forms of $\omega(\phi)$ and $U(\phi)$ are given by, we find,
\begin{align}
\label{fRsclromegaPhimdl2}
\omega(\phi)
=&\, \frac{2\left( \alpha_0 \alpha_1 \e^{\alpha_1 \phi} + 4\beta_0 \e^{4\phi} \right)}{1 + \alpha_0 \e^{\alpha_1 \phi} + \beta_0 \e^{4\phi}}
f_R \left( \frac{12{H_0}^2}{1 + \alpha_0 \e^{\alpha_1 \phi} + \beta_0 \e^{4\phi}}
 - \frac{3{H_0}^2\left( \alpha_0 \alpha_1 \e^{\alpha_1 \phi} + 4\beta_0 \e^{4\phi} \right)}{\left(1 + \alpha_0 \e^{\alpha_1 \phi} + \beta_0 \e^{4\phi}\right)^2} \right) \nonumber \\
%%%%%%%%
&\, + 12 {H_0}^2 \left[
\frac{\alpha_0 \left( -4 \alpha_1 + 3 {\alpha_1}^2 + {\alpha_1}^3 \right)\e^{\alpha_1 \phi} + 96\beta_0 \e^{4\phi} }{2\left(1 + \alpha_0 \e^{\alpha_1 \phi} + \beta_0 \e^{4\phi}\right)^2} \right. \nonumber \\
&\, + \frac{ - 16\left( \alpha_0 \alpha_1 \e^{\alpha_1 \phi} + 4\beta_0 \e^{4\phi} \right)^2
+ 3\left( \alpha_0 \alpha_1 \e^{\alpha_1 \phi} + 4\beta_0 \e^{4\phi} \right)\left( \alpha_0 {\alpha_1}^2 \e^{\alpha_1 \phi} + 16\beta_0 \e^{4\phi} \right)
 - 2 \left( \alpha_0 {\alpha_1}^2 \e^{\alpha_1 \phi} + 16\beta_0 \e^{4\phi} \right)^2
}{4\left(1 + \alpha_0 \e^{\alpha_1 \phi} + \beta_0 \e^{4\phi}\right)^3} \nonumber \\
&\, \left. + \frac{19\left( \alpha_0 \alpha_1 \e^{\alpha_1 \phi} + 4\beta_0 \e^{4\phi} \right)^3}{8\left(1 + \alpha_0 \e^{\alpha_1 \phi} + \beta_0 \e^{4\phi}\right)^4} \right]
f_{RR} \left( \frac{12{H_0}^2}{1 + \alpha_0 \e^{\alpha_1 \phi} + \beta_0 \e^{4\phi}}
 - \frac{3{H_0}^2\left( \alpha_0 \alpha_1 \e^{\alpha_1 \phi} + 4\beta_0 \e^{4\phi} \right)}{\left(1 + \alpha_0 \e^{\alpha_1 \phi} + \beta_0 \e^{4\phi}\right)^2} \right) \nonumber \\
%%%%%%%%%%%%
&\, - 72 \left[ - \frac{\alpha_0 \left( 4 \alpha_1 + {\alpha_1}^2 \right) \e^{\alpha_1 \phi} + 32\beta_0 \e^{4\phi}}{2\left(1 + \alpha_0 \e^{\alpha_1 \phi} + \beta_0 \e^{4\phi}\right)}
+ \frac{3\left( \alpha_0 \alpha_1 \e^{\alpha_1 \phi} + 4\beta_0 \e^{4\phi} \right)^2}{4\left(1 + \alpha_0 \e^{\alpha_1 \phi} + \beta_0 \e^{4\phi}\right)^2} \right] \nonumber \\
&\, \times f_{RRR} \left( \frac{12{H_0}^2}{1 + \alpha_0 \e^{\alpha_1 \phi} + \beta_0 \e^{4\phi}}
 - \frac{3{H_0}^2\left( \alpha_0 \alpha_1 \e^{\alpha_1 \phi} + 4\beta_0 \e^{4\phi} \right)}{\left(1 + \alpha_0 \e^{\alpha_1 \phi} + \beta_0 \e^{4\phi}\right)^2} \right)
\, , \\
\label{fRsclrVPhimdl2}
U\left( \phi \right)
=&\, f \left( \frac{12{H_0}^2}{1 + \alpha_0 \e^{\alpha_1 \phi} + \beta_0 \e^{4\phi}}
 - \frac{3{H_0}^2\left( \alpha_0 \alpha_1 \e^{\alpha_1 \phi} + 4\beta_0 \e^{4\phi} \right)}{\left(1 + \alpha_0 \e^{\alpha_1 \phi} + \beta_0 \e^{4\phi}\right)^2} \right) \nonumber \\
&\, + {H_0}^2 \left( - \frac{\alpha_0 \alpha_1 \e^{\alpha_1 \phi} + 4\beta_0 \e^{4\phi}}{\left(1 + \alpha_0 \e^{\alpha_1 \phi} + \beta_0 \e^{4\phi}\right)^2}
+ \frac{6}{1 + \alpha_0 \e^{\alpha_1 \phi} + \beta_0 \e^{4\phi}} \right) \nonumber \\
&\, \times
f_R \left( \frac{12{H_0}^2}{1 + \alpha_0 \e^{\alpha_1 \phi} + \beta_0 \e^{4\phi}}
 - \frac{3{H_0}^2\left( \alpha_0 \alpha_1 \e^{\alpha_1 \phi} + 4\beta_0 \e^{4\phi} \right)}{\left(1 + \alpha_0 \e^{\alpha_1 \phi} + \beta_0 \e^{4\phi}\right)^2} \right) \nonumber \\
&\, + 6 {H_0}^4\left\{ - \frac{\alpha_0 \left( 20 \alpha_1 + {\alpha_1}^3 \right) \e^{\alpha_1 \phi} + 144\beta_0 \e^{4\phi}}{2\left(1 + \alpha_0 \e^{\alpha_1 \phi} + \beta_0 \e^{4\phi}\right)^3}
+ \frac{13\left( \alpha_0 \alpha_1 \e^{\alpha_1 \phi} + 4\beta_0 \e^{4\phi} \right)^2}{4\left(1 + \alpha_0 \e^{\alpha_1 \phi} + \beta_0 \e^{4\phi}\right)^4} \right. \nonumber \\
&\, + \frac{\left( \alpha_0 {\alpha_1}^2 \e^{\alpha_1 \phi} + 16\beta_0 \e^{4\phi} \right)^2}{2\left(1 + \alpha_0 \e^{\alpha_1 \phi} + \beta_0 \e^{4\phi}\right)^4}
+ \frac{5\left( \alpha_0 \alpha_1 \e^{\alpha_1 \phi} + 4\beta_0 \e^{4\phi} \right) \left( \alpha_0 {\alpha_1}^2 \e^{\alpha_1 \phi} + 16\beta_0 \e^{4\phi} \right)}
{4\left(1 + \alpha_0 \e^{\alpha_1 \phi} + \beta_0 \e^{4\phi}\right)^4} \nonumber \\
&\, \left. - \frac{9\left( \alpha_0 \alpha_1 \e^{\alpha_1 \phi} + 4\beta_0 \e^{4\phi} \right)^3}{4\left(1 + \alpha_0 \e^{\alpha_1 \phi} + \beta_0 \e^{4\phi}\right)^5} \right\}
f_{RR} \left( \frac{12{H_0}^2}{1 + \alpha_0 \e^{\alpha_1 \phi} + \beta_0 \e^{4\phi}}
 - \frac{3{H_0}^2\left( \alpha_0 \alpha_1 \e^{\alpha_1 \phi} + 4\beta_0 \e^{4\phi} \right)}{\left(1 + \alpha_0 \e^{\alpha_1 \phi} + \beta_0 \e^{4\phi}\right)^2} \right) \nonumber \\
&\, + \frac{36{H_0}^6}{1 + \alpha_0 \e^{\alpha_1 \phi} + \beta_0 \e^{4\phi}} \left[
- \frac{\alpha_0 \left(\alpha_1 + {\alpha_1}^2 \right) \e^{\alpha_1 \phi} + 20\beta_0 \e^{4\phi}}{2\left(1 + \alpha_0 \e^{\alpha_1 \phi} + \beta_0 \e^{4\phi}\right)^2}
+ \frac{3\left( \alpha_0 \alpha_1 \e^{\alpha_1 \phi} + 4\beta_0 \e^{4\phi} \right)^2}{4\left(1 + \alpha_0 \e^{\alpha_1 \phi} + \beta_0 \e^{4\phi}\right)^3}
\right]^2 \nonumber \\
&\, \times f_{RRR} \left( \frac{12{H_0}^2}{1 + \alpha_0 \e^{\alpha_1 \phi} + \beta_0 \e^{4\phi}}
 - \frac{3{H_0}^2\left( \alpha_0 \alpha_1 \e^{\alpha_1 \phi} + 4\beta_0 \e^{4\phi} \right)}{\left(1 + \alpha_0 \e^{\alpha_1 \phi} + \beta_0 \e^{4\phi}\right)^2} \right) \, .
\end{align}
In \eqref{fRsclromegaPhimdl2Ap} and \eqref{fRsclrVPhimdl2}, we derive \eqref{fRsclromegaPhimdl2} and \eqref{fRsclrVPhimdl2}.

Because $f(R)$ gravity includes a scalar mode, as in \eqref{JGRG23}, introducing a scalar field as in \eqref{fRsclr} yields two scalar modes.
There are several models including two scalar fields, as in the curvaton model~\cite{Enqvist:2001zp, Lyth:2001nq, Moroi:2001ct} (see also \cite{Lyth:2002my, Sasaki:2006kq, Mazumdar:2010sa}).
In the curvaton model, the curvaton dominantly creates the overall primordial density fluctuation and also contributes to the reheating.
One of the scalar modes in our model might take the role of the curvaton.

Thus, by adding the scalar field to $f(R)$ gravity, we have obtained two models consistent with the observations by ACT.
The first model, namely, Model I, is given by 
\begin{align}
\label{sumModelI}
H^2 =&\, \frac{{H_0}^2}{\left(1 + A_0 \e^{4\alpha_0 \left( N - N_0 \right)}\right)^\frac{1}{\alpha_0}}\, , \nonumber \\
S_{F(R)-\mathrm{scalar}} =&\, \int d^4 x \sqrt{-g} \left\{ f(R) - \frac{1}{2}\omega(\phi)\partial_\mu \phi \partial^\mu\phi - U\left( \phi \right)\right\}\, , \nonumber \\
\omega \left( \phi \right) =&\, \frac{8 A_0 \e^{4 \alpha_0 \left( \phi - N_0 \right)}}{1 + A_0 \e^{4 \alpha_0 \left( \phi - N_0 \right)}}
f_R \left( \frac{12{H_0}^2}{\left(1 + A_0 \e^{4\alpha_0 \left( \phi - N_0 \right)}\right)^{\frac{1}{\alpha_0}+1}} \right)
+ \frac{12{H_0}^2}{\left(1 + A_0 \e^{4\alpha_0 \left( \phi - N_0 \right)}\right)^\frac{1}{\alpha_0}}
\left\{ - \frac{8 A_0 \e^{4 \alpha_0 \left( \phi - N_0 \right)}}{1 + A_0 \e^{4 \alpha_0 \left( \phi - N_0 \right)}} \right. \nonumber \\
&\, + \frac{\left( 24\alpha_0 + 32{\alpha_0}^2 \right) A_0 \e^{4 \alpha_0 \left( \phi - N_0 \right)}}{\left( 1 + A_0 \e^{4 \alpha_0 \left( \phi - N_0 \right)} \right)^2}
 - \frac{\left(40- 32 \alpha_0 \right) {A_0}^2 \e^{8 \alpha_0 \left( \phi - N_0 \right)}}{\left( 1 + A_0 \e^{4 \alpha_0 \left( \phi - N_0 \right)} \right)^2} \nonumber \\
&\, \left. - \frac{\left( 64 {\alpha_0}^2 + 80 \alpha_0 \right){A_0}^2 \e^{8 \alpha_0 \left( \phi - N_0 \right)}}{\left( 1 + A_0 \e^{4 \alpha_0 \left( \phi - N_0 \right)} \right)^3}
+ \frac{\left(32 \alpha_0 + 48 \right) {A_0}^3 \e^{12 \alpha_0 \left( \phi - N_0 \right)}}{\left( 1 + A_0 \e^{4 \alpha_0 \left( \phi - N_0 \right)} \right)^3}
\right\} f_{RR} \left( \frac{12{H_0}^2}{\left(1 + A_0 \e^{4\alpha_0 \left( \phi - N_0 \right)}\right)^{\frac{1}{\alpha_0}+1}} \right) \nonumber \\
&\, - \frac{72{H_0}^2}{\left(1 + A_0 \e^{4\alpha_0 \left( \phi - N_0 \right)}\right)^\frac{1}{\alpha_0}} \left[
 - \frac{2 A_0 \e^{4 \alpha_0 \left( \phi - N_0 \right)}}{1 + A_0 \e^{4 \alpha_0 \left( \phi - N_0 \right)}}
 - \frac{8\alpha_0 A_0 \e^{4 \alpha_0 \left( \phi - N_0 \right)}}{\left( 1 + A_0 \e^{4 \alpha_0 \left( \phi - N_0 \right)} \right)^2} \right. \nonumber \\
&\, \left. + \frac{8 {A_0}^2 \e^{8 \alpha_0 \left( \phi - N_0 \right)}}{\left( 1 + A_0 \e^{4 \alpha_0 \left( \phi - N_0 \right)} \right)^2} \right]^2
f_{RRR} \left( \frac{12{H_0}^2}{\left(1 + A_0 \e^{4\alpha_0 \left( \phi - N_0 \right)}\right)^{\frac{1}{\alpha_0}+1}} \right) \, , \nonumber \\
U\left( \phi \right) =&\, f \left( \frac{12{H_0}^2}{\left(1 + A_0 \e^{4\alpha_0 \left( \phi - N_0 \right)}\right)^{\frac{1}{\alpha_0}+1}} \right)
+ \frac{2{H_0}^2 \left( 3 + A_0 \e^{4\alpha_0 \left( \phi - N_0 \right)}\right)}{\left(1 + A_0 \e^{4\alpha_0 \left( \phi - N_0 \right)}\right)^{\frac{1}{\alpha_0} +1}}
f_R \left( \frac{12{H_0}^2}{\left(1 + A_0 \e^{4\alpha_0 \left( \phi - N_0 \right)}\right)^{\frac{1}{\alpha_0}+1}} \right) \nonumber \\
&\, + \frac{6{H_0}^4}{\left(1 + A_0 \e^{4\alpha_0 \left( \phi - N_0 \right)}\right)^\frac{2}{\alpha_0}}
\left[ - \frac{40 A_0 \e^{4 \alpha_0 \left( \phi - N_0 \right)}}{1 + A_0 \e^{4 \alpha_0 \left( \phi - N_0 \right)}}
 - \frac{32{\alpha_0}^2 A_0 \e^{4 \alpha_0 \left( \phi - N_0 \right)}}{\left( 1 + A_0 \e^{4 \alpha_0 \left( \phi - N_0 \right)} \right)^2}
+ \frac{\left(32 \alpha_0 + 52 \right){A_0}^2 \e^{8 \alpha_0 \left( \phi - N_0 \right)}}{\left( 1 + A_0 \e^{4 \alpha_0 \left( \phi - N_0 \right)} \right)^2}
\right. \nonumber \\
&\, \left.
+ \frac{\left( 64 {\alpha_0}^2 + 80 \alpha_0 \right){A_0}^2 \e^{8 \alpha_0 \left( \phi - N_0 \right)}}{\left( 1 + A_0 \e^{4 \alpha_0 \left( \phi - N_0 \right)} \right)^3}
 - \frac{\left(32 \alpha_0 + 40 \right) {A_0}^3 \e^{12 \alpha_0 \left( \phi - N_0 \right)}}{\left( 1 + A_0 \e^{4 \alpha_0 \left( \phi - N_0 \right)} \right)^3} \right]
f_{RR} \left( \frac{12{H_0}^2}{\left(1 + A_0 \e^{4\alpha_0 \left( \phi - N_0 \right)}\right)^{\frac{1}{\alpha_0}+1}} \right) \nonumber \\
&\, + \frac{{H_0}^6}{\left(1 + A_0 \e^{4\alpha_0 \left( \phi - N_0 \right)}\right)^\frac{3}{\alpha_0}}
\left[ - \frac{8 A_0 \e^{4 \alpha_0 \left( \phi - N_0 \right)}}{1 + A_0 \e^{4 \alpha_0 \left( \phi - N_0 \right)}}
- \frac{8\alpha_0 A_0 \e^{4 \alpha_0 \left( \phi - N_0 \right)}}{\left( 1 + A_0 \e^{4 \alpha_0 \left( \phi - N_0 \right)} \right)^2} \right. \nonumber \\
&\, \left. + \frac{8 {A_0}^2 \e^{8 \alpha_0 \left( \phi - N_0 \right)}}{\left( 1 + A_0 \e^{4 \alpha_0 \left( \phi - N_0 \right)} \right)^2}
\right]^2
f_{RRR} \left( \frac{12{H_0}^2}{\left(1 + A_0 \e^{4\alpha_0 \left( \phi - N_0 \right)}\right)^{\frac{1}{\alpha_0}+1}} \right) \, , \nonumber \\
\alpha_0=&\, 0.006\, , \quad A_0 \e^{4 \left( N - N_0 \right)} = 0.0005\, ,
\end{align}
and Model II is given by,
\begin{align}
\label{sumModelII}
H^2 =&\, \frac{{H_0}^2}{1 + \alpha_0 \e^{\alpha_1 N} + \beta_0 \e^{4N}}\, , \nonumber \\
S_{F(R)-\mathrm{scalar}} =&\, \int d^4 x \sqrt{-g} \left\{ f(R) - \frac{1}{2}\omega(\phi)\partial_\mu \phi \partial^\mu\phi - U\left( \phi \right)\right\}\, , \nonumber \\
\omega \left( \phi \right) =&\, \frac{2\left( \alpha_0 \alpha_1 \e^{\alpha_1 \phi} + 4\beta_0 \e^{4\phi} \right)}{1 + \alpha_0 \e^{\alpha_1 \phi} + \beta_0 \e^{4\phi}}
f_R \left( \frac{12{H_0}^2}{1 + \alpha_0 \e^{\alpha_1 \phi} + \beta_0 \e^{4\phi}}
 - \frac{3{H_0}^2\left( \alpha_0 \alpha_1 \e^{\alpha_1 \phi} + 4\beta_0 \e^{4\phi} \right)}{\left(1 + \alpha_0 \e^{\alpha_1 \phi} + \beta_0 \e^{4\phi}\right)^2} \right) \nonumber \\
&\, + 12 {H_0}^2 \left[
\frac{\alpha_0 \left( -4 \alpha_1 + 3 {\alpha_1}^2 + {\alpha_1}^3 \right)\e^{\alpha_1 \phi} + 96\beta_0 \e^{4\phi} }{2\left(1 + \alpha_0 \e^{\alpha_1 \phi} + \beta_0 \e^{4\phi}\right)^2} \right. \nonumber \\
&\, + \frac{ - 16\left( \alpha_0 \alpha_1 \e^{\alpha_1 \phi} + 4\beta_0 \e^{4\phi} \right)^2
+ 3\left( \alpha_0 \alpha_1 \e^{\alpha_1 \phi} + 4\beta_0 \e^{4\phi} \right)\left( \alpha_0 {\alpha_1}^2 \e^{\alpha_1 \phi} + 16\beta_0 \e^{4\phi} \right)
 - 2 \left( \alpha_0 {\alpha_1}^2 \e^{\alpha_1 \phi} + 16\beta_0 \e^{4\phi} \right)^2
}{4\left(1 + \alpha_0 \e^{\alpha_1 \phi} + \beta_0 \e^{4\phi}\right)^3} \nonumber \\
&\, \left. + \frac{19\left( \alpha_0 \alpha_1 \e^{\alpha_1 \phi} + 4\beta_0 \e^{4\phi} \right)^3}{8\left(1 + \alpha_0 \e^{\alpha_1 \phi} + \beta_0 \e^{4\phi}\right)^4} \right]
f_{RR} \left( \frac{12{H_0}^2}{1 + \alpha_0 \e^{\alpha_1 \phi} + \beta_0 \e^{4\phi}}
 - \frac{3{H_0}^2\left( \alpha_0 \alpha_1 \e^{\alpha_1 \phi} + 4\beta_0 \e^{4\phi} \right)}{\left(1 + \alpha_0 \e^{\alpha_1 \phi} + \beta_0 \e^{4\phi}\right)^2} \right) \nonumber \\
&\, - 72 \left[ - \frac{\alpha_0 \left( 4 \alpha_1 + {\alpha_1}^2 \right) \e^{\alpha_1 \phi} + 32\beta_0 \e^{4\phi}}{2\left(1 + \alpha_0 \e^{\alpha_1 \phi} + \beta_0 \e^{4\phi}\right)}
+ \frac{3\left( \alpha_0 \alpha_1 \e^{\alpha_1 \phi} + 4\beta_0 \e^{4\phi} \right)^2}{4\left(1 + \alpha_0 \e^{\alpha_1 \phi} + \beta_0 \e^{4\phi}\right)^2} \right] \nonumber \\
&\, \times f_{RRR} \left( \frac{12{H_0}^2}{1 + \alpha_0 \e^{\alpha_1 \phi} + \beta_0 \e^{4\phi}}
 - \frac{3{H_0}^2\left( \alpha_0 \alpha_1 \e^{\alpha_1 \phi} + 4\beta_0 \e^{4\phi} \right)}{\left(1 + \alpha_0 \e^{\alpha_1 \phi} + \beta_0 \e^{4\phi}\right)^2} \right) \, , \nonumber \\
U\left( \phi \right) =&\, f \left( \frac{12{H_0}^2}{1 + \alpha_0 \e^{\alpha_1 \phi} + \beta_0 \e^{4\phi}}
 - \frac{3{H_0}^2\left( \alpha_0 \alpha_1 \e^{\alpha_1 \phi} + 4\beta_0 \e^{4\phi} \right)}{\left(1 + \alpha_0 \e^{\alpha_1 \phi} + \beta_0 \e^{4\phi}\right)^2} \right) \nonumber \\
&\, + {H_0}^2 \left( - \frac{\alpha_0 \alpha_1 \e^{\alpha_1 \phi} + 4\beta_0 \e^{4\phi}}{\left(1 + \alpha_0 \e^{\alpha_1 \phi} + \beta_0 \e^{4\phi}\right)^2}
+ \frac{6}{1 + \alpha_0 \e^{\alpha_1 \phi} + \beta_0 \e^{4\phi}} \right) \nonumber \\
&\, \times
f_R \left( \frac{12{H_0}^2}{1 + \alpha_0 \e^{\alpha_1 \phi} + \beta_0 \e^{4\phi}}
 - \frac{3{H_0}^2\left( \alpha_0 \alpha_1 \e^{\alpha_1 \phi} + 4\beta_0 \e^{4\phi} \right)}{\left(1 + \alpha_0 \e^{\alpha_1 \phi} + \beta_0 \e^{4\phi}\right)^2} \right) \nonumber \\
&\, + 6 {H_0}^4\left\{ - \frac{\alpha_0 \left( 20 \alpha_1 + {\alpha_1}^3 \right) \e^{\alpha_1 \phi} + 144\beta_0 \e^{4\phi}}{2\left(1 + \alpha_0 \e^{\alpha_1 \phi} + \beta_0 \e^{4\phi}\right)^3}
+ \frac{13\left( \alpha_0 \alpha_1 \e^{\alpha_1 \phi} + 4\beta_0 \e^{4\phi} \right)^2}{4\left(1 + \alpha_0 \e^{\alpha_1 \phi} + \beta_0 \e^{4\phi}\right)^4} \right. \nonumber \\
&\, + \frac{\left( \alpha_0 {\alpha_1}^2 \e^{\alpha_1 \phi} + 16\beta_0 \e^{4\phi} \right)^2}{2\left(1 + \alpha_0 \e^{\alpha_1 \phi} + \beta_0 \e^{4\phi}\right)^4}
+ \frac{5\left( \alpha_0 \alpha_1 \e^{\alpha_1 \phi} + 4\beta_0 \e^{4\phi} \right) \left( \alpha_0 {\alpha_1}^2 \e^{\alpha_1 \phi} + 16\beta_0 \e^{4\phi} \right)}
{4\left(1 + \alpha_0 \e^{\alpha_1 \phi} + \beta_0 \e^{4\phi}\right)^4} \nonumber \\
&\, \left. - \frac{9\left( \alpha_0 \alpha_1 \e^{\alpha_1 \phi} + 4\beta_0 \e^{4\phi} \right)^3}{4\left(1 + \alpha_0 \e^{\alpha_1 \phi} + \beta_0 \e^{4\phi}\right)^5} \right\}
f_{RR} \left( \frac{12{H_0}^2}{1 + \alpha_0 \e^{\alpha_1 \phi} + \beta_0 \e^{4\phi}}
 - \frac{3{H_0}^2\left( \alpha_0 \alpha_1 \e^{\alpha_1 \phi} + 4\beta_0 \e^{4\phi} \right)}{\left(1 + \alpha_0 \e^{\alpha_1 \phi} + \beta_0 \e^{4\phi}\right)^2} \right) \nonumber \\
&\, + \frac{36{H_0}^6}{1 + \alpha_0 \e^{\alpha_1 \phi} + \beta_0 \e^{4\phi}} \left[
- \frac{\alpha_0 \left(\alpha_1 + {\alpha_1}^2 \right) \e^{\alpha_1 \phi} + 20\beta_0 \e^{4\phi}}{2\left(1 + \alpha_0 \e^{\alpha_1 \phi} + \beta_0 \e^{4\phi}\right)^2}
+ \frac{3\left( \alpha_0 \alpha_1 \e^{\alpha_1 \phi} + 4\beta_0 \e^{4\phi} \right)^2}{4\left(1 + \alpha_0 \e^{\alpha_1 \phi} + \beta_0 \e^{4\phi}\right)^3}
\right]^2 \nonumber \\
&\, \times f_{RRR} \left( \frac{12{H_0}^2}{1 + \alpha_0 \e^{\alpha_1 \phi} + \beta_0 \e^{4\phi}}
 - \frac{3{H_0}^2\left( \alpha_0 \alpha_1 \e^{\alpha_1 \phi} + 4\beta_0 \e^{4\phi} \right)}{\left(1 + \alpha_0 \e^{\alpha_1 \phi} + \beta_0 \e^{4\phi}\right)^2} \right) \, , \nonumber \\
\alpha_1 \sim&\, 0.024 \, , 6 \alpha_0 \alpha_1 \e^{\alpha_1 N_*}<0.038 \, .
\end{align}
We may choose $f(R)$ to be that of the Starobinsky model~\eqref{Strbnskymdl} or the model in \eqref{Rmmdl}.

\section{Possible Curvaton Scenario}\label{SecVI}

As we mentioned, $f(R)$ gravity includes a scalar mode~\eqref{JGRG23}, and therefore, if we couple another scalar field as in \eqref{fRsclr}, we have two scalar modes as in the curvaton model~\cite{Enqvist:2001zp, Lyth:2001nq, Moroi:2001ct, Lyth:2002my, Sasaki:2006kq, Mazumdar:2010sa}.
The overall primordial density fluctuation can be dominantly created by the curvaton.
We consider the possibility that one of the scalar modes in our model might take the role of the curvaton.
In the curvaton scenario, in addition to the inflaton, there appears another scalar field, the curvaton.
During inflation, the curvaton is light, but after that the fluctuation becomes the curvature fluctuation, including the non-Gaussianity.

For the action~\eqref{fRsclr}, by the transformation~\eqref{JGRG22}, we obtain the Einstein frame action,
\begin{align}
\label{JGRG23sclr}
S_E =&\, \frac{1}{2\kappa^2}\int d^4 x \sqrt{-g} \left\{ R - \frac{3}{2}g^{\rho\sigma} \partial_\rho \sigma \partial_\sigma \sigma - V(\sigma)
 - \frac{1}{2}\e^\sigma \omega(\phi)\partial_\mu \phi \partial^\mu\phi - \e^{2\sigma} U\left( \phi \right)
\right\} \, ,\nonumber \\
V(\sigma) =&\, \e^\sigma g\left(\e^{-\sigma}\right)
 - \e^{2\sigma} f\left(g\left(\e^{-\sigma}\right)\right) = \frac{A}{f'(A)} - \frac{f(A)}{f'(A)^2}\, .
\end{align}
Then in the Einstein frame, we have Einstein's gravity coupled with two scalar fields.
We may define the effective potential $U_\mathrm{eff} \left( \sigma, \phi \right)$ by,
\begin{align}
\label{effptl}
U_\mathrm{eff} \left( \sigma, \phi \right) \equiv V(\sigma) + \e^{2\sigma} U\left( \phi \right) \, .
\end{align}
By the transformation~\eqref{JGRG22}, the metric \eqref{FLRW} in the spatially flat FLRW spacetime is also changed by
\begin{align}
\label{FLRWchng}
ds^2 = - dt^2 + a(t)^2 \sum_{i=1,2,3} \left( dx^i \right)^2 \ \Rightarrow \ d{s_\mathrm{E}}^2 = \e^\sigma ds^2 = - \e^\sigma dt^2 + \e^\sigma a(t)^2 \sum_{i=1,2,3} \left( dx^i \right)^2 \, .
\end{align}
If $\sigma$ only depends on the cosmological time $t$, we often like to redefine the cosmological time to a new one $\tilde t$ by defining,
\begin{align}
\label{Ect}
d\tilde t = \e^\frac{\sigma}{2} dt\, .
\end{align}
We should note, however, that the cosmological time in the Jordan frame, which corresponds to Eq.~\eqref{FLRW}, is physically observed time.
Therefore, we often use the cosmological time $t$ and corresponding $e$-folding number $N$ even in the Einstein
frame, although it is convenient to use the time coordinate $\tilde t$ when we consider the perturbation.
In terms of $\tilde t$, we obtain,
\begin{align}
\label{tlddrvtv}
\frac{d}{d\tilde t} = \e^{-\sigma}\frac{d}{dt}\, , \quad
\frac{d^2}{d\tilde t^2} = \e^{-2\sigma} \left( \frac{d^2}{dt^2} - \frac{d\sigma}{dt} \frac{d}{dt} \right)\, , \quad \mbox{etc.}
\end{align}
We use the formulae in \eqref{tlddrvtv} when we consider the process in the Einstein frame.

%$\overset{\circ}{x}$ $\overset{\circ\circ}{x}$

By using the action~\eqref{JGRG23sclr}, we define the metric $G_{IJ}$ $\left( I,J=\sigma, \phi\right.$ or $\left. \left( \Phi^I \right) = \left(\sigma,\phi \right) \right)$ in the scalar fields as follows,
\begin{align}
\label{sclrmtrc}
G_{\sigma\sigma} = 3\, , \quad G_{\phi\phi}=\e^\sigma \omega(\phi)\, , \quad G_{\sigma\phi}=G_{\phi\sigma}=0\, .
\end{align}
In terms of $G_{IJ}$, we define the speed $v$ of the scalar field space as,
\begin{align}
\label{sp}
v \equiv \sqrt{G_{\sigma\sigma} {\overset{\circ}{\sigma}}^2 + G_{\phi\phi} {\overset{\circ}{\phi}}^2}
= \sqrt{ 3{\overset{\circ}{\sigma}}^2 + \e^\sigma \omega(\phi) {\overset{\circ}{\phi}}^2} \, .
\end{align}
Here $\overset{\circ}{\sigma} \equiv \frac{d\sigma}{d\tilde t}$.
The unit vector $e_t^I$ tangent to the orbit of the scalar field space is given by,
\begin{align}
\label{tv}
\left( e_t^I \right) = \left( \frac{\overset{\circ}{\sigma}}{v}, \frac{\overset{\circ}{\phi}}{v} \right) \, .
\end{align}
The direction of $e_t^I$ is called the adiabatic direction.
When we consider the perturbation of two scalar fields $\left(\Phi^I\right) = \left(\sigma,\phi \right)$, the perturbation of the two scalar fields in the adiabatic direction, which is called the adiabatic mode, generates the fluctuation of the spatial curvature in the comoving time slice.
The fluctuation is called the curvature mode.

We may also define a unit vector $e_p^I$ perpendicular to $e_t^I$,
\begin{align}
\label{pv}
\sum_{I,J=\sigma,\phi} G_{IJ}e_t^I e_p^J = 0 \, , \quad \sum_{I,J=\sigma,\phi} G_{IJ}e_p^I e_p^J = 1\, ,
\end{align}
or explicitly,
\begin{align}
\label{pv2}
\left( e_p^I \right) = \frac{1}{\sqrt{\frac{{\overset{\circ}{\phi}}^2}{G_{\sigma\sigma}} +\frac{{\overset{\circ}{\sigma}}^2}{G_{\phi\phi}}}}\left( \frac{\overset{\circ}{\phi}}{G_{\sigma\sigma}}, -\frac{\overset{\circ}{\sigma}}{G_{\phi\phi}} \right)
= \frac{1}{\sqrt{\frac{{\overset{\circ}{\phi}}^2}{3} +\frac{{\overset{\circ}{\sigma}}^2}{\e^\sigma \omega(\phi)}}}
\left( \frac{\overset{\circ}{\phi}}{3}, -\frac{\overset{\circ}{\sigma}}{\e^\sigma \omega(\phi)} \right) \, .
\end{align}
The direction of $e_p^I$ is called an entropy direction.
The fluctuation in the entropy direction, which is called the entropy mode, does not contribute to the fluctuation of the curvature.
Such a fluctuation is called an isocurvature mode.
In the curvaton scenario, the direction of the adiabatic direction expressed by \eqref{tv} changes after inflation and therefore the fluctuation in the entropy direction in the inflation epoch is converted to the fluctuation in the adiabatic direction, which contributes to the curvature mode.
The rotation of the direction is described by a tune-rate vector, which is given later in \eqref{trv} or \eqref{trv2}.

For the metric~\eqref{sclrmtrc} in the field space, the inverse metric is given by,
\begin{align}
\label{sclrmtrcB}
G^{\sigma\sigma} = \frac{1}{3}\, , \quad G^{\phi\phi}=\frac{\e^{-\sigma}}{\omega(\phi)}\, , \quad G^{\sigma\phi}=G^{\phi\sigma}=0\, .
\end{align}
Then the Levi-Civita connections in the field space are given by,
\begin{align}
\label{fscnnctn}
\Gamma^\sigma_{\sigma\sigma}=\Gamma^\sigma_{\sigma\phi} = \Gamma^\sigma_{\phi\sigma}=\Gamma^\phi_{\sigma\sigma}= 0 \, , \quad
\Gamma^\sigma_{\phi\phi} = - \frac{1}{6} \e^\sigma \omega(\phi) \, , \quad
\Gamma^\phi_{\sigma\phi} = \Gamma^\phi_{\phi\sigma} = \frac{1}{2} \, , \quad
\Gamma^\phi_{\phi\phi} = \frac{\omega'(\phi)}{2\omega(\phi)} \, .
\end{align}
By using the metric $G_{IJ}$ of the fields space in \eqref{sclrmtrc}, we can rewrite the action in \eqref{JGRG23sclr} as follows,
\begin{align}
\label{JGRG23sclrfsmtrc}
S_E =&\, \frac{1}{2\kappa^2}\int d^4 x \sqrt{-g} \left\{ R - \frac{1}{2}\sum_{I,J=\sigma, \phi} G_{IJ} \left( \Phi^K \right) g^{\rho\sigma} \partial_\rho \Phi^I \partial_\sigma \Phi^J
 - U_\mathrm{eff} \left( \Phi^I \right) \right\} \, .
\end{align}
By the variation of the action \eqref{JGRG23sclrfsmtrc} with respect to $\Phi^I$, we obtain the following equation,
\begin{align}
\label{eq}
0 =&\, - \frac{1}{2} \sum_{J,K=\sigma,\phi} G_{JK,I} g^{\rho\sigma} \partial_\rho \Phi^J \partial_\sigma \Phi^K
+ \frac{1}{\sqrt{-g}} \partial_\rho \left( \sum_{J=\sigma,\phi} G_{IJ} \sqrt{-g} g^{\rho\sigma} \partial_\sigma \Phi^J \right) - U_{\mathrm{eff}, I} \nonumber \\
=&\, - \frac{1}{2} \sum_{J,K=\sigma,\phi} G_{JK,I} g^{\rho\sigma} \partial_\rho \Phi^J \partial_\sigma \Phi^K
+ \left( \sum_{J,K=\sigma,\phi} G_{IJ,K} g^{\rho\sigma} \partial_\rho \Phi^K \partial_\sigma \Phi^J \right)
+ \sum_{J=\sigma,\phi} G_{IJ} \frac{1}{\sqrt{-g}} \partial_\rho \left( \sqrt{-g} g^{\rho\sigma} \partial_\sigma \Phi^J \right) \nonumber \\
&\, - U_{\mathrm{eff}, I} \, .
\end{align}
Here $G_{JK,I} = \frac{\partial G_{JK}}{\partial \Phi^I}$ and $U_{\mathrm{eff}, I} \equiv \frac{\partial U_\mathrm{eff}}{\partial \Phi^I}$.
Eq.~\eqref{eq} can be further rewritten as,
\begin{align}
\label{eq2}
0 =&\, \frac{1}{\sqrt{-g}} \partial_\rho \left( \sqrt{-g} g^{\rho\sigma} \partial_\sigma \Phi^I \right)
+ \frac{1}{2} \sum_{J,K,L=\sigma,\phi} G^{IL} \left( G_{LJ,K} + G_{LK,J} - G_{JK,L} \right) g^{\rho\sigma} \partial_\rho \Phi^K \partial_\sigma \Phi^J
 - \sum_{J=\sigma,\phi} G^{IJ} U_{\mathrm{eff}, J} \nonumber \\
=&\, \frac{1}{\sqrt{-g}} \partial_\rho \left( \sqrt{-g} g^{\rho\sigma} \partial_\sigma \Phi^I \right)
+ \sum_{J,K=\sigma,\phi} \Gamma^I_{JK} g^{\rho\sigma} \partial_\rho \Phi^K \partial_\sigma \Phi^J
 - \sum_{J=\sigma,\phi} G^{IJ} U_{\mathrm{eff}, J} \, .
\end{align}
We define a tune-rate vector $\left( \omega^I \right)$ by,
\begin{align}
\label{trv}
\omega^I = D_{\tilde t} e_t^I \equiv {\overset{\circ}{e}}_t^I + \sum_{J,K=\sigma.\phi} \Gamma^I_{JK} \Phi^J e_t^K \, ,
\end{align}
that is,
\begin{align}
\label{trv2}
\left( \omega^I \right) = \left( \frac{\overset{\circ\circ}{\sigma}}{v} - \frac{\overset{\circ}{\sigma} \dot v}{v^2} - \frac{\e^\sigma \omega(\phi) {\overset{\circ}{\phi}}^2}{6v},
\frac{\overset{\circ\circ}{\phi}}{v} - \frac{\overset{\circ}{\phi} \dot v}{v^2} + \frac{\overset{\circ}{\sigma}\overset{\circ}{\phi}}{v} + \frac{\omega'(\phi) {\overset{\circ}{\phi}}^2}{\omega(\phi)v} \right)\, .
\end{align}
The tune-rate vector $\left( \omega^I \right)$ expresses the rotation of the directions corresponding to the adiabatic direction~\eqref{tv} and the entropy or isocurvature direction.
Therefore, if the tune-rate vector $\left( \omega^I \right)$ is non-trivial, the fluctuation in the entropy or isocurvature direction is converted into the curvature fluctuation.

In the case of the Starobinsky model~\eqref{Strbnskymdl}, for the Model I~\eqref{sumModelI}, $\omega(\phi)$ and $V(\phi)$ have the following forms,
\begin{align}
\label{sumModelI0}
\omega \left( \phi \right)
=&\, \frac{8 A_0 \e^{4 \alpha_0 \left( \phi - N_0 \right)}}{1 + A_0 \e^{4 \alpha_0 \left( \phi - N_0 \right)}}
+ \frac{24\alpha {H_0}^2}{\left(1 + A_0 \e^{4\alpha_0 \left( \phi - N_0 \right)}\right)^{\frac{1}{\alpha_0}+3}}
\left\{ 8A_0
+ \left(- 8 + 32 A_0 + 32{\alpha_0}^2 \right) A_0 \e^{4 \alpha_0 \left( \phi - N_0 \right)} \right. \nonumber \\
&\, \left. + \left(- 56 - 24\alpha_0 - 32{\alpha_0}^2 \right) {A_0}^2 \e^{8 \alpha_0 \left( \phi - N_0 \right)}
+ 64 \alpha_0 {A_0}^3 \e^{12 \alpha_0 \left( \phi - N_0 \right)}
\right\} \, , \nonumber \\
%%%%%%%%%%%
U\left( \phi \right)
=&\, \frac{2{H_0}^2 \left( 9 + A_0 \e^{4\alpha_0 \left( \phi - N_0 \right)}\right)}{\left(1 + A_0 \e^{4\alpha_0 \left( \phi - N_0 \right)}\right)^{\frac{1}{\alpha_0} +1}}
+ \frac{96 \alpha {H_0}^4}{\left(1 + A_0 \e^{4\alpha_0 \left( \phi - N_0 \right)}\right)^{\frac{2}{\alpha_0}+ 3}}
\left\{ 6 + \left( - 3 - 8 {\alpha_0}^2 \right) A_0 \e^{4 \alpha_0 \left( \phi - N_0 \right)} \right. \nonumber \\
&\, \left. + \left( - 6 + 28 \alpha_0 + 8 {\alpha_0}^2 \right) {A_0}^2 \e^{8 \alpha_0 \left( \phi - N_0 \right)}
 - 7 {A_0}^3 \e^{12 \alpha_0 \left( \phi - N_0 \right)} \right\} \, .
\end{align}
The derivation of \eqref{sumModelI0} is given in \eqref{sumModelI0Ap}.
Then we obtain,
\begin{align}
\label{circsgmph}
\overset{\circ}{\sigma}
= \frac{96\alpha \alpha_0 A_0 {H_0}^3\left(\frac{1}{\alpha_0}+1\right)\e^{4\alpha_0 \left( \phi - N_0 \right)}}{\left(1 + A_0 \e^{4\alpha_0 \left( \phi - N_0 \right)}\right)^{\frac{3}{2\alpha_0}+2}}
\, , \quad
\overset{\circ}{\phi}
= \frac{H_0\left( 1 + \frac{24\alpha{H_0}^2}{\left(1 + A_0 \e^{4\alpha_0 \left( \phi - N_0 \right)}\right)^{\frac{1}{\alpha_0}+1}} \right)}
{\left(1 + A_0 \e^{4\alpha_0 \left( N - N_0 \right)}\right)^\frac{1}{2\alpha_0}}\, .
\end{align}
The derivation of \eqref{circsgmph} is given in \eqref{circsgmphAp}.
We also find that $v^2$ is given by
\begin{align}
\label{vsqrt}
v^2
=&\, \frac{27648\alpha^2 {\alpha_0}^2 {A_0}^2 {H_0}^6\left(\frac{1}{\alpha_0}+1\right)^2 \e^{8\alpha_0 \left( \phi - N_0 \right)}}
{\left(1 + A_0 \e^{4\alpha_0 \left( \phi - N_0 \right)}\right)^{\frac{6}{2\alpha_0}+4}} \nonumber \\
&\, +\left[ \frac{8 A_0 \e^{4 \alpha_0 \left( \phi - N_0 \right)}}{2\kappa^2 \left( 1 + A_0 \e^{4 \alpha_0 \left( \phi - N_0 \right)} \right)}
+ \frac{12\alpha {H_0}^2}{\kappa^2 \left(1 + A_0 \e^{4\alpha_0 \left( \phi - N_0 \right)}\right)^{\frac{1}{\alpha_0}+3}}
\left\{ 8A_0
+ \left(- 8 + 32 A_0 + 32{\alpha_0}^2 \right) A_0 \e^{4 \alpha_0 \left( \phi - N_0 \right)} \right. \right. \nonumber \\
&\, \left. \left. + \left(- 56 - 24\alpha_0 - 32{\alpha_0}^2 \right) {A_0}^2 \e^{8 \alpha_0 \left( \phi - N_0 \right)}
+ 64 \alpha_0 {A_0}^3 \e^{12 \alpha_0 \left( \phi - N_0 \right)} \right\} \right]
%\nonumber \\
%&\, \times
\frac{{H_0}^2\left( 1 + \frac{24\alpha{H_0}^2}{\left(1 + A_0 \e^{4\alpha_0 \left( \phi - N_0 \right)}\right)^{\frac{1}{\alpha_0}+1}} \right)}
{\left(1 + A_0 \e^{4\alpha_0 \left( N - N_0 \right)}\right)^\frac{1}{\alpha_0}} \, .
\end{align}
When $\phi=N\to -\infty$, $\overset{\circ}{\sigma}$ and $\overset{\circ}{\phi}$ behaves as,
\begin{align}
\label{circsgmphNmnsinf}
\overset{\circ}{\sigma}
\to &\, - 96\alpha \alpha_0 A_0 {H_0}^3\left(\frac{1}{\alpha_0}+1\right)\e^{4\alpha_0 \left( N - N_0 \right)} \, , \nonumber \\
\overset{\circ}{\phi} \to &\, H_0\left(1 + 24\alpha{H_0}^2\right) \, .
\end{align}
Therefore, we find $\left| \overset{\circ}{\sigma} \right| \ll \left| \overset{\circ}{\phi}\right|$, which tells us that $\phi$ plays the role of the inflaton and $\sigma$ could play the role of the curvaton.
On the other hand, at the end of the inflation defined by \eqref{einf}, that is, when $A_0 \e^{4 \alpha_0 \left(
N - N_0 \right)}=1$, the values of $\overset{\circ}{\sigma}$ and $\overset{\circ}{\phi}$ are given by,
\begin{align}
\label{circsgmphend}
\overset{\circ}{\sigma} = - \frac{96\alpha \alpha_0 {H_0}^3\left(\frac{1}{\alpha_0}+1\right)}{2^{\frac{3}{2\alpha_0}+2}} \, , \quad
\overset{\circ}{\phi} = \frac{H_0}{2^\frac{1}{2\alpha_0}}\left( 1 + \frac{24\alpha{H_0}^2}{2^{\frac{1}{\alpha_0}+1}} \right)
\, .
\end{align}
If $\alpha$ is large enough, $\overset{\circ}{\sigma}$ cannot be neglected compared with $\overset{\circ}{\phi}$, which tells us that the tune-rate vector $\omega^I$ in \eqref{trv} plays a non-trivial role for the fluctuation in the early Universe after inflation.
The perturbation of $\sigma$ in the isocurvature direction can be transformed to the curvature perturbation by the tune-rate vector $\omega^I$ as in \eqref{circsgmphend}.

In order to show that the scalar mode $\sigma$ can play the role of the curvaton, we need more detailed checks for the complicated model in \eqref{sumModelI0}, which may require numerical calculations.

We now investigate the curvaton scenario in more detail.
In order to investigate the primordial fluctuations, we consider the perturbation,
\begin{align}
\label{prtbtn}
\left( \Phi^I \right) = \left( \sigma, \phi \right) \ \Rightarrow \
\left( \Phi^I + \delta\Phi^I \right) = \left( \sigma + \delta\sigma, \phi + \delta \phi \right) \, .
\end{align}
Then Eq.~\eqref{eq2} has the following form,
\begin{align}
\label{eq2prtbtn}
0 =&\, \frac{1}{\sqrt{-g}} \partial_\rho \left( \sqrt{-g} g^{\rho\sigma} \partial_\sigma \delta \Phi^I \right)
+ \sum_{J,K,L=\sigma,\phi} \Gamma^I_{JK,L} g^{\rho\sigma} \partial_\rho \Phi^K \partial_\sigma \Phi^J \delta \Phi^L
+ \sum_{J,K=\sigma,\phi} \Gamma^I_{JK} g^{\rho\sigma} \partial_\rho \Phi^K \partial_\sigma \delta \Phi^J \nonumber \\
&\, - \sum_{J,K=\sigma,\phi} G^{IJ}_{,K} U_{\mathrm{eff}, J} \delta \Phi^K
 - \sum_{J,K=\sigma,\phi} G^{IJ} U_{\mathrm{eff}, JK} \delta \Phi^K \, .
\end{align}
In the Einstein frame metric~\eqref{FLRWchng}, instead of $t$ or $\tilde t$, we use the $e$-folding number $N$ in the background in the Jordan frame.
By the transformation~\eqref{JGRG22}, the metric \eqref{FLRW} in the spatially flat FLRW spacetime is also changed by
\begin{align}
\label{FLRWchngB}
d{s_\mathrm{E}}^2 =- \frac{\e^\sigma}{H^2} dN^2 + \e^\sigma a(t)^2 \sum_{i=1,2,3} \left( dx^i \right)^2 \, .
\end{align}
Here $H$ is the Hubble rate given in \eqref{mdl2} in the Jordan frame.
We rewrote the metric \eqref{FLRWchng} in terms of $N$ instead of the time coordinate $t$ or $\tilde t$.
We now assume,
\begin{align}
\label{adassmptn}
\delta \Phi^I = \Phi^I_0 \e^{-i \Omega(N) N + i \bm{k}\cdot\bm{r}}\, .
\end{align}
Here $\left( \bm{r} \right) = \left( x^1, x^2, x^3 \right)$ and due to the translational invariance in the spatial directions, $\bm{k}$, which is a spatial three-dimensional vector $\left( \bm{k} \right) = \left( k_1, k_2, k_3 \right)$, is a constant vector.
We may also assume the Bunch-Davies vacuum as initial conditions.

Although $\delta\Phi^I$ is not a gauge-invariant quantity without introducing the contribution from the scalar perturbation of the metric, we may assume a spatially flat gauge where we choose the time coordinate so that the scalar perturbation of the metric vanishes.
We also use an adiabatic approximation, assuming $\left| \Omega(N) \right| \gg 1$, although this approximation cannot be applied when the horizon crossing occurs.
Then Eq.~\eqref{eq2prtbtn} can be rewritten as,
\begin{align}
\label{eq2prtbtn2}
0 =&\, \e^{-2\sigma} a^{-3} H \left( \frac{\e^{2\sigma} a^3}{H} H^2 \e^{-\sigma} \Omega(N)^2 - i \frac{ d \left( \frac{\e^{2\sigma} a^3}{H} H^2 \e^{-\sigma} \right)}{dN} \Omega(N)
- \frac{\e^{2\sigma} a^3}{H} \e^{-\sigma} a^{-2} k^2 \right) \delta \Phi^I \nonumber \\
&\, - \sum_{J,K,L=\sigma,\phi} ijuk \e^{-\sigma} H^2 \frac{d\Phi^K}{dN} \frac{d\Phi^J}{dN} \delta \Phi^L
+ i\Omega(N) \sum_{J,K=\sigma,\phi} \Gamma^I_{JK} \e^{-\sigma} H^2 \frac{d \Phi^K}{dN} \delta \Phi^J \nonumber \\
&\, - \sum_{J,K=\sigma,\phi} G^{IJ}_{,K} U_{\mathrm{eff}, J} \delta \Phi^K
 - \sum_{J,K=\sigma,\phi} G^{IJ} U_{\mathrm{eff}, JK} \delta \Phi^K \nonumber \\
=&\, \left\{ H^2 \e^{-\sigma} \Omega(N)^2 - i \left( \frac{d\e^\sigma}{dN} H^2 \e^{-2\sigma} + 3 H^2 \e^{-\sigma} + HH' \e^{-\sigma} \right) \Omega(N) - \e^{-\sigma} a^{-2} k^2 \right\} \delta \Phi^I \nonumber \\
&\, - \sum_{J,K,L=\sigma,\phi} \Gamma^I_{JK,L} \e^{-\sigma} H^2 \frac{d\Phi^K}{dN} \frac{d\Phi^J}{dN} \delta \Phi^L
+ i\Omega(N) \sum_{J,K=\sigma,\phi} \Gamma^I_{JK} \e^{-\sigma} H^2 \frac{d \Phi^K}{dN} \delta \Phi^J \nonumber \\
&\, - \sum_{J,K=\sigma,\phi} G^{IJ}_{,K} U_{\mathrm{eff}, J} \delta \Phi^K
 - \sum_{J,K=\sigma,\phi} G^{IJ} U_{\mathrm{eff}, JK} \delta \Phi^K \, .
\end{align}
In order that Eq.~\eqref{eq2prtbtn2} has a non-trivial solution for $\delta\Phi^I$, the following equation must be satisfied,
\begin{align}
\label{det}
0 =&\, \left[ H^2 \e^{-\sigma} \Omega(N)^2 - i \left( \frac{d\e^\sigma}{dN} H^2 \e^{-2\sigma} + 3 H^2 \e^{-\sigma} + HH' \e^{-\sigma} \right) \Omega(N) - \e^{-\sigma} a^{-2} k^2 \right. \nonumber \\
&\, \left. - \sum_{J,K=\sigma,\phi} \Gamma^1_{JK,1} \e^{-\sigma} H^2 \frac{d\Phi^K}{dN} \frac{d\Phi^J}{dN}
+ \sum_{J=\sigma,\phi} \Gamma^1_{1J} \left( i\Omega(N) \e^{-\sigma} H^2 \frac{d \Phi^J}{dN}
 - G^{1J}_{,1} U_{\mathrm{eff}, J} - G^{1J} U_{\mathrm{eff}, J1} \right) \right] \nonumber \\
&\, \times \left[ H^2 \e^{-\sigma} \Omega(N)^2 - i \left( \frac{d\e^\sigma}{dN} H^2 \e^{-2\sigma} + 3 H^2 \e^{-\sigma} + HH' \e^{-\sigma} \right) \Omega(N) - \e^{-\sigma} a^{-2} k^2 \right. \nonumber \\
&\, \left. - \sum_{J,K=\sigma,\phi} \Gamma^2_{JK,2} \e^{-\sigma} H^2 \frac{d\Phi^K}{dN} \frac{d\Phi^J}{dN}
+ \sum_{J=\sigma,\phi} \Gamma^2_{2J} \left( i\Omega(N) \e^{-\sigma} H^2 \frac{d \Phi^J}{dN}
 - G^{2J}_{,2} U_{\mathrm{eff}, J} - G^{2J} U_{\mathrm{eff}, J2} \right) \right] \nonumber \\
&\, - \left[ - \sum_{J,K=\sigma,\phi} \Gamma^1_{JK,2} \e^{-\sigma} H^2 \frac{d\Phi^K}{dN} \frac{d\Phi^J}{dN}
+ \sum_{J=\sigma,\phi} \Gamma^1_{2J} \left( i\Omega(N) \e^{-\sigma} H^2 \frac{d \Phi^J}{dN}
 - G^{1J}_{,2} U_{\mathrm{eff}, J} - G^{1J} U_{\mathrm{eff}, J2} \right) \right] \nonumber \\
&\, \times \left[ - \sum_{J,K=\sigma,\phi} \Gamma^2_{JK,1} \e^{-\sigma} H^2 \frac{d\Phi^K}{dN} \frac{d\Phi^J}{dN}
+ \sum_{J=\sigma,\phi} \Gamma^2_{1J} \left( i\Omega(N) \e^{-\sigma} H^2 \frac{d \Phi^J}{dN}
 - G^{2J}_{,1} U_{\mathrm{eff}, J} - G^{2J} U_{\mathrm{eff}, J1} \right) \right] \, .
\end{align}
Eq.~\eqref{det} is a fourth-degree algebraic equation in $\omega$ and yields the dispersion relation $\omega = \omega(k)$.
When Eq.~\eqref{det} is satisfied, the ratio of $\delta \Phi^1 = \delta \sigma$ and $\delta \Phi^2 = \delta \phi$ is given by,
\begin{align}
\label{ratio}
\delta \Phi^1 : \delta \Phi^2
=&\, \left[ H^2 \e^{-\sigma} \Omega(N)^2 - i \left( \frac{d\e^\sigma}{dN} H^2 \e^{-2\sigma} + 3 H^2 \e^{-\sigma} + HH' \e^{-\sigma} \right) \Omega(N) - \e^{-\sigma} a^{-2} k^2 \right. \nonumber \\
&\, \left. - \sum_{J,K=\sigma,\phi} \Gamma^2_{JK,2} \e^{-\sigma} H^2 \frac{d\Phi^K}{dN} \frac{d\Phi^J}{dN}
+ \sum_{J=\sigma,\phi} \Gamma^2_{2J} \left( i\Omega(N) \e^{-\sigma} H^2 \frac{d \Phi^J}{dN}
 - G^{2J}_{,2} U_{\mathrm{eff}, J} - G^{2J} U_{\mathrm{eff}, J2} \right) \right] \nonumber \\
:&\, \left[ \sum_{J,K=\sigma,\phi} \Gamma^2_{JK,1} \e^{-\sigma} H^2 \frac{d\Phi^K}{dN} \frac{d\Phi^J}{dN}
 - \sum_{J=\sigma,\phi} \Gamma^2_{1J} \left( i\Omega(N) \e^{-\sigma} H^2 \frac{d \Phi^J}{dN}
 - G^{2J}_{,1} U_{\mathrm{eff}, J} - G^{2J} U_{\mathrm{eff}, J1} \right) \right] \nonumber \\
\mbox{or} & \nonumber \\
\delta \Phi^1 : \delta \Phi^2
=&\, \left[ \sum_{J,K=\sigma,\phi} \Gamma^1_{JK,2} \e^{-\sigma} H^2 \frac{d\Phi^K}{dN} \frac{d\Phi^J}{dN}
 - \sum_{J=\sigma,\phi} \Gamma^1_{2J} \left( i\Omega(N) \e^{-\sigma} H^2 \frac{d \Phi^J}{dN}
 - G^{1J}_{,2} U_{\mathrm{eff}, J} - G^{1J} U_{\mathrm{eff}, J2} \right) \right] \nonumber \\
:&\, \left[ H^2 \e^{-\sigma} \Omega(N)^2 - i \left( \frac{d\e^\sigma}{dN} H^2 \e^{-2\sigma} + 3 H^2 \e^{-\sigma} + HH' \e^{-\sigma} \right) \Omega(N) - \e^{-\sigma} a^{-2} k^2 \right. \nonumber \\
&\, \left. - \sum_{J,K=\sigma,\phi} \Gamma^1_{JK,1} \e^{-\sigma} H^2 \frac{d\Phi^K}{dN} \frac{d\Phi^J}{dN}
+ \sum_{J=\sigma,\phi} \Gamma^1_{1J} \left( i\Omega(N) \e^{-\sigma} H^2 \frac{d \Phi^J}{dN}
 - G^{1J}_{,1} U_{\mathrm{eff}, J} - G^{1J} U_{\mathrm{eff}, J1} \right) \right] \, .
\end{align}
If the ratio is not trivial, the curvaton scenario could be realized, that is, the fluctuation of the curvaton is converted to the curvature fluctuation.

The metric in the field space in \eqref{sclrmtrc} is explicitly given by,
\begin{align}
\label{sclrmtrcexplct}
G_{\sigma\sigma} =&\, 3\, , \quad G_{\sigma\phi}=G_{\phi\sigma}=0\, , \nonumber \\
G_{\phi\phi}=&\, \e^\sigma \omega(\phi) \nonumber \\
=&\, \e^\sigma \left[ \frac{8 A_0 \e^{4 \alpha_0 \left( \phi - N_0 \right)}}{1 + A_0 \e^{4 \alpha_0 \left( \phi - N_0 \right)}}
+ \frac{24\alpha {H_0}^2}{\left(1 + A_0 \e^{4\alpha_0 \left( \phi - N_0 \right)}\right)^{\frac{1}{\alpha_0}+3}}
\left\{ 8A_0 + \left(- 8 + 32 A_0 + 32{\alpha_0}^2 \right) A_0 \e^{4 \alpha_0 \left( \phi - N_0 \right)} \right. \right. \nonumber \\
&\, \left. \left. + \left(- 56 - 24\alpha_0 - 32{\alpha_0}^2 \right) {A_0}^2 \e^{8 \alpha_0 \left( \phi - N_0 \right)}
+ 64 \alpha_0 {A_0}^3 \e^{12 \alpha_0 \left( \phi - N_0 \right)} \right\} \right] \, .
\end{align}
In terms of $N$, we obtain,
\begin{align}
\label{sclrmtrcexplctN}
G_{\sigma\sigma} =&\, 3\, , \quad G_{\sigma\phi}=G_{\phi\sigma}=0\, , \nonumber \\
G_{\phi\phi}=&\, \left\{ 1 + \frac{24\alpha{H_0}^2}{\left(1 + A_0 \e^{4\alpha_0 \left( N - N_0 \right)}\right)^{\frac{1}{\alpha_0}+1}} \right\}^{-1}
\left[ \frac{8 A_0 \e^{4 \alpha_0 \left( N - N_0 \right)}}{1 + A_0 \e^{4 \alpha_0 \left( N - N_0 \right)}} \right. \nonumber \\
&\, + \frac{24\alpha {H_0}^2}{\left(1 + A_0 \e^{4\alpha_0 \left( N - N_0 \right)}\right)^{\frac{1}{\alpha_0}+3}}
\left\{ 8A_0 + \left(- 8 + 32 A_0 + 32{\alpha_0}^2 \right) A_0 \e^{4 \alpha_0 \left( N - N_0 \right)} \right. \nonumber \\
&\, \left. \left. + \left(- 56 - 24\alpha_0 - 32{\alpha_0}^2 \right) {A_0}^2 \e^{8 \alpha_0 \left( N - N_0 \right)}
+ 64 \alpha_0 {A_0}^3 \e^{12 \alpha_0 \left( N - N_0 \right)} \right\} \right] \, .
\end{align}
The connections in the field space in \eqref{fscnnctn} are given by,
\begin{align}
\label{fscnnctnexplct}
\Gamma^\sigma_{\sigma\sigma}=&\, \Gamma^\sigma_{\sigma\phi} = \Gamma^\sigma_{\phi\sigma}=\Gamma^\phi_{\sigma\sigma}= 0 \, , \quad
\Gamma^\phi_{\sigma\phi} = \Gamma^\phi_{\phi\sigma} = \frac{1}{2} \, , \nonumber \\
\Gamma^\sigma_{\phi\phi} =&\, - \frac{1}{6} \e^\sigma \omega(\phi) \nonumber \\
=&\, \frac{\e^\sigma}{6} \left[ \frac{8 A_0 \e^{4 \alpha_0 \left( \phi - N_0 \right)}}{1 + A_0 \e^{4 \alpha_0 \left( \phi - N_0 \right)}}
+ \frac{24\alpha {H_0}^2}{\left(1 + A_0 \e^{4\alpha_0 \left( \phi - N_0 \right)}\right)^{\frac{1}{\alpha_0}+3}}
\left\{ 8A_0
+ \left(- 8 + 32 A_0 + 32{\alpha_0}^2 \right) A_0 \e^{4 \alpha_0 \left( \phi - N_0 \right)} \right. \right. \nonumber \\
&\, \left. \left. + \left(- 56 - 24\alpha_0 - 32{\alpha_0}^2 \right) {A_0}^2 \e^{8 \alpha_0 \left( \phi - N_0 \right)}
+ 64 \alpha_0 {A_0}^3 \e^{12 \alpha_0 \left( \phi - N_0 \right)} \right\} \right] \, , \nonumber \\
\Gamma^\phi_{\phi\phi} =&\, \frac{\omega'(\phi)}{2\omega(\phi)} \nonumber \\
=&\, \left[ \frac{16 A_0 \alpha_0 \e^{4 \alpha_0 \left( \phi - N_0 \right)}}{\left( 1 + A_0 \e^{4 \alpha_0 \left( \phi - N_0 \right)} \right)^2}
+ \frac{384\alpha A_0 {H_0}^2\e^{4\alpha_0 \left( \phi - N_0 \right)}}{\left(1 + A_0 \e^{4\alpha_0 \left( \phi - N_0 \right)}\right)^{\frac{1}{\alpha_0}+4}}
\left\{ - \alpha_0 + 3 A_0 - 3\alpha_0 A_0 + 4{\alpha_0}^3 \right. \right. \nonumber \\
&\, + \left( 1 - 12 \alpha_0 - 4 A_0 - 10 {\alpha_0}^2 - 8 \alpha_0 A_0 - 16 {\alpha_0}^3 \right) A_0 \e^{4 \alpha_0 \left( \phi - N_0 \right)} \nonumber \\
&\, \left. \left. + \left( 7 + 10 \alpha_0 + 31 {\alpha_0}^2 + 4 {\alpha_0}^3 \right) {A_0}^2 \e^{8 \alpha_0 \left( \phi - N_0 \right)}
+ \left( 8 \alpha_0 + 48 {\alpha_0}^2 \right) {A_0}^3 \e^{12 \alpha_0 \left( \phi - N_0 \right)} \right\} \right] \nonumber \\
&\, \times \left[ \frac{8 A_0 \e^{4 \alpha_0 \left( \phi - N_0 \right)}}{1 + A_0 \e^{4 \alpha_0 \left( \phi - N_0 \right)}}
+ \frac{24\alpha {H_0}^2}{\left(1 + A_0 \e^{4\alpha_0 \left( \phi - N_0 \right)}\right)^{\frac{1}{\alpha_0}+3}}
\left\{ 8A_0 + \left(- 8 + 32 A_0 + 32{\alpha_0}^2 \right) A_0 \e^{4 \alpha_0 \left( \phi - N_0 \right)} \right. \right. \nonumber \\
&\, \left. \left. + \left(- 56 - 24\alpha_0 - 32{\alpha_0}^2 \right) {A_0}^2 \e^{8 \alpha_0 \left( \phi - N_0 \right)}
+ 64 \alpha_0 {A_0}^3 \e^{12 \alpha_0 \left( \phi - N_0 \right)} \right\} \right]^{-1} \, .
\end{align}
By using the $e$-folding number $N$, we can rewrite \eqref{fscnnctnexplct} as follows,
\begin{align}
\label{fscnnctnexplctN}
\Gamma^\sigma_{\sigma\sigma}=&\, \Gamma^\sigma_{\sigma\phi} = \Gamma^\sigma_{\phi\sigma}=\Gamma^\phi_{\sigma\sigma}= 0 \, , \quad
\Gamma^\phi_{\sigma\phi} = \Gamma^\phi_{\phi\sigma} = \frac{1}{2} \, , \nonumber \\
\Gamma^\sigma_{\phi\phi} =&\, - \frac{1}{6} \left\{ 1 + \frac{24\alpha{H_0}^2}{\left(1 + A_0 \e^{4\alpha_0 \left( N - N_0 \right)}\right)^{\frac{1}{\alpha_0}+1}} \right\}^{-1}
\left[
%%%
\frac{8 A_0 \e^{4 \alpha_0 \left( N - N_0 \right)}}{1 + A_0 \e^{4 \alpha_0 \left( N - N_0 \right)}} \right. \nonumber \\
&\, + \frac{24\alpha {H_0}^2}{\left(1 + A_0 \e^{4\alpha_0 \left( N - N_0 \right)}\right)^{\frac{1}{\alpha_0}+3}}
\left\{ 8A_0 + \left(- 8 + 32 A_0 + 32{\alpha_0}^2 \right) A_0 \e^{4 \alpha_0 \left( N - N_0 \right)} \right. \nonumber \\
&\, \left. \left. + \left(- 56 - 24\alpha_0 - 32{\alpha_0}^2 \right) {A_0}^2 \e^{8 \alpha_0 \left( N - N_0 \right)}
+ 64 \alpha_0 {A_0}^3 \e^{12 \alpha_0 \left( N - N_0 \right)} \right\}
%%%%
 \right] \, , \nonumber \\
\Gamma^\phi_{\phi\phi} =&\, \left[ \frac{16 A_0 \alpha_0 \e^{4 \alpha_0 \left( N - N_0 \right)}}{\left( 1 + A_0 \e^{4 \alpha_0 \left( N - N_0 \right)} \right)^2}
+ \frac{384\alpha A_0 {H_0}^2\e^{4\alpha_0 \left( N - N_0 \right)}}{\left(1 + A_0 \e^{4\alpha_0 \left( N - N_0 \right)}\right)^{\frac{1}{\alpha_0}+4}}
\left\{ - \alpha_0 + 3 A_0 - 3\alpha_0 A_0 + 4{\alpha_0}^3 \right. \right. \nonumber \\
&\, + \left( 1 - 12 \alpha_0 - 4 A_0 - 10 {\alpha_0}^2 - 8 \alpha_0 A_0 - 16 {\alpha_0}^3 \right) A_0 \e^{4 \alpha_0 \left( N - N_0 \right)} \nonumber \\
&\, \left. \left. + \left( 7 + 10 \alpha_0 + 31 {\alpha_0}^2 + 4 {\alpha_0}^3 \right) {A_0}^2 \e^{8 \alpha_0 \left( N - N_0 \right)}
+ \left( 8 \alpha_0 + 48 {\alpha_0}^2 \right) {A_0}^3 \e^{12 \alpha_0 \left( N - N_0 \right)} \right\} \right] \nonumber \\
&\, \times \left[
%%%
\frac{8 A_0 \e^{4 \alpha_0 \left( N - N_0 \right)}}{1 + A_0 \e^{4 \alpha_0 \left( N - N_0 \right)}}
+ \frac{24\alpha {H_0}^2}{\left(1 + A_0 \e^{4\alpha_0 \left( N - N_0 \right)}\right)^{\frac{1}{\alpha_0}+3}}
\left\{ 8A_0 + \left(- 8 + 32 A_0 + 32{\alpha_0}^2 \right) A_0 \e^{4 \alpha_0 \left( N - N_0 \right)} \right. \right. \nonumber \\
&\, \left. \left. + \left(- 56 - 24\alpha_0 - 32{\alpha_0}^2 \right) {A_0}^2 \e^{8 \alpha_0 \left( N - N_0 \right)}
+ 64 \alpha_0 {A_0}^3 \e^{12 \alpha_0 \left( N - N_0 \right)} \right\}
%%%
\right]^{-1} \, .
\end{align}
We use the above calculations to check the curvaton scenario in the following.

We first investigate the epoch of inflation, which corresponds to $N\to -\infty$,
The behaviors of the relevant quantities are given by,
\begin{align}
\label{Ninf}
H^2 \to &\, {H_0}^2 \, , \quad
H H' \to 2{H_0}^2 A_0 \e^{4\alpha_0 \left( N - N_0 \right)}\, , \quad
\e^{-\sigma} \to 1 + 24\alpha {H_0}^2\, , \quad
 - \frac{d\e^{-\sigma}}{dN}
\to 96\alpha{H_0}^2 A_0 \left( 1 + \alpha_0 \right) \e^{4\alpha_0 \left( N - N_0 \right)}
\, , \nonumber \\
G_{\sigma\sigma} =&\, 3\, , \quad G_{\sigma\phi}=G_{\phi\sigma}=0\, , \quad
G_{\phi\phi}
= \frac{18{H_0}^2 \left( 1 + 32 \alpha {H_0}^2\right)}{1 + 24\alpha{H_0}^2}
\, , \nonumber \\
\Gamma^\sigma_{\sigma\sigma}=&\, \Gamma^\sigma_{\sigma\phi} = \Gamma^\sigma_{\phi\sigma}=\Gamma^\phi_{\sigma\sigma}= 0 \, , \quad
\Gamma^\phi_{\sigma\phi} = \Gamma^\phi_{\phi\sigma} = \frac{1}{2} \, , \nonumber \\
\Gamma^\sigma_{\phi\phi,\sigma} =&\, \Gamma^\sigma_{\phi\phi}
\to - \frac{32 \alpha A_0 {H_0}^2}{1 + 24\alpha{H_0}^2} \, , \nonumber \\
\Gamma^\sigma_{\phi\phi,\phi} \to &\, - \frac{A_0 \e^{4\alpha_0 \left( N - N_0 \right)}}{3 \left( 1 + 24\alpha{H_0}^2 \right)}
\left\{ 16 \alpha_0 + 384\alpha {H_0}^2 \left( - \alpha_0 + 3 A_0 - 3\alpha_0 A_0 + 4{\alpha_0}^3 \right) \right\} \nonumber \\
%%%
\Gamma^\phi_{\phi\phi} \to&\, \frac{\left( \alpha_0 - 24 \alpha A_0 \alpha_0 {H_0}^2\right)\e^{4 \alpha_0 \left( \phi - N_0 \right)}}{12\alpha{H_0}^2} \, , \nonumber \\
\Gamma^\phi_{\phi\phi, \phi}
\to &\, \frac{ \left\{ {\alpha_0}^2 + 24\alpha {H_0}^2 \left( - {\alpha_0}^2 + 3 A_0 \alpha_0 - 3{\alpha_0}^2 A_0 + 4{\alpha_0}^4 \right) \right\} \e^{4 \alpha_0 \left( \phi - N_0 \right)}}{6 \alpha{H_0}^2}
\, , \nonumber \\
&\, \mbox{other components}=0 \, .
\end{align}
We derive \eqref{Ninf} in \eqref{NinfAp}.
Then the dispersion relation Eq.~\eqref{det} has the following form,
\begin{align}
\label{detNminfty}
0 =&\, \left[ {H_0}^2 \left( 1 + 24\alpha {H_0}^2 \right) \Omega(N)^2 - 3 i {H_0}^2 \left( 1 + 24\alpha {H_0}^2 \right) \Omega(N) - \left( 1 + 24\alpha {H_0}^2 \right) \e^{-2N} k^2 \right] \nonumber \\
&\, \times \left[ {H_0}^2 \left( 1 + 24\alpha {H_0}^2 \right) \Omega(N)^2 - 3 i {H_0}^2 \left( 1 + 24\alpha {H_0}^2 \right) \Omega(N) - \left( 1 + 24\alpha {H_0}^2 \right) \e^{-2N} k^2 \right] \nonumber \\
&\, - \left[ - \frac{32 \alpha A_0 {H_0}^2}{1 + 24\alpha{H_0}^2} i \Omega(N) {H_0}^2 \left( 1 + 24\alpha {H_0}^2 \right) \right]
\left[ \frac{1}{2} i \Omega(N) {H_0}^2 \left( 1 + 24\alpha {H_0}^2 \right) \right] \, ,
\end{align}
which gives,
\begin{align}
\label{detNminfty0}
0 =&\,
\left[ {H_0}^2 \Omega(N)^2 - 3 i {H_0}^2 \Omega(N) - \e^{-2N} k^2 \right]^2
 - \frac{16 \alpha A_0 {H_0}^6}{1 + 24\alpha{H_0}^2} \Omega(N)^2 \, ,
\end{align}
that is,
\begin{align}
\label{detNminfty0B}
0 = {H_0}^2 \Omega(N)^2 + \left( \pm 4{H_0}^3 \sqrt{ \frac{\alpha A_0}{1 + 24\alpha{H_0}^2}} - 3 i {H_0}^2 \right) \Omega(N) - \e^{-2N} k^2 \, ,
\end{align}
which can be solved with respect to $\Omega(N)$ as follows,
\begin{align}
\label{Omgdsprsn}
\Omega (N) = \frac{ - \left( \pm 4{H_0}^3 \sqrt{ \frac{\alpha A_0}{1 + 24\alpha{H_0}^2}} - 3 i {H_0}^2 \right) \pm
\sqrt{ \left( \pm 4{H_0}^3 \sqrt{ \frac{\alpha A_0}{1 + 24\alpha{H_0}^2}} - 3 i {H_0}^2 \right)^2 + 4 {H_0}^2 k^2 \e^{-2N} k^2 }}{2{H_0}^2}\, .
\end{align}
The $\pm$ in front of the square root is independent of the two $\pm$ in front of $4{H_0}^3$.

The ratio in \eqref{ratio} is given by,
\begin{align}
\label{ratioinfty}
\delta\sigma: \delta\phi =&\, \delta \Phi^1 : \delta \Phi^2 \nonumber \\
=&\, \pm 4{H_0}^3 \sqrt{ \frac{\alpha A_0}{1 + 24\alpha{H_0}^2}} \nonumber \\
&\, : \frac{1}{2} i {H_0}^2 \frac{ - \left( \pm 4{H_0}^3 \sqrt{ \frac{\alpha A_0}{1 + 24\alpha{H_0}^2}} - 3 i {H_0}^2 \right)
\pm \sqrt{ \left( \pm 4{H_0}^3 \sqrt{ \frac{\alpha A_0}{1 + 24\alpha{H_0}^2}} - 3 i {H_0}^2 \right)^2 + 4 {H_0}^2 k^2 \e^{-2N} k^2 }}{2{H_0}^2} \nonumber \\
=&\, \pm 4{H_0}^3 \sqrt{ \frac{\alpha A_0}{1 + 24\alpha{H_0}^2}} \nonumber \\
&\, : \frac{i}{4} \left\{ - \left( \pm 4{H_0}^3 \sqrt{ \frac{\alpha A_0}{1 + 24\alpha{H_0}^2}} - 3 i {H_0}^2 \right) \right. \nonumber \\
&\, \qquad \qquad \qquad \qquad \left. \pm \sqrt{ \left( \pm 4{H_0}^3 \sqrt{ \frac{\alpha A_0}{1 + 24\alpha{H_0}^2}} - 3 i {H_0}^2 \right)^2 + 4 {H_0}^2 \e^{-2N} k^2 } \right\} \, . \nonumber \\
\mbox{or} & \nonumber \\
\delta\sigma: \delta\phi =&\, \delta \Phi^1 : \delta \Phi^2 \nonumber \\
=&\, - \frac{32 i \alpha A_0 {H_0}^2}{1 + 24\alpha{H_0}^2} \frac{ - \left( \pm 4{H_0}^3 \sqrt{ \frac{\alpha A_0}{1 + 24\alpha{H_0}^2}} - 3 i {H_0}^2 \right)
\pm \sqrt{ \left( \pm 4{H_0}^3 \sqrt{ \frac{\alpha A_0}{1 + 24\alpha{H_0}^2}} - 3 i {H_0}^2 \right)^2 + 4 {H_0}^2 k^2 \e^{-2N} k^2 }}{2{H_0}^2} \nonumber \\
&\, : \pm 4{H_0}^3 \sqrt{ \frac{\alpha A_0}{1 + 24\alpha{H_0}^2}} \nonumber \\
=&\, - \frac{ 16 i \alpha A_0 {H_0}^2 }{1 + 24\alpha{H_0}^2} \left\{- \left( \pm 4{H_0}^3 \sqrt{ \frac{\alpha A_0}{1 + 24\alpha{H_0}^2}} - 3 i {H_0}^2 \right) \right. \nonumber \\
&\, \qquad \qquad \qquad \qquad \left. \pm \sqrt{ \left( \pm 4{H_0}^3 \sqrt{ \frac{\alpha A_0}{1 + 24\alpha{H_0}^2}} - 3 i {H_0}^2 \right)^2 + 4 {H_0}^2 k^2 \e^{-2N} k^2 }\right\} \nonumber \\
&\, : \pm 4{H_0}^3 \sqrt{ \frac{\alpha A_0}{1 + 24\alpha{H_0}^2}} \nonumber \\
\end{align}
If $k$ is finite, the term $\e^{-2N} k^2$ dominates when $N\to -\infty$.
Therefore in the limit, $\delta\sigma$ and $\delta\phi$ do not mix with each other,
\begin{align}
\label{ratioinftylmt}
\delta\sigma: \delta\phi = \delta \Phi^1 : \delta \Phi^2 = 1:0\ \mbox{or}\ 0:1\, .
\end{align}
After that, $\delta\sigma$ and $\delta\phi$ begin to mix with each other.
Both of the dispersion relations \eqref{Omgdsprsn} of $\delta\sigma$ and $\delta\phi$ have the following standard forms,
\begin{align}
\label{Omgdsprsnlmt}
\Omega (N) = \pm \frac{k}{H_0} \e^{-N}\, .
\end{align}
Here $k=\sqrt{\bm{k}\cdot\bm{k}}$.

On the other hand, at the end of the inflation defined by \eqref{einf}, that is, when $A_0 \e^{4 \alpha_0 \left( N - N_0 \right)}=1$, by using \eqref{Ex1}, \eqref{Ex2}, \eqref{Ex3}, and \eqref{Ex4}, we obtain the dispersion relation \eqref{det} as
follows,
\begin{align}
\label{dsprsnlt}
0=&\, \left[ 2^{-\frac{1}{\alpha_0}} {H_0}^2 \left( 1 + 12 \cdot 2^{-\frac{1}{\alpha_0}} \alpha {H_0}^2 \right) \Omega(N)^2
 - i \left\{ 24 \cdot 2^{-\frac{2}{\alpha_0}} \alpha {H_0}^4 \left( 1 + \alpha_0 \right) \right. \right. \nonumber \\
&\, \left. + 4 \cdot 2^{-\frac{1}{\alpha_0}} {H_0}^2 \left( 1 + 12 \cdot 2^{-\frac{1}{\alpha_0}} \alpha {H_0}^2 \right) \right\} \Omega(N) - \e^{-\sigma} a^{-2} k^2 \nonumber \\
&\, \left. + \frac{2^{-\frac{1}{\alpha_0}} {H_0}^2 }{3} \left\{ 2 + 3 \cdot 2^{-\frac{1}{\alpha_0}} \alpha {H_0}^2 \left(-32 + 20 \alpha_0 + 20 A_0 \right)\right\} \right] \nonumber \\
&\, \times \left[ 2^{-\frac{1}{\alpha_0}} {H_0}^2 \left( 1 + 12 \cdot 2^{-\frac{1}{\alpha_0}} \alpha {H_0}^2 \right) \Omega(N)^2
 - i \left\{ 24 \cdot 2^{-\frac{2}{\alpha_0}} \alpha {H_0}^4 \left( 1 + \alpha_0 \right) \right. \right. \nonumber \\
&\, \left. + 4 \cdot 2^{-\frac{1}{\alpha_0}} {H_0}^2 \left( 1 + 12 \cdot 2^{-\frac{1}{\alpha_0}} \alpha {H_0}^2 \right) \right\} \Omega(N) - \e^{-\sigma} a^{-2} k^2 \nonumber \\
&\, - \left[ \frac{2 {\alpha_0}^2 + 24 \cdot 2^{-\frac{1}{\alpha_0}} \alpha {H_0}^2 \left( - 8 - 30 \alpha_0 - 107 {\alpha_0}^2 + 396 {\alpha_0}^3 - 32 {\alpha_0}^4 - 3 A_0
+ 9 \alpha_0 A_0 + 14 {\alpha_0}^2 A_0 \right) }
{1 + 6 \cdot 2^{-\frac{1}{\alpha_0}} \alpha {H_0}^2 \left( - 8 + 5 \alpha_0 + 5 A_0 \right) }
\right. \nonumber \\
&\, \left. - \frac{ 2 \left\{ 1 + 6 \cdot 2^{-\frac{1}{\alpha_0}} \alpha {H_0}^2
\left( 8 + 5 \alpha_0 + 69 {\alpha_0}^2 - 8 {\alpha_0}^3 - A_0 - 11 \alpha_0 A_0 \right) \right\}^2}
{\left\{ 1 + 6 \cdot 2^{-\frac{1}{\alpha_0}} \alpha {H_0}^2 \left( - 8 + 5 \alpha_0 + 5 A_0 \right) \right\}^2} \right] 2^{-\frac{1}{\alpha_0}} {H_0}^2 \nonumber \\
&\, + \frac{1}{2} \left[ i\Omega(N) 2^{-\frac{1}{\alpha_0}} \alpha {H_0}^2 24 \cdot 2^{-\frac{1}{\alpha_0}} \alpha {H_0}^2 \left( 1 + \alpha_0 \right) \right. \nonumber \\
&\, + \frac{8 + 48\cdot 2^{-\frac{1}{\alpha_0}} \alpha {H_0}^2 \left( 8 + 5 \alpha_0 + 69 {\alpha_0}^2 - 8 {\alpha_0}^3 - A_0 - 11 \alpha_0 A_0 \right)}
{\left\{10 \cdot 2^{-\frac{1}{\alpha_0}} {H_0}^2 + 12 \cdot 2^{-\frac{1}{\alpha_0}} \alpha {H_0}^4 \left( - 10 + 28 \alpha_0 \right)\right\}^2} \nonumber \\
&\, \times \left\{ - 2 \cdot 2^{-\frac{1}{\alpha_0}} {H_0}^2 \left( 9 + 8 \alpha_0 \right)
+ 24 \cdot 2^{-\frac{2}{\alpha_0}} {H_0}^4 \left( 20 - 98 \alpha_0 + 26 {\alpha_0}^2 + 16{\alpha_0}^3 \right) \right\} \nonumber \\
&\, \left. - \frac{4 \cdot 2^{-\frac{1}{\alpha_0}} {H_0}^2 \left( 9 + 8 \alpha_0 \right)
+ 48 \cdot 2^{-\frac{2}{\alpha_0}} \alpha {H_0}^4 \left( - 36 + 291 \alpha_0 + 146 {\alpha_0}^2 - 128 {\alpha_0}^3 \right)}
{10 \cdot 2^{-\frac{1}{\alpha_0}} {H_0}^2 + 12 \cdot 2^{-\frac{1}{\alpha_0}} \alpha {H_0}^4 \left( - 10 + 28 \alpha_0 \right)} \right] \nonumber \\
&\, - \left[ \frac{2^{-\frac{1}{\alpha_0}} {H_0}^2 \left\{ 2 + 3 \cdot 2^{-\frac{1}{\alpha_0}} \alpha {H_0}^2 \left(-32 + 20 \alpha_0 + 20 A_0 \right)\right\} }{3} \right. \nonumber \\
&\, - \frac{2 + 3 \cdot 2^{-\frac{1}{\alpha_0}} \alpha {H_0}^2 \left(-32 + 20 \alpha_0 + 20 A_0 \right)}{3 \left( 1 + 12 \cdot 2^{-\frac{1}{\alpha_0}} \alpha{H_0}^2 \right)}
\left\{ 2^{-\frac{1}{\alpha_0}} {H_0}^2 \left( 1 + 12 \cdot 2^{-\frac{1}{\alpha_0}} \alpha {H_0}^2 \right) i\Omega(N) \right. \nonumber \\
&\, \left. \left. \frac{ - 4 \cdot 2^{-\frac{1}{\alpha_0}} {H_0}^2 \left( 9 + 8 \alpha_0 \right) + 48 \cdot 2^{-\frac{2}{\alpha_0}} {H_0}^4 \left( 20 - 98 \alpha_0 + 26 {\alpha_0}^2 + 16{\alpha_0}^3 \right)}
{3\left( 1 + 12 \cdot 2^{-\frac{1}{\alpha_0}} \alpha {H_0}^2 \right)^2} \right\} \right] \nonumber \\
&\, \times \left[ i\Omega(N) 2^{-1 -\frac{1}{\alpha_0}} {H_0}^2 \left( 1 + 12 \cdot 2^{-\frac{1}{\alpha_0}} \alpha {H_0}^2 \right) \right] \, .
\end{align}
On the other hand, the ratio of $\delta \Phi^1 = \delta \sigma$ and $\delta \Phi^2 = \delta \phi$ in \eqref{ratio} is given by,
\begin{align}
\label{ratiolt}
\delta\sigma: \delta\phi =&\, \delta \Phi^1 : \delta \Phi^2 \nonumber \\
&\, 2^{-\frac{1}{\alpha_0}} {H_0}^2 \left( 1 + 12 \cdot 2^{-\frac{1}{\alpha_0}} \alpha {H_0}^2 \right) \Omega(N)^2
 - i \left\{ 24 \cdot 2^{-\frac{2}{\alpha_0}} \alpha {H_0}^4 \left( 1 + \alpha_0 \right) \right. \nonumber \\
&\, \left. + 4 \cdot 2^{-\frac{1}{\alpha_0}} {H_0}^2 \left( 1 + 12 \cdot 2^{-\frac{1}{\alpha_0}} \alpha {H_0}^2 \right) \right\} \Omega(N) - \e^{-\sigma} a^{-2} k^2 \nonumber \\
&\, - \left[ \frac{2 {\alpha_0}^2 + 24 \cdot 2^{-\frac{1}{\alpha_0}} \alpha {H_0}^2 \left( - 8 - 30 \alpha_0 - 107 {\alpha_0}^2 + 396 {\alpha_0}^3 - 32 {\alpha_0}^4 - 3 A_0
+ 9 \alpha_0 A_0 + 14 {\alpha_0}^2 A_0 \right) }
{1 + 6 \cdot 2^{-\frac{1}{\alpha_0}} \alpha {H_0}^2 \left( - 8 + 5 \alpha_0 + 5 A_0 \right) }
\right. \nonumber \\
&\, \left. - \frac{ 2 \left\{ 1 + 6 \cdot 2^{-\frac{1}{\alpha_0}} \alpha {H_0}^2
\left( 8 + 5 \alpha_0 + 69 {\alpha_0}^2 - 8 {\alpha_0}^3 - A_0 - 11 \alpha_0 A_0 \right) \right\}^2}
{\left\{ 1 + 6 \cdot 2^{-\frac{1}{\alpha_0}} \alpha {H_0}^2 \left( - 8 + 5 \alpha_0 + 5 A_0 \right) \right\}^2} \right] 2^{-\frac{1}{\alpha_0}} {H_0}^2 \nonumber \\
&\, + \frac{1}{2} \left[ i\Omega(N) 2^{-\frac{1}{\alpha_0}} \alpha {H_0}^2 24 \cdot 2^{-\frac{1}{\alpha_0}} \alpha {H_0}^2 \left( 1 + \alpha_0 \right) \right. \nonumber \\
&\, + \frac{8 + 48\cdot 2^{-\frac{1}{\alpha_0}} \alpha {H_0}^2 \left( 8 + 5 \alpha_0 + 69 {\alpha_0}^2 - 8 {\alpha_0}^3 - A_0 - 11 \alpha_0 A_0 \right)}
{\left\{10 \cdot 2^{-\frac{1}{\alpha_0}} {H_0}^2 + 12 \cdot 2^{-\frac{1}{\alpha_0}} \alpha {H_0}^4 \left( - 10 + 28 \alpha_0 \right)\right\}^2} \nonumber \\
&\, \times \left\{ - 2 \cdot 2^{-\frac{1}{\alpha_0}} {H_0}^2 \left( 9 + 8 \alpha_0 \right)
+ 24 \cdot 2^{-\frac{2}{\alpha_0}} {H_0}^4 \left( 20 - 98 \alpha_0 + 26 {\alpha_0}^2 + 16{\alpha_0}^3 \right) \right\} \nonumber \\
&\, - \frac{4 \cdot 2^{-\frac{1}{\alpha_0}} {H_0}^2 \left( 9 + 8 \alpha_0 \right)
+ 48 \cdot 2^{-\frac{2}{\alpha_0}} \alpha {H_0}^4 \left( - 36 + 291 \alpha_0 + 146 {\alpha_0}^2 - 128 {\alpha_0}^3 \right)}
{10 \cdot 2^{-\frac{1}{\alpha_0}} {H_0}^2 + 12 \cdot 2^{-\frac{1}{\alpha_0}} \alpha {H_0}^4 \left( - 10 + 28 \alpha_0 \right)} \nonumber \\
: &\, i\Omega(N) 2^{-1 -\frac{1}{\alpha_0}} {H_0}^2 \left( 1 + 12 \cdot 2^{-\frac{1}{\alpha_0}} \alpha {H_0}^2 \right) \nonumber \\
\mbox{or} & \nonumber \\
\delta\sigma: \delta\phi =&\, \delta \Phi^1 : \delta \Phi^2 \nonumber \\
&\, \frac{2^{-\frac{1}{\alpha_0}} {H_0}^2 \left\{ 2 + 3 \cdot 2^{-\frac{1}{\alpha_0}} \alpha {H_0}^2 \left(-32 + 20 \alpha_0 + 20 A_0 \right)\right\} }{3} \nonumber \\
&\, - \frac{2 + 3 \cdot 2^{-\frac{1}{\alpha_0}} \alpha {H_0}^2 \left(-32 + 20 \alpha_0 + 20 A_0 \right)}{3 \left( 1 + 12 \cdot 2^{-\frac{1}{\alpha_0}} \alpha{H_0}^2 \right)}
\left\{ 2^{-\frac{1}{\alpha_0}} {H_0}^2 \left( 1 + 12 \cdot 2^{-\frac{1}{\alpha_0}} \alpha {H_0}^2 \right) i\Omega(N) \right. \nonumber \\
&\, \left. \frac{ - 4 \cdot 2^{-\frac{1}{\alpha_0}} {H_0}^2 \left( 9 + 8 \alpha_0 \right) + 48 \cdot 2^{-\frac{2}{\alpha_0}} {H_0}^4 \left( 20 - 98 \alpha_0 + 26 {\alpha_0}^2 + 16{\alpha_0}^3 \right)}
{3\left( 1 + 12 \cdot 2^{-\frac{1}{\alpha_0}} \alpha {H_0}^2 \right)^2} \right\} \nonumber \\
: &\, 2^{-\frac{1}{\alpha_0}} {H_0}^2 \left( 1 + 12 \cdot 2^{-\frac{1}{\alpha_0}} \alpha {H_0}^2 \right) \Omega(N)^2
 - i \left\{ 24 \cdot 2^{-\frac{2}{\alpha_0}} \alpha {H_0}^4 \left( 1 + \alpha_0 \right) \right. \nonumber \\
&\, \left. + 4 \cdot 2^{-\frac{1}{\alpha_0}} {H_0}^2 \left( 1 + 12 \cdot 2^{-\frac{1}{\alpha_0}} \alpha {H_0}^2 \right) \right\} \Omega(N) - \e^{-\sigma} a^{-2} k^2 \nonumber \\
&\, - \left[ \frac{2 {\alpha_0}^2 + 24 \cdot 2^{-\frac{1}{\alpha_0}} \alpha {H_0}^2 \left( - 8 - 30 \alpha_0 - 107 {\alpha_0}^2 + 396 {\alpha_0}^3 - 32 {\alpha_0}^4 - 3 A_0
+ 9 \alpha_0 A_0 + 14 {\alpha_0}^2 A_0 \right) }
{1 + 6 \cdot 2^{-\frac{1}{\alpha_0}} \alpha {H_0}^2 \left( - 8 + 5 \alpha_0 + 5 A_0 \right) }
\right. \nonumber \\
&\, \left. - \frac{ 2 \left\{ 1 + 6 \cdot 2^{-\frac{1}{\alpha_0}} \alpha {H_0}^2
\left( 8 + 5 \alpha_0 + 69 {\alpha_0}^2 - 8 {\alpha_0}^3 - A_0 - 11 \alpha_0 A_0 \right) \right\}^2}
{\left\{ 1 + 6 \cdot 2^{-\frac{1}{\alpha_0}} \alpha {H_0}^2 \left( - 8 + 5 \alpha_0 + 5 A_0 \right) \right\}^2} \right] 2^{-\frac{1}{\alpha_0}} {H_0}^2 \nonumber \\
&\, + \frac{1}{2} \left[ i\Omega(N) 2^{-\frac{1}{\alpha_0}} \alpha {H_0}^2 24 \cdot 2^{-\frac{1}{\alpha_0}} \alpha {H_0}^2 \left( 1 + \alpha_0 \right) \right. \nonumber \\
&\, + \frac{8 + 48\cdot 2^{-\frac{1}{\alpha_0}} \alpha {H_0}^2 \left( 8 + 5 \alpha_0 + 69 {\alpha_0}^2 - 8 {\alpha_0}^3 - A_0 - 11 \alpha_0 A_0 \right)}
{\left\{10 \cdot 2^{-\frac{1}{\alpha_0}} {H_0}^2 + 12 \cdot 2^{-\frac{1}{\alpha_0}} \alpha {H_0}^4 \left( - 10 + 28 \alpha_0 \right)\right\}^2} \nonumber \\
&\, \times \left\{ - 2 \cdot 2^{-\frac{1}{\alpha_0}} {H_0}^2 \left( 9 + 8 \alpha_0 \right)
+ 24 \cdot 2^{-\frac{2}{\alpha_0}} {H_0}^4 \left( 20 - 98 \alpha_0 + 26 {\alpha_0}^2 + 16{\alpha_0}^3 \right) \right\} \nonumber \\
&\, - \frac{4 \cdot 2^{-\frac{1}{\alpha_0}} {H_0}^2 \left( 9 + 8 \alpha_0 \right)
+ 48 \cdot 2^{-\frac{2}{\alpha_0}} \alpha {H_0}^4 \left( - 36 + 291 \alpha_0 + 146 {\alpha_0}^2 - 128 {\alpha_0}^3 \right)}
{10 \cdot 2^{-\frac{1}{\alpha_0}} {H_0}^2 + 12 \cdot 2^{-\frac{1}{\alpha_0}} \alpha {H_0}^4 \left( - 10 + 28 \alpha_0 \right)} \, .
\end{align}
Although the dispersion relation~\eqref{dsprsnlt} cannot be explicitly solved with respect to $\Omega(N)$, Eq.~\eqref{ratiolt} tells us that there should be a mixing between $\delta \Phi^1 = \delta \sigma$ and $\delta \Phi^2 = \delta \phi$.
Therefore, the curvaton scenario could be realized, that is, the isocurvature fluctuations in the early Universe could be converted to curvature fluctuations when inflation ends.

\section{Summary and Conclusions}

In this paper, the models compatible with the constraints from the ACT data~\cite{AtacamaCosmologyTelescope:2025vnj, AtacamaCosmologyTelescope:2025nti, CosmoVerseNetwork:2025alb}, Planck, ACT, lensing, BAO, BK18 (BICEP/Keck), and others have been constructed in the framework of the $f\left(Q, \tilde R\right)$ gravity and the $f(R)$ gravity coupled with a scalar field.
As we explained, the curvature in Einstein's gravity constructed from the Levi-Civita connection can be compatibly included in the non-metricity gravity because the conservation law is usually given in terms of the covariant derivative defined by the Levi-Civita connection, and the curvature in Einstein's gravity appears in the commutator of two covariant derivatives \cite{Nojiri:2024hau}.

Two kinds of behaviors of the Hubble rate $H$, which satisfy the constraints, have been proposed as Model I~\eqref{mdl2} and Model II~\eqref{mdl5}.
The models that realize the behaviors of the Hubble rate $H$ have been constructed in the framework of $f\left(Q, \tilde R\right)$ gravity as in \eqref{rmfqmdl1} and \eqref{rmfqmdl1RmR}.
We also constructed the models of the $f(R)$ gravity coupled with a scalar field as in \eqref{sumModelI} and \eqref{sumModelII}.
The models where the $f(Q)$ part of the gravity only plays a role in the inflation epoch, and the $f(Q)$ part can be neglected in the present universe, have been proposed in \eqref{late}

Although we discussed the reheating after inflation, we have found that the reheating in the $f\left(Q, \tilde R\right)$ gravity is problematic, although the case of the $f(R)$ gravity coupled with a scalar field is well-known because the model can be rewritten in the form of Einstein's gravity coupled with two scalar fields.
The problems in $f\left(Q, \tilde R\right)$ come from the problem of the number of dynamical degrees of freedom in $f(Q)$ gravity, which is not settled~\cite{Hu:2022anq, DAmbrosio:2023asf, Heisenberg:2023lru, Hu:2023gui}.
We used the arguments in the perturbation around the flat background.
As in \eqref{fQfR}, if we consider the model of $f\left(Q, R \right) = f_{(Q)}\left(Q\right) + f_{(R)}\left( R \right)$, the $f_{(R)}\left( R \right)$ gives the free Lagrangian density of the massless spin 2 mode
corresponding to the graviton and the massive scalar mode corresponding to the scalaron, but $f_{(Q)}\left(Q \right)$ only shifts the Lagrangian density of the massless spin 2 mode, as shown in \cite{Capozziello:2024vix}.
Then the resulting perturbative theory includes only the graviton and scalaron, and other degrees of freedom do not appear. Although in the epoch of reheating, the extra modes might appear, we might be able to assume that the reheating might be generated only by the scalaron.
Under this assumption, which could not be strongly supported, we have shown that the reheating mechanism could not be so different from that in the standard $f(R)$ gravity.

We also investigated the possibility of the curvaton scenario~\cite{Enqvist:2001zp, Lyth:2001nq, Moroi:2001ct, Lyth:2002my, Sasaki:2006kq, Mazumdar:2010sa} in the framework of $f(R)$ gravity coupled with a scalar field by using the model in this paper.
In the curvaton scenario, there appears one more scalar field called the curvaton in addition to the inflaton.
On the other hand, it is well-known that $f(R)$ gravity coupled with a scalar field can be rewritten in the form of Einstein gravity coupled with two scalar fields in the Einstein frame.
For our specific model of $f(R)$ gravity coupled with a scalar field in this paper, this curvaton scenario could be realized.
We have investigated the scenario by using the asymptotic behaviors of a tune-rate vector and the perturbation in the adiabatic approximation.
We have found that the additional scalar field plays the role of the inflaton, and the scalar mode in the $f(R)$ sector works as a curvaton.

\newpage

\appendix

\section{Equations in Sec.~\ref{SecIV}}\label{SecIVA}

We now derive Eqs.~ \eqref{R}, \eqref{rhototalN2}, \eqref{arbQtotalmdl2}, \eqref{mpotential}, and \eqref{mpotentialprime}.
For the model~\eqref{mdl2}, we find,
\begin{align}
\label{HdHdd}
\frac{H'}{H} =&\, - \frac{1}{2} \left( \ln \left( H^2 \right) \right)'
= - \frac{2 A_0 \e^{4 \alpha_0 \left( N - N_0 \right)}}{1 + A_0 \e^{4 \alpha_0 \left( N - N_0 \right)}}
= - 2 + \frac{2}{\left( - \frac{6{H_0}^2}{Q} \right)^{\alpha_0}}
= - 2 + 2\left( - \frac{Q}{6{H_0}^2} \right)^{\alpha_0} \, , \nonumber \\
\frac{H''}{H} =&\, \left( \frac{H'}{H} \right)' + \frac{{H'}^2}{H^2} = - \frac{8\alpha_0 A_0 \e^{4 \alpha_0 \left( N - N_0 \right)}}{\left( 1 + A_0 \e^{4 \alpha_0 \left( N - N_0 \right)} \right)^2}
+ \frac{4 {A_0}^2 \e^{8 \alpha_0 \left( N - N_0 \right)}}{\left( 1 + A_0 \e^{4 \alpha_0 \left( N - N_0 \right)} \right)^2} \nonumber \\
=&\, \left( - \frac{Q}{6{H_0}^2} \right)^{2\alpha_0} \left\{ 8\alpha_0 - 8\alpha_0 \left( - \frac{6{H_0}^2}{Q} \right)^{\alpha_0}
+ 4 \left( - 1 + \left( - \frac{6{H_0}^2}{Q} \right)^{\alpha_0} \right)^2 \right\} \nonumber \\
=&\, \left( - \frac{Q}{6{H_0}^2} \right)^{2\alpha_0} \left\{ 4\left(2 \alpha_0 + 1 \right) - 8\left(\alpha_0 + 1 \right)\left( - \frac{6{H_0}^2}{Q} \right)^{\alpha_0}
+ 4 \left( - \frac{6{H_0}^2}{Q} \right)^{2\alpha_0} \right\} \nonumber \\
=&\, 4\left(2 \alpha_0 + 1 \right) \left( - \frac{Q}{6{H_0}^2} \right)^{2\alpha_0} - 8\left(\alpha_0 + 1 \right) \left( - \frac{Q}{6{H_0}^2} \right)^{\alpha_0} + 4 \, .
\end{align}
Then we find \eqref{R}
\begin{align}
\label{RAp}
R =&\, 12 H^2 + 6 HH'
= - 2 Q - Q \left\{ - 2 + 2\left( - \frac{Q}{6{H_0}^2} \right)^{\alpha_0} \right\}
= - 2 Q\left( - \frac{Q}{6{H_0}^2} \right)^{\alpha_0}
= 12 {H_0}^2 \left( - \frac{Q}{6{H_0}^2} \right)^{\alpha_0+1} \, ,
\end{align}
and \eqref{rhototalN2}
\begin{align}
\label{rhototalN2Ap}
\rho_\mathrm{total} =&\, \frac{1}{2} \left[ - f_{(R)} + 6\left(H^2 + H H'\right) f_{(R)}' - 36 \left( 4H^3 H' + H^2 {H'}^2 + H^3 H'' \right) f_{(R)}'' \right] + \rho \nonumber \\
=&\, \frac{1}{2} \left[ - f_{(R)} - Q\left(1 + \frac{H'}{H} \right) f_{(R)}' - Q^2 \left( 4\frac{H'}{H} + \left(\frac{H'}{H} \right)^2 + \frac{H''}{H} \right) f_{(R)}'' \right] + \rho \nonumber \\
=&\, \frac{1}{2} \left[ - f_{(R)} - Q\left(- 1 + 2\left( - \frac{Q}{6{H_0}^2} \right)^{\alpha_0} \right) f_{(R)}' \right. \nonumber \\
&\, - Q^2 \left( - 8 + 8 \left( - \frac{Q}{6{H_0}^2} \right)^{\alpha_0} + \left( - 2 + 2\left( - \frac{Q}{6{H_0}^2} \right)^{\alpha_0} \right)^2 \right. \nonumber \\
&\, \left. \left. \qquad + 4\left(2 \alpha_0 + 1 \right) \left( - \frac{Q}{6{H_0}^2} \right)^{2\alpha_0} - 8\left(\alpha_0 + 1 \right) \left( - \frac{Q}{6{H_0}^2} \right)^{\alpha_0} + 4 \right) f_{(R)}'' \right]
+ \rho \nonumber \\
=&\, \frac{1}{2} \left[ - f_{(R)} - Q\left(- 1 + 2\left( - \frac{Q}{6{H_0}^2} \right)^{\alpha_0} \right) f_{(R)}' \right. \nonumber \\
&\, \left. - Q^2 \left( 8\left( \alpha_0 + 1 \right) \left( - \frac{Q}{6{H_0}^2} \right)^{2\alpha_0} - 8\left(\alpha_0 + 1 \right) \left( - \frac{Q}{6{H_0}^2} \right)^{\alpha_0} + 4 \right) f_{(R)}'' \right]
+ \rho \nonumber \\
=&\, \frac{1}{2} \left[ - f_{(R)} + 6{H_0}^2\left\{- \left( - \frac{Q}{6{H_0}^2} \right) + 2\left( - \frac{Q}{6{H_0}^2} \right)^{\alpha_0+1} \right\} f_{(R)}'
- \left( 6{H_0}^2 \right)^2 \left\{ 8\left( \alpha_0 + 1 \right) \left( - \frac{Q}{6{H_0}^2} \right)^{2\alpha_0+2} \right. \right. \nonumber \\
&\, \left. \left. \left. \qquad \qquad - 8\left(\alpha_0 + 1 \right) \left( - \frac{Q}{6{H_0}^2} \right)^{\alpha_0+2}
+ 4 \left( - \frac{Q}{6{H_0}^2} \right) \right\} f_{(R)}'' \right] \right|_{R = 12 {H_0}^2 \left( - \frac{Q}{6{H_0}^2} \right)^{\alpha_0+1}} \nonumber \\
&\, + \rho \, .
\end{align}
The derivation of \eqref{arbQtotalmdl2} is the following,
\begin{align}
\label{arbQtotalmdl2Ap}
f_{(Q)}
=&\, \frac{1}{2} \left( -\frac{Q}{6{H_0}^2} \right)^\frac{1}{2} \int^Q \left( - \frac{dq}{6{H_0}^2} \right) \left( - \frac{q}{6{H_0}^2} \right)^{-\frac{3}{2}} \nonumber \\
&\, \times \left[ - f_{(R)} + 6{H_0}^2\left\{- \left( - \frac{q}{6{H_0}^2} \right) + 2\left( - \frac{q}{6{H_0}^2} \right)^{\alpha_0+1} \right\} f_{(R)}' \right. \nonumber \\
&\, \left. \left. - \left( 6{H_0}^2 \right)^2 \left\{ 8\left( \alpha_0 + 1 \right) \left( - \frac{q}{6{H_0}^2} \right)^{2\alpha_0+2} - 8\left(\alpha_0 + 1 \right) \left( - \frac{q}{6{H_0}^2} \right)^{\alpha_0+2}
+ 4 \left( - \frac{q}{6{H_0}^2} \right) \right\} f_{(R)}'' \right] \right|_{R = 12 {H_0}^2 \left( - \frac{q}{6{H_0}^2} \right)^{\alpha_0+1}} \nonumber \\
=&\, \frac{1}{2} \left( -\frac{Q}{6{H_0}^2} \right)^\frac{1}{2} \int^{ -\frac{Q}{6{H_0}^2}} dx x^{-\frac{3}{2}}
\left[ - f_{(R)} + 6{H_0}^2\left\{- x + 2 x^{\alpha_0+1} \right\} f_{(R)}' \right. \nonumber \\
&\, \left. \left. - \left( 6{H_0}^2 \right)^2 \left\{ 8\left( \alpha_0 + 1 \right) x^{2\alpha_0+2} - 8\left(\alpha_0 + 1 \right) x^{\alpha_0+2}
+ 4 \right\} f_{(R)}'' \right] \right|_{R = 12 {H_0}^2 x^{\alpha_0+1}} \nonumber \\
=&\, \frac{f_m}{2} \left( -\frac{Q}{6{H_0}^2} \right)^\frac{1}{2} \int^{ -\frac{Q}{6{H_0}^2}} dx x^{-\frac{3}{2}}
\left[ - \left(12 {H_0}^2\right)^m x^{m\left(\alpha_0+1\right)} + 6m{H_0}^2\left\{- x + 2 x^{\alpha_0+1} \right\} \left(12 {H_0}^2\right)^{m-1} x^{\left(m-1\right)\left(\alpha_0+1\right)} \right. \nonumber \\
&\, \left. \qquad \qquad - \left( 6{H_0}^2 \right)^2 \left\{ 8\left( \alpha_0 + 1 \right) x^{2\alpha_0+2} - 8\left(\alpha_0 + 1 \right) x^{\alpha_0+2}
+ 4 m \left( m - 1 \right) \right\} \left(12 {H_0}^2\right)^{m-2} x^{\left(m-2\right)\left(\alpha_0+1\right)} \right] \nonumber \\
=&\, \frac{f_m}{2} \left(12 {H_0}^2\right)^m \left( -\frac{Q}{6{H_0}^2} \right)^\frac{1}{2} \int^{ -\frac{Q}{6{H_0}^2}} dx x^{-\frac{3}{2}}
\left[ - x^{m\left(\alpha_0+1\right)} + \frac{1}{2} \left\{- x + 2 x^{\alpha_0+1} \right\} x^{\left(m-1\right)\left(\alpha_0+1\right)} \right. \nonumber \\
&\, \left. \qquad \qquad - \left\{ 2\left( \alpha_0 + 1 \right) x^{2\alpha_0+2} - 2 \left(\alpha_0 + 1 \right) x^{\alpha_0+2}
+ m \left( m - 1 \right) \right\} x^{\left(m-2\right)\left(\alpha_0+1\right)} \right] \nonumber \\
=&\, \frac{f_m}{2} \left(12 {H_0}^2\right)^m \left[ - \frac{1}{m\left(\alpha_0+1\right) - \frac{1}{2}} \left( -\frac{Q}{6{H_0}^2} \right)^{m\left(\alpha_0+1\right)}
 - \frac{1}{2\left\{\left(m-1\right)\left(\alpha_0+1\right) + \frac{1}{2}\right\}} \left( -\frac{Q}{6{H_0}^2} \right)^{\left(m-1\right)\left(\alpha_0+1\right) + 1} \right. \nonumber \\
&\, + \frac{1}{m \left(\alpha_0+1\right) - \frac{1}{2}} \left( -\frac{Q}{6{H_0}^2} \right)^{m \left(\alpha_0+1\right)}
 - \frac{2\left( \alpha_0 + 1 \right)}{m \left(\alpha_0+1\right) - \frac{1}{2}} \left( -\frac{Q}{6{H_0}^2} \right)^{m\left(\alpha_0+1\right)} \nonumber \\
&\, \left. + \frac{2\left( \alpha_0 + 1 \right)}{\left(m-2\right)\left(\alpha_0+1\right) + \alpha_0 + \frac{3}{2}} \left( -\frac{Q}{6{H_0}^2} \right)^{\left(m-2\right)\left(\alpha_0+1\right) + \alpha_0 + 2}
 - \frac{m \left( m - 1 \right)}{\left(m-2\right)\left(\alpha_0+1\right) - \frac{1}{2}} \left( -\frac{Q}{6{H_0}^2} \right)^{\left(m-2\right)\left(\alpha_0+1\right)} \right] \nonumber \\
=&\, \frac{f_m}{2} \left(12 {H_0}^2\right)^m \left[
\frac{2\alpha_0 + 1}{2\left\{\left(m-1\right)\left(\alpha_0+1\right) + \frac{1}{2}\right\}} \left( -\frac{Q}{6{H_0}^2} \right)^{\left(m-1\right)\left(\alpha_0+1\right) + 1}
 - \frac{2 \left( \alpha_0 + 1\right)}{m \left(\alpha_0+1\right) - \frac{1}{2}} \left( -\frac{Q}{6{H_0}^2} \right)^{m\left(\alpha_0+1\right)} \right. \nonumber \\
&\, \left. - \frac{m \left( m - 1 \right)}{\left(m-2\right)\left(\alpha_0+1\right) - \frac{1}{2}} \left( -\frac{Q}{6{H_0}^2} \right)^{\left(m-2\right)\left(\alpha_0+1\right)}
+ C \left( -\frac{Q}{6{H_0}^2} \right)^\frac{1}{2} \right] \nonumber \\
=&\, \frac{f_m}{2} \left(12 {H_0}^2\right)^m \left( -\frac{Q}{6{H_0}^2} \right)^{m\left(\alpha_0+1\right)} \left[
\frac{2\alpha_0 + 1}{2\left\{\left(m-1\right)\left(\alpha_0+1\right) + \frac{1}{2}\right\}} \left( -\frac{Q}{6{H_0}^2} \right)^{-\alpha_0}
 - \frac{2 \left( \alpha_0 + 1\right)}{m \left(\alpha_0+1\right) - \frac{1}{2}} \right. \nonumber \\
&\, \left. \qquad \qquad - \frac{m \left( m - 1 \right)}{\left(m-2\right)\left(\alpha_0+1\right) - \frac{1}{2}} \left( -\frac{Q}{6{H_0}^2} \right)^{-2\left(\alpha_0+1\right)} \right]
+ C \left( -\frac{Q}{6{H_0}^2} \right)^\frac{1}{2} \, .
\end{align}
The derivation of \eqref{mpotential} is the following,
\begin{align}
\label{mpotentialAp}
V(\sigma) =&\, \e^\sigma\left( \frac{\e^{-\sigma} -1}{m f_m} \right)^\frac{1}{m-1} - \e^{2\sigma} \left\{ \left( \frac{\e^{-\sigma} -1}{m f_m} \right)^\frac{1}{m-1}
+ f_m \left( \frac{\e^{-\sigma} -1}{m f_m} \right)^\frac{m}{m-1} \right\} \nonumber \\
=&\, mf_m \e^{2\sigma} \left( \frac{\e^{-\sigma} -1}{m f_m} \right)^\frac{m}{m-1} - f_m \e^{2\sigma} \left( \frac{\e^{-\sigma} -1}{m f_m} \right)^\frac{m}{m-1} \nonumber \\
=&\, \left(m-1\right) f_m \e^{2\sigma} \left( \frac{\e^{-\sigma} -1}{m f_m} \right)^\frac{m}{m-1} \, .
\end{align}
The derivation of Eq.~\eqref{mpotentialprime} is given by,
\begin{align}
\label{mpotentialprimeAp}
V'(\sigma)
=&\, 2\left(m-1\right) f_m \e^{2\sigma} \left( \frac{\e^{-\sigma} -1}{m f_m} \right)^\frac{m}{m-1}
 - \frac{m f_m}{mf_m} \e^\sigma \left( \frac{\e^{-\sigma} -1}{m f_m} \right)^\frac{1}{m-1} \nonumber \\
=&\, 2\left(m-1\right) f_m \e^{2\sigma} \left( \frac{\e^{-\sigma} -1}{m f_m} \right)^\frac{m}{m-1}
 - \e^\sigma \left( \frac{\e^{-\sigma} -1}{m f_m} \right)^\frac{1}{m-1} \nonumber \\
=&\, \left\{ 2\left(m-1\right) f_m \frac{1 - \e^{\sigma}}{m f_m} -1 \right\} \e^\sigma \left( \frac{\e^{-\sigma} -1}{m f_m} \right)^\frac{1}{m-1} \nonumber \\
=&\, \left\{m -2 - 2\left(m-1\right) \e^{\sigma} \right\} \frac{\e^\sigma}{m} \left( \frac{\e^{-\sigma} -1}{m f_m} \right)^\frac{1}{m-1} \, .
\end{align}

\section{Derivation of the equations in Sec.~\ref{SecV}}\label{SecVA}

We now derive \eqref{omgmdl1} and {Vmdl1}.
Because,
\begin{align}
\label{HdHddsclrmdl2}
\frac{\Phi'}{\Phi} =&\, - \frac{1}{2} \left( \ln \left( \Phi^2 \right) \right)'
= - \frac{2 A_0 \e^{4 \alpha_0 \left( N - N_0 \right)}}{1 + A_0 \e^{4 \alpha_0 \left( N - N_0 \right)}} \, , \nonumber \\
\frac{\Phi''}{\Phi} =&\, \left( \frac{\Phi'}{\Phi} \right)' + \frac{{\Phi'}^2}{\Phi^2} = - \frac{8\alpha_0 A_0 \e^{4 \alpha_0 \left( N - N_0 \right)}}{\left( 1 + A_0 \e^{4 \alpha_0 \left( N - N_0 \right)} \right)^2}
+ \frac{4 {A_0}^2 \e^{8 \alpha_0 \left( N - N_0 \right)}}{\left( 1 + A_0 \e^{4 \alpha_0 \left( N - N_0 \right)} \right)^2} \, , \nonumber \\
\frac{\Phi'''}{\Phi} =&\, \left( \frac{\Phi''}{\Phi} \right)' + \frac{\Phi'' \Phi'}{\Phi^2} \nonumber \\
=&\, - \frac{32{\alpha_0}^2 A_0 \e^{4 \alpha_0 \left( N - N_0 \right)}}{\left( 1 + A_0 \e^{4 \alpha_0 \left( N - N_0 \right)} \right)^2}
+ \frac{32 \alpha_0 {A_0}^2 \e^{8 \alpha_0 \left( N - N_0 \right)}}{\left( 1 + A_0 \e^{4 \alpha_0 \left( N - N_0 \right)} \right)^2}
%%%%%%%%%
+ \frac{64 {\alpha_0}^2 {A_0}^2 \e^{8 \alpha_0 \left( N - N_0 \right)}}{\left( 1 + A_0 \e^{4 \alpha_0 \left( N - N_0 \right)} \right)^3}
 - \frac{32 \alpha_0 {A_0}^3 \e^{12 \alpha_0 \left( N - N_0 \right)}}{\left( 1 + A_0 \e^{4 \alpha_0 \left( N - N_0 \right)} \right)^3} \nonumber \\
&\, - \left\{ - \frac{8\alpha_0 A_0 \e^{4 \alpha_0 \left( N - N_0 \right)}}{\left( 1 + A_0 \e^{4 \alpha_0 \left( N - N_0 \right)} \right)^2}
+ \frac{4 {A_0}^2 \e^{8 \alpha_0 \left( N - N_0 \right)}}{\left( 1 + A_0 \e^{4 \alpha_0 \left( N - N_0 \right)} \right)^2} \right\}
\frac{2 A_0 \e^{4 \alpha_0 \left( N - N_0 \right)}}{1 + A_0 \e^{4 \alpha_0 \left( N - N_0 \right)}} \nonumber \\
=&\, - \frac{32{\alpha_0}^2 A_0 \e^{4 \alpha_0 \left( N - N_0 \right)}}{\left( 1 + A_0 \e^{4 \alpha_0 \left( N - N_0 \right)} \right)^2}
+ \frac{32 \alpha_0 {A_0}^2 \e^{8 \alpha_0 \left( N - N_0 \right)}}{\left( 1 + A_0 \e^{4 \alpha_0 \left( N - N_0 \right)} \right)^2} \nonumber \\
%%%%%%%%%
&\, + \frac{\left( 64 {\alpha_0}^2 + 16 \alpha_0 \right){A_0}^2 \e^{8 \alpha_0 \left( N - N_0 \right)}}{\left( 1 + A_0 \e^{4 \alpha_0 \left( N - N_0 \right)} \right)^3}
 - \frac{\left(32 \alpha_0 + 8 \right) {A_0}^3 \e^{12 \alpha_0 \left( N - N_0 \right)}}{\left( 1 + A_0 \e^{4 \alpha_0 \left( N - N_0 \right)} \right)^3} \nonumber \\
R=&\, 12 \Phi^2 + 6\Phi\Phi' = \frac{{H_0}^2}{\left(1 + A_0 \e^{4\alpha_0 \left( \phi - N_0 \right)}\right)^\frac{1}{\alpha_0}}
\left( 12 - \frac{12 A_0 \e^{4 \alpha_0 \left( N - N_0 \right)}}{1 + A_0 \e^{4 \alpha_0 \left( N - N_0 \right)}} \right)
= \frac{12{H_0}^2}{\left(1 + A_0 \e^{4\alpha_0 \left( \phi - N_0 \right)}\right)^{\frac{1}{\alpha_0}+1}} \, .
\, ,
\end{align}
we have,
\begin{align}
\label{omgmdl1Ap}
\omega\left( \phi \right) =&\, \frac{8 A_0 \e^{4 \alpha_0 \left( \phi - N_0 \right)}}{1 + A_0 \e^{4 \alpha_0 \left( \phi - N_0 \right)}}
f_R \left( \frac{12{H_0}^2}{\left(1 + A_0 \e^{4\alpha_0 \left( \phi - N_0 \right)}\right)^{\frac{1}{\alpha_0}+1}} \right) \nonumber \\
&\, + \frac{12{H_0}^2}{\left(1 + A_0 \e^{4\alpha_0 \left( \phi - N_0 \right)}\right)^\frac{1}{\alpha_0}} \left[ - \frac{8 A_0 \e^{4 \alpha_0 \left( \phi - N_0 \right)}}{1 + A_0 \e^{4 \alpha_0 \left( \phi - N_0 \right)}}
 - 7 \left\{ - \frac{2 A_0 \e^{4 \alpha_0 \left( \phi - N_0 \right)}}{1 + A_0 \e^{4 \alpha_0 \left( \phi - N_0 \right)}} \right\}^2 \right. \nonumber \\
&\, - 3 \left\{ - \frac{8\alpha_0 A_0 \e^{4 \alpha_0 \left( \phi - N_0 \right)}}{\left( 1 + A_0 \e^{4 \alpha_0 \left( \phi - N_0 \right)} \right)^2}
+ \frac{4 {A_0}^2 \e^{8 \alpha_0 \left( \phi - N_0 \right)}}{\left( 1 + A_0 \e^{4 \alpha_0 \left( \phi - N_0 \right)} \right)^2} \right\}
 - \left\{ - \frac{2 A_0 \e^{4 \alpha_0 \left( \phi - N_0 \right)}}{1 + A_0 \e^{4 \alpha_0 \left( \phi - N_0 \right)}} \right\}^3 \nonumber \\
&\, + \frac{8 A_0 \e^{4 \alpha_0 \left( \phi - N_0 \right)}}{1 + A_0 \e^{4 \alpha_0 \left( \phi - N_0 \right)}}
\left\{ - \frac{8\alpha_0 A_0 \e^{4 \alpha_0 \left( \phi - N_0 \right)}}{\left( 1 + A_0 \e^{4 \alpha_0 \left( \phi - N_0 \right)} \right)^2}
+ \frac{4 {A_0}^2 \e^{8 \alpha_0 \left( \phi - N_0 \right)}}{\left( 1 + A_0 \e^{4 \alpha_0 \left( \phi - N_0 \right)} \right)^2} \right\} \nonumber \\
&\, - \left\{ - \frac{32{\alpha_0}^2 A_0 \e^{4 \alpha_0 \left( \phi - N_0 \right)}}{\left( 1 + A_0 \e^{4 \alpha_0 \left( \phi - N_0 \right)} \right)^2}
+ \frac{32 \alpha_0 {A_0}^2 \e^{8 \alpha_0 \left( \phi - N_0 \right)}}{\left( 1 + A_0 \e^{4 \alpha_0 \left( \phi - N_0 \right)} \right)^2}
+ \frac{\left( 64 {\alpha_0}^2 + 16 \alpha_0 \right){A_0}^2 \e^{8 \alpha_0 \left( \phi - N_0 \right)}}{\left( 1 + A_0 \e^{4 \alpha_0 \left( \phi - N_0 \right)} \right)^3} \right. \nonumber \\
&\, \left. \left. - \frac{\left(32 \alpha_0 + 8 \right) {A_0}^3 \e^{12 \alpha_0 \left( \phi - N_0 \right)}}{\left( 1 + A_0 \e^{4 \alpha_0 \left( \phi - N_0 \right)} \right)^3} \right\}
\right] f_{RR} \left( \frac{12{H_0}^2}{\left(1 + A_0 \e^{4\alpha_0 \left( \phi - N_0 \right)}\right)^{\frac{1}{\alpha_0}+1}} \right) \nonumber \\
&\, - \frac{72{H_0}^2}{\left(1 + A_0 \e^{4\alpha_0 \left( \phi - N_0 \right)}\right)^\frac{1}{\alpha_0}} \left[
 - \frac{2 A_0 \e^{4 \alpha_0 \left( \phi - N_0 \right)}}{1 + A_0 \e^{4 \alpha_0 \left( \phi - N_0 \right)}}
+ \left\{ - \frac{2 A_0 \e^{4 \alpha_0 \left( \phi - N_0 \right)}}{1 + A_0 \e^{4 \alpha_0 \left( \phi - N_0 \right)}} \right\}^2
 - \frac{8\alpha_0 A_0 \e^{4 \alpha_0 \left( \phi - N_0 \right)}}{\left( 1 + A_0 \e^{4 \alpha_0 \left( \phi - N_0 \right)} \right)^2} \right. \nonumber \\
&\, \left. + \frac{4 {A_0}^2 \e^{8 \alpha_0 \left( \phi - N_0 \right)}}{\left( 1 + A_0 \e^{4 \alpha_0 \left( \phi - N_0 \right)} \right)^2} \right]^2
f_{RRR} \left( \frac{12{H_0}^2}{\left(1 + A_0 \e^{4\alpha_0 \left( \phi - N_0 \right)}\right)^{\frac{1}{\alpha_0}+1}} \right) \nonumber \\
%%%%%%%%%%%%%%
=&\, \frac{8 A_0 \e^{4 \alpha_0 \left( \phi - N_0 \right)}}{1 + A_0 \e^{4 \alpha_0 \left( \phi - N_0 \right)}}
f_R \left( \frac{12{H_0}^2}{\left(1 + A_0 \e^{4\alpha_0 \left( \phi - N_0 \right)}\right)^{\frac{1}{\alpha_0}+1}} \right)
+ \frac{12{H_0}^2}{\left(1 + A_0 \e^{4\alpha_0 \left( \phi - N_0 \right)}\right)^\frac{1}{\alpha_0}}
\left\{ - \frac{8 A_0 \e^{4 \alpha_0 \left( \phi - N_0 \right)}}{1 + A_0 \e^{4 \alpha_0 \left( \phi - N_0 \right)}} \right. \nonumber \\
&\, + \frac{\left( 24\alpha_0 + 32{\alpha_0}^2 \right) A_0 \e^{4 \alpha_0 \left( \phi - N_0 \right)}}{\left( 1 + A_0 \e^{4 \alpha_0 \left( \phi - N_0 \right)} \right)^2}
 - \frac{\left(40- 32 \alpha_0 \right) {A_0}^2 \e^{8 \alpha_0 \left( \phi - N_0 \right)}}{\left( 1 + A_0 \e^{4 \alpha_0 \left( \phi - N_0 \right)} \right)^2} \nonumber \\
&\, \left. - \frac{\left( 64 {\alpha_0}^2 + 80 \alpha_0 \right){A_0}^2 \e^{8 \alpha_0 \left( \phi - N_0 \right)}}{\left( 1 + A_0 \e^{4 \alpha_0 \left( \phi - N_0 \right)} \right)^3}
+ \frac{\left(32 \alpha_0 + 48 \right) {A_0}^3 \e^{12 \alpha_0 \left( \phi - N_0 \right)}}{\left( 1 + A_0 \e^{4 \alpha_0 \left( \phi - N_0 \right)} \right)^3}
\right\} f_{RR} \left( \frac{12{H_0}^2}{\left(1 + A_0 \e^{4\alpha_0 \left( \phi - N_0 \right)}\right)^{\frac{1}{\alpha_0}+1}} \right) \nonumber \\
&\, - \frac{72{H_0}^2}{\left(1 + A_0 \e^{4\alpha_0 \left( \phi - N_0 \right)}\right)^\frac{1}{\alpha_0}} \left[
 - \frac{2 A_0 \e^{4 \alpha_0 \left( \phi - N_0 \right)}}{1 + A_0 \e^{4 \alpha_0 \left( \phi - N_0 \right)}}
 - \frac{8\alpha_0 A_0 \e^{4 \alpha_0 \left( \phi - N_0 \right)}}{\left( 1 + A_0 \e^{4 \alpha_0 \left( \phi - N_0 \right)} \right)^2} \right. \nonumber \\
&\, \left. + \frac{8 {A_0}^2 \e^{8 \alpha_0 \left( \phi - N_0 \right)}}{\left( 1 + A_0 \e^{4 \alpha_0 \left( \phi - N_0 \right)} \right)^2} \right]^2
f_{RRR} \left( \frac{12{H_0}^2}{\left(1 + A_0 \e^{4\alpha_0 \left( \phi - N_0 \right)}\right)^{\frac{1}{\alpha_0}+1}} \right) \, , \\
\label{Vmdl1Ap}
U\left( \phi \right) =&\, f \left( \frac{12{H_0}^2}{\left(1 + A_0 \e^{4\alpha_0 \left( \phi - N_0 \right)}\right)^{\frac{1}{\alpha_0}+1}} \right) \nonumber \\
&\, + 2 \frac{{H_0}^2}{\left(1 + A_0 \e^{4\alpha_0 \left( \phi - N_0 \right)}\right)^\frac{1}{\alpha_0}}
\left\{ - \frac{2 A_0 \e^{4 \alpha_0 \left( \phi - N_0 \right)}}{1 + A_0 \e^{4 \alpha_0 \left( \phi - N_0 \right)}} + 3 \right\}
f_R \left( \frac{12{H_0}^2}{\left(1 + A_0 \e^{4\alpha_0 \left( \phi - N_0 \right)}\right)^{\frac{1}{\alpha_0}+1}} \right) \nonumber \\
&\, + 6 \frac{{H_0}^4}{\left(1 + A_0 \e^{4\alpha_0 \left( \phi - N_0 \right)}\right)^\frac{2}{\alpha_0}}
\left[ - 20 \frac{2 A_0 \e^{4 \alpha_0 \left( \phi - N_0 \right)}}{1 + A_0 \e^{4 \alpha_0 \left( \phi - N_0 \right)}}
+ 13 \left\{ - \frac{2 A_0 \e^{4 \alpha_0 \left( \phi - N_0 \right)}}{1 + A_0 \e^{4 \alpha_0 \left( \phi - N_0 \right)}} \right\}^2 \right. \nonumber \\
&\, - 4 \frac{2 A_0 \e^{4 \alpha_0 \left( \phi - N_0 \right)}}{1 + A_0 \e^{4 \alpha_0 \left( \phi - N_0 \right)}}
\left\{ - \frac{8\alpha_0 A_0 \e^{4 \alpha_0 \left( \phi - N_0 \right)}}{\left( 1 + A_0 \e^{4 \alpha_0 \left( \phi - N_0 \right)} \right)^2}
+ \frac{4 {A_0}^2 \e^{8 \alpha_0 \left( \phi - N_0 \right)}}{\left( 1 + A_0 \e^{4 \alpha_0 \left( \phi - N_0 \right)} \right)^2} \right\}
 - \frac{32{\alpha_0}^2 A_0 \e^{4 \alpha_0 \left( \phi - N_0 \right)}}{\left( 1 + A_0 \e^{4 \alpha_0 \left( \phi - N_0 \right)} \right)^2} \nonumber \\
&\, \left. + \frac{32 \alpha_0 {A_0}^2 \e^{8 \alpha_0 \left( \phi - N_0 \right)}}{\left( 1 + A_0 \e^{4 \alpha_0 \left( \phi - N_0 \right)} \right)^2}
+ \frac{\left( 64 {\alpha_0}^2 + 16 \alpha_0 \right){A_0}^2 \e^{8 \alpha_0 \left( \phi - N_0 \right)}}{\left( 1 + A_0 \e^{4 \alpha_0 \left( \phi - N_0 \right)} \right)^3}
 - \frac{\left(32 \alpha_0 + 8 \right) {A_0}^3 \e^{12 \alpha_0 \left( \phi - N_0 \right)}}{\left( 1 + A_0 \e^{4 \alpha_0 \left( \phi - N_0 \right)} \right)^3}
\right] \nonumber \\
&\, \times f_{RR} \left( \frac{12{H_0}^2}{\left(1 + A_0 \e^{4\alpha_0 \left( \phi - N_0 \right)}\right)^{\frac{1}{\alpha_0}+1}} \right) \nonumber \\
&\, + \frac{{H_0}^6}{\left(1 + A_0 \e^{4\alpha_0 \left( \phi - N_0 \right)}\right)^\frac{3}{\alpha_0}}
\left[ - 4 \frac{2 A_0 \e^{4 \alpha_0 \left( \phi - N_0 \right)}}{1 + A_0 \e^{4 \alpha_0 \left( \phi - N_0 \right)}}
+ \left\{ - \frac{2 A_0 \e^{4 \alpha_0 \left( \phi - N_0 \right)}}{1 + A_0 \e^{4 \alpha_0 \left( \phi - N_0 \right)}} \right\}^2
- \frac{8\alpha_0 A_0 \e^{4 \alpha_0 \left( \phi - N_0 \right)}}{\left( 1 + A_0 \e^{4 \alpha_0 \left( \phi - N_0 \right)} \right)^2} \right. \nonumber \\
&\, \left. + \frac{4 {A_0}^2 \e^{8 \alpha_0 \left( \phi - N_0 \right)}}{\left( 1 + A_0 \e^{4 \alpha_0 \left( \phi - N_0 \right)} \right)^2}
\right]^2
f_{RRR} \left( \frac{12{H_0}^2}{\left(1 + A_0 \e^{4\alpha_0 \left( \phi - N_0 \right)}\right)^{\frac{1}{\alpha_0}+1}} \right) \nonumber \\
%%%%%%%%%%%%%%%%%%%%%
=&\, f \left( \frac{12{H_0}^2}{\left(1 + A_0 \e^{4\alpha_0 \left( \phi - N_0 \right)}\right)^{\frac{1}{\alpha_0}+1}} \right)
+ \frac{2{H_0}^2 \left( 3 + A_0 \e^{4\alpha_0 \left( \phi - N_0 \right)}\right)}{\left(1 + A_0 \e^{4\alpha_0 \left( \phi - N_0 \right)}\right)^{\frac{1}{\alpha_0} +1}}
f_R \left( \frac{12{H_0}^2}{\left(1 + A_0 \e^{4\alpha_0 \left( \phi - N_0 \right)}\right)^{\frac{1}{\alpha_0}+1}} \right) \nonumber \\
&\, + \frac{6{H_0}^4}{\left(1 + A_0 \e^{4\alpha_0 \left( \phi - N_0 \right)}\right)^\frac{2}{\alpha_0}}
\left[ - \frac{40 A_0 \e^{4 \alpha_0 \left( \phi - N_0 \right)}}{1 + A_0 \e^{4 \alpha_0 \left( \phi - N_0 \right)}}
 - \frac{32{\alpha_0}^2 A_0 \e^{4 \alpha_0 \left( \phi - N_0 \right)}}{\left( 1 + A_0 \e^{4 \alpha_0 \left( \phi - N_0 \right)} \right)^2}
+ \frac{\left(32 \alpha_0 + 52 \right){A_0}^2 \e^{8 \alpha_0 \left( \phi - N_0 \right)}}{\left( 1 + A_0 \e^{4 \alpha_0 \left( \phi - N_0 \right)} \right)^2}
\right. \nonumber \\
&\, \left.
+ \frac{\left( 64 {\alpha_0}^2 + 80 \alpha_0 \right){A_0}^2 \e^{8 \alpha_0 \left( \phi - N_0 \right)}}{\left( 1 + A_0 \e^{4 \alpha_0 \left( \phi - N_0 \right)} \right)^3}
 - \frac{\left(32 \alpha_0 + 40 \right) {A_0}^3 \e^{12 \alpha_0 \left( \phi - N_0 \right)}}{\left( 1 + A_0 \e^{4 \alpha_0 \left( \phi - N_0 \right)} \right)^3} \right]
f_{RR} \left( \frac{12{H_0}^2}{\left(1 + A_0 \e^{4\alpha_0 \left( \phi - N_0 \right)}\right)^{\frac{1}{\alpha_0}+1}} \right) \nonumber \\
&\, + \frac{{H_0}^6}{\left(1 + A_0 \e^{4\alpha_0 \left( \phi - N_0 \right)}\right)^\frac{3}{\alpha_0}}
\left[ - \frac{8 A_0 \e^{4 \alpha_0 \left( \phi - N_0 \right)}}{1 + A_0 \e^{4 \alpha_0 \left( \phi - N_0 \right)}}
- \frac{8\alpha_0 A_0 \e^{4 \alpha_0 \left( \phi - N_0 \right)}}{\left( 1 + A_0 \e^{4 \alpha_0 \left( \phi - N_0 \right)} \right)^2} \right. \nonumber \\
&\, \left. + \frac{8 {A_0}^2 \e^{8 \alpha_0 \left( \phi - N_0 \right)}}{\left( 1 + A_0 \e^{4 \alpha_0 \left( \phi - N_0 \right)} \right)^2}
\right]^2
f_{RRR} \left( \frac{12{H_0}^2}{\left(1 + A_0 \e^{4\alpha_0 \left( \phi - N_0 \right)}\right)^{\frac{1}{\alpha_0}+1}} \right) \, .
\end{align}

We now derive \eqref{fRsclromegaPhimdl2} and \eqref{fRsclrVPhimdl2}.
Because,
\begin{align}
\label{mdl2cal}
\Phi^2 =&\, \frac{{H_0}^2}{1 + \alpha_0 \e^{\alpha_1 N} + \beta_0 \e^{4N}}\, , \nonumber \\
\Phi \Phi' =&\, - \frac{{H_0}^2\left( \alpha_0 \alpha_1 \e^{\alpha_1 N} + 4\beta_0 \e^{4N} \right)}{2\left(1 + \alpha_0 \e^{\alpha_1 N} + \beta_0 \e^{4N}\right)^2} \, , \nonumber \\
\Phi \Phi'' + {\Phi'}^2 =&\, - \frac{{H_0}^2\left( \alpha_0 {\alpha_1}^2 \e^{\alpha_1 N} + 16\beta_0 \e^{4N} \right)}{2\left(1 + \alpha_0 \e^{\alpha_1 N} + \beta_0 \e^{4N}\right)^2}
+ \frac{{H_0}^2\left( \alpha_0 \alpha_1 \e^{\alpha_1 N} + 4\beta_0 \e^{4N} \right)^2}{\left(1 + \alpha_0 \e^{\alpha_1 N} + \beta_0 \e^{4N}\right)^3} \, , \nonumber \\
\Phi \Phi'' =&\, - \frac{{H_0}^2\left( \alpha_0 {\alpha_1}^2 \e^{\alpha_1 N} + 16\beta_0 \e^{4N} \right)}{2\left(1 + \alpha_0 \e^{\alpha_1 N} + \beta_0 \e^{4N}\right)^2}
+ \frac{{H_0}^2\left( \alpha_0 \alpha_1 \e^{\alpha_1 N} + 4\beta_0 \e^{4N} \right)^2}{\left(1 + \alpha_0 \e^{\alpha_1 N} + \beta_0 \e^{4N}\right)^3}
 - \frac{{H_0}^2\left( \alpha_0 \alpha_1 \e^{\alpha_1 N} + 4\beta_0 \e^{4N} \right)^2}{2\left(1 + \alpha_0 \e^{\alpha_1 N} + \beta_0 \e^{4N}\right)^3} \nonumber \\
=&\, - \frac{{H_0}^2\left( \alpha_0 {\alpha_1}^2 \e^{\alpha_1 N} + 16\beta_0 \e^{4N} \right)}{2\left(1 + \alpha_0 \e^{\alpha_1 N} + \beta_0 \e^{4N}\right)^2}
+ \frac{{H_0}^2\left( \alpha_0 \alpha_1 \e^{\alpha_1 N} + 4\beta_0 \e^{4N} \right)^2}{2\left(1 + \alpha_0 \e^{\alpha_1 N} + \beta_0 \e^{4N}\right)^3} \, , \nonumber \\
\Phi\Phi''' =&\, - \frac{{H_0}^2\left( \alpha_0 {\alpha_1}^3 \e^{\alpha_1 N} + 64 \beta_0 \e^{4N} \right)}{2\left(1 + \alpha_0 \e^{\alpha_1 N} + \beta_0 \e^{4N}\right)^2}
+ \frac{{H_0}^2\left( \alpha_0 {\alpha_1}^2 \e^{\alpha_1 N} + 16\beta_0 \e^{4N} \right)^2}{2\left(1 + \alpha_0 \e^{\alpha_1 N} + \beta_0 \e^{4N}\right)^3} \nonumber \\
&\, + \frac{{H_0}^2\left( \alpha_0 \alpha_1 \e^{\alpha_1 N} + 4\beta_0 \e^{4N} \right) \left( \alpha_0 {\alpha_1}^2 \e^{\alpha_1 N} + 16\beta_0 \e^{4N} \right)}
{2\left(1 + \alpha_0 \e^{\alpha_1 N} + \beta_0 \e^{4N}\right)^3}
 - \frac{3{H_0}^2\left( \alpha_0 \alpha_1 \e^{\alpha_1 N} + 4\beta_0 \e^{4N} \right)^3}{2\left(1 + \alpha_0 \e^{\alpha_1 N} + \beta_0 \e^{4N}\right)^4} \, , \nonumber \\
\Phi\Phi''' =&\, - \frac{{H_0}^2\left( \alpha_0 {\alpha_1}^3 \e^{\alpha_1 N} + 64 \beta_0 \e^{4N} \right)}{2\left(1 + \alpha_0 \e^{\alpha_1 N} + \beta_0 \e^{4N}\right)^2}
+ \frac{{H_0}^2\left( \alpha_0 {\alpha_1}^2 \e^{\alpha_1 N} + 16\beta_0 \e^{4N} \right)^2}{2\left(1 + \alpha_0 \e^{\alpha_1 N} + \beta_0 \e^{4N}\right)^3} \nonumber \\
&\, + \frac{{H_0}^2\left( \alpha_0 \alpha_1 \e^{\alpha_1 N} + 4\beta_0 \e^{4N} \right) \left( \alpha_0 {\alpha_1}^2 \e^{\alpha_1 N} + 16\beta_0 \e^{4N} \right)}
{2\left(1 + \alpha_0 \e^{\alpha_1 N} + \beta_0 \e^{4N}\right)^3}
 - \frac{3{H_0}^2\left( \alpha_0 \alpha_1 \e^{\alpha_1 N} + 4\beta_0 \e^{4N} \right)^3}{2\left(1 + \alpha_0 \e^{\alpha_1 N} + \beta_0 \e^{4N}\right)^4} \nonumber \\
&\, + \frac{\left( \alpha_0 \alpha_1 \e^{\alpha_1 N} + 4\beta_0 \e^{4N} \right)}{2\left(1 + \alpha_0 \e^{\alpha_1 N} + \beta_0 \e^{4N}\right)}
\left\{ - \frac{{H_0}^2\left( \alpha_0 {\alpha_1}^2 \e^{\alpha_1 N} + 16\beta_0 \e^{4N} \right)}{2\left(1 + \alpha_0 \e^{\alpha_1 N} + \beta_0 \e^{4N}\right)^2}
+ \frac{{H_0}^2\left( \alpha_0 \alpha_1 \e^{\alpha_1 N} + 4\beta_0 \e^{4N} \right)^2}{2\left(1 + \alpha_0 \e^{\alpha_1 N} + \beta_0 \e^{4N}\right)^3} \right\} \nonumber \\
=&\, - \frac{{H_0}^2\left( \alpha_0 {\alpha_1}^3 \e^{\alpha_1 N} + 64 \beta_0 \e^{4N} \right)}{2\left(1 + \alpha_0 \e^{\alpha_1 N} + \beta_0 \e^{4N}\right)^2}
+ \frac{{H_0}^2\left( \alpha_0 {\alpha_1}^2 \e^{\alpha_1 N} + 16\beta_0 \e^{4N} \right)^2}{2\left(1 + \alpha_0 \e^{\alpha_1 N} + \beta_0 \e^{4N}\right)^3} \nonumber \\
&\, + \frac{{H_0}^2\left( \alpha_0 \alpha_1 \e^{\alpha_1 N} + 4\beta_0 \e^{4N} \right) \left( \alpha_0 {\alpha_1}^2 \e^{\alpha_1 N} + 16\beta_0 \e^{4N} \right)}
{4\left(1 + \alpha_0 \e^{\alpha_1 N} + \beta_0 \e^{4N}\right)^3}
 - \frac{5{H_0}^2\left( \alpha_0 \alpha_1 \e^{\alpha_1 N} + 4\beta_0 \e^{4N} \right)^3}{4\left(1 + \alpha_0 \e^{\alpha_1 N} + \beta_0 \e^{4N}\right)^4} \, , \nonumber \\
R =&\, 12 \Phi^2 + 6 \Phi \Phi'
= \frac{12{H_0}^2}{1 + \alpha_0 \e^{\alpha_1 N} + \beta_0 \e^{4N}}
 - \frac{3{H_0}^2\left( \alpha_0 \alpha_1 \e^{\alpha_1 N} + 4\beta_0 \e^{4N} \right)}{\left(1 + \alpha_0 \e^{\alpha_1 N} + \beta_0 \e^{4N}\right)^2}
\end{align}
we find,
\begin{align}
\label{fRsclromegaPhimdl2Ap}
\omega(\phi)
=&\, \frac{4\left( \alpha_0 \alpha_1 \e^{\alpha_1 \phi} + 4\beta_0 \e^{4\phi} \right)}{2\left(1 + \alpha_0 \e^{\alpha_1 \phi} + \beta_0 \e^{4\phi}\right)}
f_R \left( \frac{12{H_0}^2}{1 + \alpha_0 \e^{\alpha_1 \phi} + \beta_0 \e^{4\phi}}
 - \frac{3{H_0}^2\left( \alpha_0 \alpha_1 \e^{\alpha_1 \phi} + 4\beta_0 \e^{4\phi} \right)}{\left(1 + \alpha_0 \e^{\alpha_1 \phi} + \beta_0 \e^{4\phi}\right)^2} \right) \nonumber \\
&\, + 12 \left[ - \frac{4{H_0}^2\left( \alpha_0 \alpha_1 \e^{\alpha_1 \phi} + 4\beta_0 \e^{4\phi} \right)}{2\left(1 + \alpha_0 \e^{\alpha_1 \phi} + \beta_0 \e^{4\phi}\right)^2}
 - 7 \frac{1 + \alpha_0 \e^{\alpha_1 \phi} + \beta_0 \e^{4\phi}}{{H_0}^2}
\left\{ - \frac{{H_0}^2\left( \alpha_0 \alpha_1 \e^{\alpha_1 \phi} + 4\beta_0 \e^{4\phi} \right)}{2\left(1 + \alpha_0 \e^{\alpha_1 \phi} + \beta_0 \e^{4\phi}\right)^2} \right\}^2 \right. \nonumber \\
&\, -3 \left\{ - \frac{{H_0}^2\left( \alpha_0 {\alpha_1}^2 \e^{\alpha_1 \phi} + 16\beta_0 \e^{4\phi} \right)}{2\left(1 + \alpha_0 \e^{\alpha_1 \phi} + \beta_0 \e^{4\phi}\right)^2}
+ \frac{{H_0}^2\left( \alpha_0 \alpha_1 \e^{\alpha_1 \phi} + 4\beta_0 \e^{4\phi} \right)^2}{2\left(1 + \alpha_0 \e^{\alpha_1 \phi} + \beta_0 \e^{4\phi}\right)^3} \right\} \nonumber \\
&\, - \frac{\left(1 + \alpha_0 \e^{\alpha_1 \phi} + \beta_0 \e^{4\phi}\right)^2}{{H_0}^4}
 \left\{ - \frac{{H_0}^2\left( \alpha_0 \alpha_1 \e^{\alpha_1 \phi} + 4\beta_0 \e^{4\phi} \right)}{2\left(1 + \alpha_0 \e^{\alpha_1 \phi} + \beta_0 \e^{4\phi}\right)^2} \right\}^3 \nonumber \\
&\, - 4 \frac{1 + \alpha_0 \e^{\alpha_1 \phi} + \beta_0 \e^{4\phi}}{{H_0}^2}
 \left\{ - \frac{{H_0}^2\left( \alpha_0 \alpha_1 \e^{\alpha_1 \phi} + 4\beta_0 \e^{4\phi} \right)}{2\left(1 + \alpha_0 \e^{\alpha_1 \phi} + \beta_0 \e^{4\phi}\right)^2} \right\} \nonumber \\
&\, \times \left\{ \frac{{H_0}^2\left( \alpha_0 {\alpha_1}^2 \e^{\alpha_1 \phi} + 16\beta_0 \e^{4\phi} \right)}{2\left(1 + \alpha_0 \e^{\alpha_1 \phi} + \beta_0 \e^{4\phi}\right)^2}
+ \frac{{H_0}^2\left( \alpha_0 \alpha_1 \e^{\alpha_1 \phi} + 4\beta_0 \e^{4\phi} \right)^2}{2\left(1 + \alpha_0 \e^{\alpha_1 \phi} + \beta_0 \e^{4\phi}\right)^3} \right\} \nonumber \\
&\, - \left\{ - \frac{{H_0}^2\left( \alpha_0 {\alpha_1}^3 \e^{\alpha_1 \phi} + 64 \beta_0 \e^{4\phi} \right)}{2\left(1 + \alpha_0 \e^{\alpha_1 \phi} + \beta_0 \e^{4\phi}\right)^2}
+ \frac{{H_0}^2\left( \alpha_0 {\alpha_1}^2 \e^{\alpha_1 \phi} + 16\beta_0 \e^{4\phi} \right)^2}{2\left(1 + \alpha_0 \e^{\alpha_1 \phi} + \beta_0 \e^{4\phi}\right)^3} \right. \nonumber \\
&\, \left. \left. + \frac{{H_0}^2\left( \alpha_0 \alpha_1 \e^{\alpha_1 \phi} + 4\beta_0 \e^{4\phi} \right) \left( \alpha_0 {\alpha_1}^2 \e^{\alpha_1 \phi} + 16\beta_0 \e^{4\phi} \right)}
{4\left(1 + \alpha_0 \e^{\alpha_1 \phi} + \beta_0 \e^{4\phi}\right)^3}
 - \frac{5{H_0}^2\left( \alpha_0 \alpha_1 \e^{\alpha_1 \phi} + 4\beta_0 \e^{4\phi} \right)^3}{4\left(1 + \alpha_0 \e^{\alpha_1 \phi} + \beta_0 \e^{4\phi}\right)^4} \right\} \right] \nonumber \\
&\, \times f_{RR} \left( \frac{12{H_0}^2}{1 + \alpha_0 \e^{\alpha_1 \phi} + \beta_0 \e^{4\phi}}
 - \frac{3{H_0}^2\left( \alpha_0 \alpha_1 \e^{\alpha_1 \phi} + 4\beta_0 \e^{4\phi} \right)}{\left(1 + \alpha_0 \e^{\alpha_1 \phi} + \beta_0 \e^{4\phi}\right)^2} \right) \nonumber \\
&\, - 72 \frac{1 + \alpha_0 \e^{\alpha_1 \phi} + \beta_0 \e^{4\phi}}{{H_0}^2} \left[
 - \frac{4{H_0}^2\left( \alpha_0 \alpha_1 \e^{\alpha_1 \phi} + 4\beta_0 \e^{4\phi} \right)}{2\left(1 + \alpha_0 \e^{\alpha_1 \phi} + \beta_0 \e^{4\phi}\right)^2} \right. \nonumber \\
&\, + \frac{1 + \alpha_0 \e^{\alpha_1 \phi} + \beta_0 \e^{4\phi}}{{H_0}^2}
\left\{ - \frac{{H_0}^2\left( \alpha_0 \alpha_1 \e^{\alpha_1 \phi} + 4\beta_0 \e^{4\phi} \right)}{2\left(1 + \alpha_0 \e^{\alpha_1 \phi} + \beta_0 \e^{4\phi}\right)^2} \right\}^2 \nonumber \\
&\, \left. - \frac{{H_0}^2\left( \alpha_0 {\alpha_1}^2 \e^{\alpha_1 \phi} + 16\beta_0 \e^{4\phi} \right)}{2\left(1 + \alpha_0 \e^{\alpha_1 \phi} + \beta_0 \e^{4\phi}\right)^2}
+ \frac{{H_0}^2\left( \alpha_0 \alpha_1 \e^{\alpha_1 \phi} + 4\beta_0 \e^{4\phi} \right)^2}{2\left(1 + \alpha_0 \e^{\alpha_1 \phi} + \beta_0 \e^{4\phi}\right)^3} \right] \nonumber \\
&\, \times f_{RRR} \left( \frac{12{H_0}^2}{1 + \alpha_0 \e^{\alpha_1 \phi} + \beta_0 \e^{4\phi}}
 - \frac{3{H_0}^2\left( \alpha_0 \alpha_1 \e^{\alpha_1 \phi} + 4\beta_0 \e^{4\phi} \right)}{\left(1 + \alpha_0 \e^{\alpha_1 \phi} + \beta_0 \e^{4\phi}\right)^2} \right) \nonumber \\
%%%%%%%%%%%%
%%%%%%%%%%%%%%%%
=&\, \frac{2\left( \alpha_0 \alpha_1 \e^{\alpha_1 \phi} + 4\beta_0 \e^{4\phi} \right)}{1 + \alpha_0 \e^{\alpha_1 \phi} + \beta_0 \e^{4\phi}}
f_R \left( \frac{12{H_0}^2}{1 + \alpha_0 \e^{\alpha_1 \phi} + \beta_0 \e^{4\phi}}
 - \frac{3{H_0}^2\left( \alpha_0 \alpha_1 \e^{\alpha_1 \phi} + 4\beta_0 \e^{4\phi} \right)}{\left(1 + \alpha_0 \e^{\alpha_1 \phi} + \beta_0 \e^{4\phi}\right)^2} \right) \nonumber \\
&\, + 12 {H_0}^2 \left[ - \frac{4 \left( \alpha_0 \alpha_1 \e^{\alpha_1 \phi} + 4\beta_0 \e^{4\phi} \right)}{2\left(1 + \alpha_0 \e^{\alpha_1 \phi} + \beta_0 \e^{4\phi}\right)^2}
 - \frac{7\left( \alpha_0 \alpha_1 \e^{\alpha_1 \phi} + 4\beta_0 \e^{4\phi} \right)^2}{4\left(1 + \alpha_0 \e^{\alpha_1 \phi} + \beta_0 \e^{4\phi}\right)^3} \right. \nonumber \\
&\, + \frac{3 \left( \alpha_0 {\alpha_1}^2 \e^{\alpha_1 \phi} + 16\beta_0 \e^{4\phi} \right)}{2\left(1 + \alpha_0 \e^{\alpha_1 \phi} + \beta_0 \e^{4\phi}\right)^2}
 - \frac{3\left( \alpha_0 \alpha_1 \e^{\alpha_1 \phi} + 4\beta_0 \e^{4\phi} \right)^2}{2\left(1 + \alpha_0 \e^{\alpha_1 \phi} + \beta_0 \e^{4\phi}\right)^3}
+ \frac{\left( \alpha_0 \alpha_1 \e^{\alpha_1 \phi} + 4\beta_0 \e^{4\phi} \right)^3}{8\left(1 + \alpha_0 \e^{\alpha_1 \phi} + \beta_0 \e^{4\phi}\right)^4} \nonumber \\
&\, + \frac{2\left( \alpha_0 \alpha_1 \e^{\alpha_1 \phi} + 4\beta_0 \e^{4\phi} \right)}{1 + \alpha_0 \e^{\alpha_1 \phi} + \beta_0 \e^{4\phi}}
\left\{ \frac{\left( \alpha_0 {\alpha_1}^2 \e^{\alpha_1 \phi} + 16\beta_0 \e^{4\phi} \right)}{2\left(1 + \alpha_0 \e^{\alpha_1 \phi} + \beta_0 \e^{4\phi}\right)^2}
+ \frac{\left( \alpha_0 \alpha_1 \e^{\alpha_1 \phi} + 4\beta_0 \e^{4\phi} \right)^2}{2\left(1 + \alpha_0 \e^{\alpha_1 \phi} + \beta_0 \e^{4\phi}\right)^3} \right\} \nonumber \\
&\, + \frac{\left( \alpha_0 {\alpha_1}^3 \e^{\alpha_1 \phi} + 64 \beta_0 \e^{4\phi} \right)}{2\left(1 + \alpha_0 \e^{\alpha_1 \phi} + \beta_0 \e^{4\phi}\right)^2}
 - \frac{\left( \alpha_0 {\alpha_1}^2 \e^{\alpha_1 \phi} + 16\beta_0 \e^{4\phi} \right)^2}{2\left(1 + \alpha_0 \e^{\alpha_1 \phi} + \beta_0 \e^{4\phi}\right)^3} \nonumber \\
&\, \left. - \frac{\left( \alpha_0 \alpha_1 \e^{\alpha_1 \phi} + 4\beta_0 \e^{4\phi} \right) \left( \alpha_0 {\alpha_1}^2 \e^{\alpha_1 \phi} + 16\beta_0 \e^{4\phi} \right)}
{4\left(1 + \alpha_0 \e^{\alpha_1 \phi} + \beta_0 \e^{4\phi}\right)^3}
+ \frac{5\left( \alpha_0 \alpha_1 \e^{\alpha_1 \phi} + 4\beta_0 \e^{4\phi} \right)^3}{4\left(1 + \alpha_0 \e^{\alpha_1 \phi} + \beta_0 \e^{4\phi}\right)^4} \right] \nonumber \\
&\, \times f_{RR} \left( \frac{12{H_0}^2}{1 + \alpha_0 \e^{\alpha_1 \phi} + \beta_0 \e^{4\phi}}
 - \frac{3{H_0}^2\left( \alpha_0 \alpha_1 \e^{\alpha_1 \phi} + 4\beta_0 \e^{4\phi} \right)}{\left(1 + \alpha_0 \e^{\alpha_1 \phi} + \beta_0 \e^{4\phi}\right)^2} \right) \nonumber \\
&\, - 72 \left[ - \frac{2\left( \alpha_0 \alpha_1 \e^{\alpha_1 \phi} + 4\beta_0 \e^{4\phi} \right)}{1 + \alpha_0 \e^{\alpha_1 \phi} + \beta_0 \e^{4\phi}}
+ \frac{\left( \alpha_0 \alpha_1 \e^{\alpha_1 \phi} + 4\beta_0 \e^{4\phi} \right)^2}{4\left(1 + \alpha_0 \e^{\alpha_1 \phi} + \beta_0 \e^{4\phi}\right)^2} \right. \nonumber \\
&\, \left. - \frac{\left( \alpha_0 {\alpha_1}^2 \e^{\alpha_1 \phi} + 16\beta_0 \e^{4\phi} \right)}{2\left(1 + \alpha_0 \e^{\alpha_1 \phi} + \beta_0 \e^{4\phi}\right)}
+ \frac{\left( \alpha_0 \alpha_1 \e^{\alpha_1 \phi} + 4\beta_0 \e^{4\phi} \right)^2}{2\left(1 + \alpha_0 \e^{\alpha_1 \phi} + \beta_0 \e^{4\phi}\right)^2} \right] \nonumber \\
&\, \times f_{RRR} \left( \frac{12{H_0}^2}{1 + \alpha_0 \e^{\alpha_1 \phi} + \beta_0 \e^{4\phi}}
 - \frac{3{H_0}^2\left( \alpha_0 \alpha_1 \e^{\alpha_1 \phi} + 4\beta_0 \e^{4\phi} \right)}{\left(1 + \alpha_0 \e^{\alpha_1 \phi} + \beta_0 \e^{4\phi}\right)^2} \right) \nonumber \\
%%%%%%%%%%%
%%%%%%%%%
=&\, \frac{2\left( \alpha_0 \alpha_1 \e^{\alpha_1 \phi} + 4\beta_0 \e^{4\phi} \right)}{1 + \alpha_0 \e^{\alpha_1 \phi} + \beta_0 \e^{4\phi}}
f_R \left( \frac{12{H_0}^2}{1 + \alpha_0 \e^{\alpha_1 \phi} + \beta_0 \e^{4\phi}}
 - \frac{3{H_0}^2\left( \alpha_0 \alpha_1 \e^{\alpha_1 \phi} + 4\beta_0 \e^{4\phi} \right)}{\left(1 + \alpha_0 \e^{\alpha_1 \phi} + \beta_0 \e^{4\phi}\right)^2} \right) \nonumber \\
%%%%%%%%
&\, + 12 {H_0}^2 \left[
\frac{\alpha_0 \left( -4 \alpha_1 + 3 {\alpha_1}^2 + {\alpha_1}^3 \right)\e^{\alpha_1 \phi} + 96\beta_0 \e^{4\phi} }{2\left(1 + \alpha_0 \e^{\alpha_1 \phi} + \beta_0 \e^{4\phi}\right)^2} \right. \nonumber \\
&\, + \frac{ - 16\left( \alpha_0 \alpha_1 \e^{\alpha_1 \phi} + 4\beta_0 \e^{4\phi} \right)^2
+ 3\left( \alpha_0 \alpha_1 \e^{\alpha_1 \phi} + 4\beta_0 \e^{4\phi} \right)\left( \alpha_0 {\alpha_1}^2 \e^{\alpha_1 \phi} + 16\beta_0 \e^{4\phi} \right)
 - 2 \left( \alpha_0 {\alpha_1}^2 \e^{\alpha_1 \phi} + 16\beta_0 \e^{4\phi} \right)^2
}{4\left(1 + \alpha_0 \e^{\alpha_1 \phi} + \beta_0 \e^{4\phi}\right)^3} \nonumber \\
&\, \left. + \frac{19\left( \alpha_0 \alpha_1 \e^{\alpha_1 \phi} + 4\beta_0 \e^{4\phi} \right)^3}{8\left(1 + \alpha_0 \e^{\alpha_1 \phi} + \beta_0 \e^{4\phi}\right)^4} \right]
f_{RR} \left( \frac{12{H_0}^2}{1 + \alpha_0 \e^{\alpha_1 \phi} + \beta_0 \e^{4\phi}}
 - \frac{3{H_0}^2\left( \alpha_0 \alpha_1 \e^{\alpha_1 \phi} + 4\beta_0 \e^{4\phi} \right)}{\left(1 + \alpha_0 \e^{\alpha_1 \phi} + \beta_0 \e^{4\phi}\right)^2} \right) \nonumber \\
%%%%%%%%%%%%
&\, - 72 \left[ - \frac{\alpha_0 \left( 4 \alpha_1 + {\alpha_1}^2 \right) \e^{\alpha_1 \phi} + 32\beta_0 \e^{4\phi}}{2\left(1 + \alpha_0 \e^{\alpha_1 \phi} + \beta_0 \e^{4\phi}\right)}
+ \frac{3\left( \alpha_0 \alpha_1 \e^{\alpha_1 \phi} + 4\beta_0 \e^{4\phi} \right)^2}{4\left(1 + \alpha_0 \e^{\alpha_1 \phi} + \beta_0 \e^{4\phi}\right)^2} \right] \nonumber \\
&\, \times f_{RRR} \left( \frac{12{H_0}^2}{1 + \alpha_0 \e^{\alpha_1 \phi} + \beta_0 \e^{4\phi}}
 - \frac{3{H_0}^2\left( \alpha_0 \alpha_1 \e^{\alpha_1 \phi} + 4\beta_0 \e^{4\phi} \right)}{\left(1 + \alpha_0 \e^{\alpha_1 \phi} + \beta_0 \e^{4\phi}\right)^2} \right)
\, , \\
\label{fRsclrVPhimdl2Ap}
U\left( \phi \right)
=&\, f \left( \frac{12{H_0}^2}{1 + \alpha_0 \e^{\alpha_1 \phi} + \beta_0 \e^{4\phi}}
 - \frac{3{H_0}^2\left( \alpha_0 \alpha_1 \e^{\alpha_1 \phi} + 4\beta_0 \e^{4\phi} \right)}{\left(1 + \alpha_0 \e^{\alpha_1 \phi} + \beta_0 \e^{4\phi}\right)^2} \right) \nonumber \\
&\, + 2 \left\{ - \frac{{H_0}^2\left( \alpha_0 \alpha_1 \e^{\alpha_1 \phi} + 4\beta_0 \e^{4\phi} \right)}{2\left(1 + \alpha_0 \e^{\alpha_1 \phi} + \beta_0 \e^{4\phi}\right)^2}
+ 3 \frac{{H_0}^2}{1 + \alpha_0 \e^{\alpha_1 \phi} + \beta_0 \e^{4\phi}} \right\} \nonumber \\
&\, \times
f_R \left( \frac{12{H_0}^2}{1 + \alpha_0 \e^{\alpha_1 \phi} + \beta_0 \e^{4\phi}}
 - \frac{3{H_0}^2\left( \alpha_0 \alpha_1 \e^{\alpha_1 \phi} + 4\beta_0 \e^{4\phi} \right)}{\left(1 + \alpha_0 \e^{\alpha_1 \phi} + \beta_0 \e^{4\phi}\right)^2} \right) \nonumber \\
&\, + 6 \left[ - 20 \frac{{H_0}^2}{1 + \alpha_0 \e^{\alpha_1 \phi} + \beta_0 \e^{4\phi}}
\frac{{H_0}^2\left( \alpha_0 \alpha_1 \e^{\alpha_1 \phi} + 4\beta_0 \e^{4\phi} \right)}{2\left(1 + \alpha_0 \e^{\alpha_1 \phi} + \beta_0 \e^{4\phi}\right)^2}
+ 13 \left\{ \frac{{H_0}^2\left( \alpha_0 \alpha_1 \e^{\alpha_1 \phi} + 4\beta_0 \e^{4\phi} \right)}{2\left(1 + \alpha_0 \e^{\alpha_1 \phi} + \beta_0 \e^{4\phi}\right)^2} \right\}^2 \right. \nonumber \\
&\, - 4 \frac{{H_0}^2\left( \alpha_0 \alpha_1 \e^{\alpha_1 \phi} + 4\beta_0 \e^{4\phi} \right)}{2\left(1 + \alpha_0 \e^{\alpha_1 \phi} + \beta_0 \e^{4\phi}\right)^2}
\left\{ - \frac{{H_0}^2\left( \alpha_0 {\alpha_1}^2 \e^{\alpha_1 \phi} + 16\beta_0 \e^{4\phi} \right)}{2\left(1 + \alpha_0 \e^{\alpha_1 \phi} + \beta_0 \e^{4\phi}\right)^2}
+ \frac{{H_0}^2\left( \alpha_0 \alpha_1 \e^{\alpha_1 \phi} + 4\beta_0 \e^{4\phi} \right)^2}{2\left(1 + \alpha_0 \e^{\alpha_1 \phi} + \beta_0 \e^{4\phi}\right)^3} \right\} \nonumber \\
&\, + \frac{{H_0}^2}{1 + \alpha_0 \e^{\alpha_1 \phi} + \beta_0 \e^{4\phi}}
\left\{- \frac{{H_0}^2\left( \alpha_0 {\alpha_1}^3 \e^{\alpha_1 \phi} + 64 \beta_0 \e^{4\phi} \right)}{2\left(1 + \alpha_0 \e^{\alpha_1 \phi} + \beta_0 \e^{4\phi}\right)^2} \right. \nonumber \\
&\, + \frac{{H_0}^2\left( \alpha_0 {\alpha_1}^2 \e^{\alpha_1 \phi} + 16\beta_0 \e^{4\phi} \right)^2}{2\left(1 + \alpha_0 \e^{\alpha_1 \phi} + \beta_0 \e^{4\phi}\right)^3}
+ \frac{{H_0}^2\left( \alpha_0 \alpha_1 \e^{\alpha_1 \phi} + 4\beta_0 \e^{4\phi} \right) \left( \alpha_0 {\alpha_1}^2 \e^{\alpha_1 \phi} + 16\beta_0 \e^{4\phi} \right)}
{4\left(1 + \alpha_0 \e^{\alpha_1 \phi} + \beta_0 \e^{4\phi}\right)^3} \nonumber \\
&\, \left. \left. - \frac{5{H_0}^2\left( \alpha_0 \alpha_1 \e^{\alpha_1 \phi} + 4\beta_0 \e^{4\phi} \right)^3}{4\left(1 + \alpha_0 \e^{\alpha_1 \phi} + \beta_0 \e^{4\phi}\right)^4}
\right\}\right]
f_{RR} \left( \frac{12{H_0}^2}{1 + \alpha_0 \e^{\alpha_1 \phi} + \beta_0 \e^{4\phi}}
 - \frac{3{H_0}^2\left( \alpha_0 \alpha_1 \e^{\alpha_1 \phi} + 4\beta_0 \e^{4\phi} \right)}{\left(1 + \alpha_0 \e^{\alpha_1 \phi} + \beta_0 \e^{4\phi}\right)^2} \right) \nonumber \\
&\, + 36 \frac{{H_0}^2}{1 + \alpha_0 \e^{\alpha_1 \phi} + \beta_0 \e^{4\phi}} \left[
- \frac{{H_0}^2\left( \alpha_0 \alpha_1 \e^{\alpha_1 \phi} + 4\beta_0 \e^{4\phi} \right)}{2\left(1 + \alpha_0 \e^{\alpha_1 \phi} + \beta_0 \e^{4\phi}\right)^2} \right. \nonumber \\
&\, + \frac{1 + \alpha_0 \e^{\alpha_1 \phi} + \beta_0 \e^{4\phi}} {{H_0}^2} \left\{
- \frac{{H_0}^2\left( \alpha_0 \alpha_1 \e^{\alpha_1 \phi} + 4\beta_0 \e^{4\phi} \right)}{2\left(1 + \alpha_0 \e^{\alpha_1 \phi} + \beta_0 \e^{4\phi}\right)^2} \right\}^2 \nonumber \\
&\, \left. - \frac{{H_0}^2\left( \alpha_0 {\alpha_1}^2 \e^{\alpha_1 \phi} + 16\beta_0 \e^{4\phi} \right)}{2\left(1 + \alpha_0 \e^{\alpha_1 \phi} + \beta_0 \e^{4\phi}\right)^2}
+ \frac{{H_0}^2\left( \alpha_0 \alpha_1 \e^{\alpha_1 \phi} + 4\beta_0 \e^{4\phi} \right)^2}{2\left(1 + \alpha_0 \e^{\alpha_1 \phi} + \beta_0 \e^{4\phi}\right)^3}
\right]^2 \nonumber \\
&\, \times f_{RRR} \left( \frac{12{H_0}^2}{1 + \alpha_0 \e^{\alpha_1 \phi} + \beta_0 \e^{4\phi}}
 - \frac{3{H_0}^2\left( \alpha_0 \alpha_1 \e^{\alpha_1 \phi} + 4\beta_0 \e^{4\phi} \right)}{\left(1 + \alpha_0 \e^{\alpha_1 \phi} + \beta_0 \e^{4\phi}\right)^2} \right) \nonumber \\
=&\, f \left( \frac{12{H_0}^2}{1 + \alpha_0 \e^{\alpha_1 \phi} + \beta_0 \e^{4\phi}}
 - \frac{3{H_0}^2\left( \alpha_0 \alpha_1 \e^{\alpha_1 \phi} + 4\beta_0 \e^{4\phi} \right)}{\left(1 + \alpha_0 \e^{\alpha_1 \phi} + \beta_0 \e^{4\phi}\right)^2} \right) \nonumber \\
&\, + {H_0}^2 \left( - \frac{\alpha_0 \alpha_1 \e^{\alpha_1 \phi} + 4\beta_0 \e^{4\phi}}{\left(1 + \alpha_0 \e^{\alpha_1 \phi} + \beta_0 \e^{4\phi}\right)^2}
+ \frac{6}{1 + \alpha_0 \e^{\alpha_1 \phi} + \beta_0 \e^{4\phi}} \right) \nonumber \\
&\, \times
f_R \left( \frac{12{H_0}^2}{1 + \alpha_0 \e^{\alpha_1 \phi} + \beta_0 \e^{4\phi}}
 - \frac{3{H_0}^2\left( \alpha_0 \alpha_1 \e^{\alpha_1 \phi} + 4\beta_0 \e^{4\phi} \right)}{\left(1 + \alpha_0 \e^{\alpha_1 \phi} + \beta_0 \e^{4\phi}\right)^2} \right) \nonumber \\
&\, + 6 {H_0}^4\left[ - \frac{10\left( \alpha_0 \alpha_1 \e^{\alpha_1 \phi} + 4\beta_0 \e^{4\phi} \right)}{\left(1 + \alpha_0 \e^{\alpha_1 \phi} + \beta_0 \e^{4\phi}\right)^3}
+ \frac{13\left( \alpha_0 \alpha_1 \e^{\alpha_1 \phi} + 4\beta_0 \e^{4\phi} \right)^2}{4\left(1 + \alpha_0 \e^{\alpha_1 \phi} + \beta_0 \e^{4\phi}\right)^4} \right. \nonumber \\
&\, + \frac{\left( \alpha_0 \alpha_1 \e^{\alpha_1 \phi} + 4\beta_0 \e^{4\phi} \right)\left( \alpha_0 {\alpha_1}^2 \e^{\alpha_1 \phi} + 16\beta_0 \e^{4\phi} \right)}{\left(1 + \alpha_0 \e^{\alpha_1 \phi} + \beta_0 \e^{4\phi}\right)^4}
 - \frac{\left( \alpha_0 \alpha_1 \e^{\alpha_1 \phi} + 4\beta_0 \e^{4\phi} \right)^3}{\left(1 + \alpha_0 \e^{\alpha_1 \phi} + \beta_0 \e^{4\phi}\right)^5} \nonumber \\
&\, - \frac{\alpha_0 {\alpha_1}^3 \e^{\alpha_1 \phi} + 64 \beta_0 \e^{4\phi}}{2\left(1 + \alpha_0 \e^{\alpha_1 \phi} + \beta_0 \e^{4\phi}\right)^3} \nonumber \\
&\, + \frac{\left( \alpha_0 {\alpha_1}^2 \e^{\alpha_1 \phi} + 16\beta_0 \e^{4\phi} \right)^2}{2\left(1 + \alpha_0 \e^{\alpha_1 \phi} + \beta_0 \e^{4\phi}\right)^4}
+ \frac{\left( \alpha_0 \alpha_1 \e^{\alpha_1 \phi} + 4\beta_0 \e^{4\phi} \right) \left( \alpha_0 {\alpha_1}^2 \e^{\alpha_1 \phi} + 16\beta_0 \e^{4\phi} \right)}
{4\left(1 + \alpha_0 \e^{\alpha_1 \phi} + \beta_0 \e^{4\phi}\right)^4} \nonumber \\
&\, \left. - \frac{5\left( \alpha_0 \alpha_1 \e^{\alpha_1 \phi} + 4\beta_0 \e^{4\phi} \right)^3}{4\left(1 + \alpha_0 \e^{\alpha_1 \phi} + \beta_0 \e^{4\phi}\right)^5} \right]
f_{RR} \left( \frac{12{H_0}^2}{1 + \alpha_0 \e^{\alpha_1 \phi} + \beta_0 \e^{4\phi}}
 - \frac{3{H_0}^2\left( \alpha_0 \alpha_1 \e^{\alpha_1 \phi} + 4\beta_0 \e^{4\phi} \right)}{\left(1 + \alpha_0 \e^{\alpha_1 \phi} + \beta_0 \e^{4\phi}\right)^2} \right) \nonumber \\
&\, + \frac{36{H_0}^6}{1 + \alpha_0 \e^{\alpha_1 \phi} + \beta_0 \e^{4\phi}} \left[
- \frac{\alpha_0 \alpha_1 \e^{\alpha_1 \phi} + 4\beta_0 \e^{4\phi}}{2\left(1 + \alpha_0 \e^{\alpha_1 \phi} + \beta_0 \e^{4\phi}\right)^2} \right. \nonumber \\
&\, + \frac{\left( \alpha_0 \alpha_1 \e^{\alpha_1 \phi} + 4\beta_0 \e^{4\phi} \right)^2}{4\left(1 + \alpha_0 \e^{\alpha_1 \phi} + \beta_0 \e^{4\phi}\right)^3} \nonumber \\
&\, \left. - \frac{\alpha_0 {\alpha_1}^2 \e^{\alpha_1 \phi} + 16\beta_0 \e^{4\phi}}{2\left(1 + \alpha_0 \e^{\alpha_1 \phi} + \beta_0 \e^{4\phi}\right)^2}
+ \frac{\left( \alpha_0 \alpha_1 \e^{\alpha_1 \phi} + 4\beta_0 \e^{4\phi} \right)^2}{2\left(1 + \alpha_0 \e^{\alpha_1 \phi} + \beta_0 \e^{4\phi}\right)^3}
\right]^2 \nonumber \\
&\, \times f_{RRR} \left( \frac{12{H_0}^2}{1 + \alpha_0 \e^{\alpha_1 \phi} + \beta_0 \e^{4\phi}}
 - \frac{3{H_0}^2\left( \alpha_0 \alpha_1 \e^{\alpha_1 \phi} + 4\beta_0 \e^{4\phi} \right)}{\left(1 + \alpha_0 \e^{\alpha_1 \phi} + \beta_0 \e^{4\phi}\right)^2} \right) \nonumber \\
=&\, f \left( \frac{12{H_0}^2}{1 + \alpha_0 \e^{\alpha_1 \phi} + \beta_0 \e^{4\phi}}
 - \frac{3{H_0}^2\left( \alpha_0 \alpha_1 \e^{\alpha_1 \phi} + 4\beta_0 \e^{4\phi} \right)}{\left(1 + \alpha_0 \e^{\alpha_1 \phi} + \beta_0 \e^{4\phi}\right)^2} \right) \nonumber \\
&\, + {H_0}^2 \left( - \frac{\alpha_0 \alpha_1 \e^{\alpha_1 \phi} + 4\beta_0 \e^{4\phi}}{\left(1 + \alpha_0 \e^{\alpha_1 \phi} + \beta_0 \e^{4\phi}\right)^2}
+ \frac{6}{1 + \alpha_0 \e^{\alpha_1 \phi} + \beta_0 \e^{4\phi}} \right) \nonumber \\
&\, \times
f_R \left( \frac{12{H_0}^2}{1 + \alpha_0 \e^{\alpha_1 \phi} + \beta_0 \e^{4\phi}}
 - \frac{3{H_0}^2\left( \alpha_0 \alpha_1 \e^{\alpha_1 \phi} + 4\beta_0 \e^{4\phi} \right)}{\left(1 + \alpha_0 \e^{\alpha_1 \phi} + \beta_0 \e^{4\phi}\right)^2} \right) \nonumber \\
&\, + 6 {H_0}^4\left\{ - \frac{\alpha_0 \left( 20 \alpha_1 + {\alpha_1}^3 \right) \e^{\alpha_1 \phi} + 144\beta_0 \e^{4\phi}}{2\left(1 + \alpha_0 \e^{\alpha_1 \phi} + \beta_0 \e^{4\phi}\right)^3}
+ \frac{13\left( \alpha_0 \alpha_1 \e^{\alpha_1 \phi} + 4\beta_0 \e^{4\phi} \right)^2}{4\left(1 + \alpha_0 \e^{\alpha_1 \phi} + \beta_0 \e^{4\phi}\right)^4} \right. \nonumber \\
&\, + \frac{\left( \alpha_0 {\alpha_1}^2 \e^{\alpha_1 \phi} + 16\beta_0 \e^{4\phi} \right)^2}{2\left(1 + \alpha_0 \e^{\alpha_1 \phi} + \beta_0 \e^{4\phi}\right)^4}
+ \frac{5\left( \alpha_0 \alpha_1 \e^{\alpha_1 \phi} + 4\beta_0 \e^{4\phi} \right) \left( \alpha_0 {\alpha_1}^2 \e^{\alpha_1 \phi} + 16\beta_0 \e^{4\phi} \right)}
{4\left(1 + \alpha_0 \e^{\alpha_1 \phi} + \beta_0 \e^{4\phi}\right)^4} \nonumber \\
&\, \left. - \frac{9\left( \alpha_0 \alpha_1 \e^{\alpha_1 \phi} + 4\beta_0 \e^{4\phi} \right)^3}{4\left(1 + \alpha_0 \e^{\alpha_1 \phi} + \beta_0 \e^{4\phi}\right)^5} \right\}
f_{RR} \left( \frac{12{H_0}^2}{1 + \alpha_0 \e^{\alpha_1 \phi} + \beta_0 \e^{4\phi}}
 - \frac{3{H_0}^2\left( \alpha_0 \alpha_1 \e^{\alpha_1 \phi} + 4\beta_0 \e^{4\phi} \right)}{\left(1 + \alpha_0 \e^{\alpha_1 \phi} + \beta_0 \e^{4\phi}\right)^2} \right) \nonumber \\
&\, + \frac{36{H_0}^6}{1 + \alpha_0 \e^{\alpha_1 \phi} + \beta_0 \e^{4\phi}} \left[
- \frac{\alpha_0 \left(\alpha_1 + {\alpha_1}^2 \right) \e^{\alpha_1 \phi} + 20\beta_0 \e^{4\phi}}{2\left(1 + \alpha_0 \e^{\alpha_1 \phi} + \beta_0 \e^{4\phi}\right)^2}
+ \frac{3\left( \alpha_0 \alpha_1 \e^{\alpha_1 \phi} + 4\beta_0 \e^{4\phi} \right)^2}{4\left(1 + \alpha_0 \e^{\alpha_1 \phi} + \beta_0 \e^{4\phi}\right)^3}
\right]^2 \nonumber \\
&\, \times f_{RRR} \left( \frac{12{H_0}^2}{1 + \alpha_0 \e^{\alpha_1 \phi} + \beta_0 \e^{4\phi}}
 - \frac{3{H_0}^2\left( \alpha_0 \alpha_1 \e^{\alpha_1 \phi} + 4\beta_0 \e^{4\phi} \right)}{\left(1 + \alpha_0 \e^{\alpha_1 \phi} + \beta_0 \e^{4\phi}\right)^2} \right) \, .
\end{align}

\section{The derivations of the equations in Sec.~\ref{SecVI}}\label{SecVIAp}

The derivation of Eq.~\eqref{sumModelI0} as given as follows,
\begin{align}
\label{sumModelI0Ap}
\omega \left( \phi \right) =&\, \frac{8 A_0 \e^{4 \alpha_0 \left( \phi - N_0 \right)}}{1 + A_0 \e^{4 \alpha_0 \left( \phi - N_0 \right)}}
+ \alpha \left[ \frac{192A_0{H_0}^2}{\left(1 + A_0 \e^{4\alpha_0 \left( \phi - N_0 \right)}\right)^{\frac{1}{\alpha_0}+2}}
+ \frac{24{H_0}^2}{\left(1 + A_0 \e^{4\alpha_0 \left( \phi - N_0 \right)}\right)^\frac{1}{\alpha_0}} \right. \nonumber \\
&\, \times \left\{ - \frac{8 A_0 \e^{4 \alpha_0 \left( \phi - N_0 \right)}}{1 + A_0 \e^{4 \alpha_0 \left( \phi - N_0 \right)}}
+ \frac{\left( 24\alpha_0 + 32{\alpha_0}^2 \right) A_0 \e^{4 \alpha_0 \left( \phi - N_0 \right)}}{\left( 1 + A_0 \e^{4 \alpha_0 \left( \phi - N_0 \right)} \right)^2}
 - \frac{\left(40- 32 \alpha_0 \right) {A_0}^2 \e^{8 \alpha_0 \left( \phi - N_0 \right)}}{\left( 1 + A_0 \e^{4 \alpha_0 \left( \phi - N_0 \right)} \right)^2} \right. \nonumber \\
&\,\left. \left. - \frac{\left( 64 {\alpha_0}^2 + 80 \alpha_0 \right){A_0}^2 \e^{8 \alpha_0 \left( \phi - N_0 \right)}}{\left( 1 + A_0 \e^{4 \alpha_0 \left( \phi - N_0 \right)} \right)^3}
+ \frac{\left(32 \alpha_0 + 48 \right) {A_0}^3 \e^{12 \alpha_0 \left( \phi - N_0 \right)}}{\left( 1 + A_0 \e^{4 \alpha_0 \left( \phi - N_0 \right)} \right)^3}
\right\} \right] \nonumber \\
=&\, \frac{8 A_0 \e^{4 \alpha_0 \left( \phi - N_0 \right)}}{1 + A_0 \e^{4 \alpha_0 \left( \phi - N_0 \right)}}
+ \frac{24\alpha {H_0}^2}{\left(1 + A_0 \e^{4\alpha_0 \left( \phi - N_0 \right)}\right)^\frac{1}{\alpha_0}} \nonumber \\
&\, \times \left\{ - \frac{8 A_0 \e^{4 \alpha_0 \left( \phi - N_0 \right)}}{1 + A_0 \e^{4 \alpha_0 \left( \phi - N_0 \right)}}
+ \frac{8 A_0 + \left( 24\alpha_0 + 32{\alpha_0}^2 \right) A_0 \e^{4 \alpha_0 \left( \phi - N_0 \right)}
 - \left(40- 32 \alpha_0 \right) {A_0}^2 \e^{8 \alpha_0 \left( \phi - N_0 \right)}}{\left( 1 + A_0 \e^{4 \alpha_0 \left( \phi - N_0 \right)} \right)^2} \right. \nonumber \\
&\,\left. + \frac{- \left( 64 {\alpha_0}^2 + 80 \alpha_0 \right){A_0}^2 \e^{8 \alpha_0 \left( \phi - N_0 \right)}
+ \left(32 \alpha_0 + 48 \right) {A_0}^3 \e^{12 \alpha_0 \left( \phi - N_0 \right)}}{\left( 1 + A_0 \e^{4 \alpha_0 \left( \phi - N_0 \right)} \right)^3}
\right\} \nonumber \\
%%%%%%%%%%%%%
=&\, \frac{8 A_0 \e^{4 \alpha_0 \left( \phi - N_0 \right)}}{1 + A_0 \e^{4 \alpha_0 \left( \phi - N_0 \right)}}
+ \frac{24\alpha {H_0}^2}{\left(1 + A_0 \e^{4\alpha_0 \left( \phi - N_0 \right)}\right)^{\frac{1}{\alpha_0}+3}} \nonumber \\
&\, \times \left\{ 8A_0
+ \left(- 8 + 8 A_0 + 24\alpha_0 + 32{\alpha_0}^2 \right) A_0 \e^{4 \alpha_0 \left( \phi - N_0 \right)} \right. \nonumber \\
&\, + \left(- 16 + 24\alpha_0 + 32{\alpha_0}^2 - 40 + 32 \alpha_0 - 64 {\alpha_0}^2 - 80 \alpha_0 \right) {A_0}^2 \e^{8 \alpha_0 \left( \phi - N_0 \right)} \nonumber \\
&\, \left. + \left(- 8 - 40 + 32 \alpha_0 + 32 \alpha_0 + 48 \right) {A_0}3 \e^{12 \alpha_0 \left( \phi - N_0 \right)}
\right\} \, , \nonumber \\
%%%%%%%%%%%%
=&\, \frac{8 A_0 \e^{4 \alpha_0 \left( \phi - N_0 \right)}}{1 + A_0 \e^{4 \alpha_0 \left( \phi - N_0 \right)}}
+ \frac{24\alpha {H_0}^2}{\left(1 + A_0 \e^{4\alpha_0 \left( \phi - N_0 \right)}\right)^{\frac{1}{\alpha_0}+3}}
\left\{ 8A_0
+ \left(- 8 + 32 A_0 + 32{\alpha_0}^2 \right) A_0 \e^{4 \alpha_0 \left( \phi - N_0 \right)} \right. \nonumber \\
&\, \left. + \left(- 56 - 24\alpha_0 - 32{\alpha_0}^2 \right) {A_0}^2 \e^{8 \alpha_0 \left( \phi - N_0 \right)}
+ 64 \alpha_0 {A_0}^3 \e^{12 \alpha_0 \left( \phi - N_0 \right)}
\right\} \, , \nonumber \\
%%%%%%%%%%%
U\left( \phi \right) =&\, \left( \frac{12{H_0}^2}{\left(1 + A_0 \e^{4\alpha_0 \left( \phi - N_0 \right)}\right)^{\frac{1}{\alpha_0}+1}}
+ \frac{2{H_0}^2 \left( 3 + A_0 \e^{4\alpha_0 \left( \phi - N_0 \right)}\right)}{\left(1 + A_0 \e^{4\alpha_0 \left( \phi - N_0 \right)}\right)^{\frac{1}{\alpha_0} +1}} \right) \nonumber \\
&\, + 2\alpha \left[ \left( \frac{12{H_0}^2}{\left(1 + A_0 \e^{4\alpha_0 \left( \phi - N_0 \right)}\right)^{\frac{1}{\alpha_0}+1}} \right)^2
+ \frac{4{H_0}^2 \left( 3 + A_0 \e^{4\alpha_0 \left( \phi - N_0 \right)}\right)}{\left(1 + A_0 \e^{4\alpha_0 \left( \phi - N_0 \right)}\right)^{\frac{1}{\alpha_0} +1}}
\left( \frac{12{H_0}^2}{\left(1 + A_0 \e^{4\alpha_0 \left( \phi - N_0 \right)}\right)^{\frac{1}{\alpha_0}+1}} \right) \right. \nonumber \\
&\, + \frac{12{H_0}^4}{\left(1 + A_0 \e^{4\alpha_0 \left( \phi - N_0 \right)}\right)^\frac{2}{\alpha_0}}
\left\{ - \frac{40 A_0 \e^{4 \alpha_0 \left( \phi - N_0 \right)}}{1 + A_0 \e^{4 \alpha_0 \left( \phi - N_0 \right)}}
 - \frac{32{\alpha_0}^2 A_0 \e^{4 \alpha_0 \left( \phi - N_0 \right)}}{\left( 1 + A_0 \e^{4 \alpha_0 \left( \phi - N_0 \right)} \right)^2}
+ \frac{\left(32 \alpha_0 + 52 \right){A_0}^2 \e^{8 \alpha_0 \left( \phi - N_0 \right)}}{\left( 1 + A_0 \e^{4 \alpha_0 \left( \phi - N_0 \right)} \right)^2}
\right. \nonumber \\
&\, \left. \left.
+ \frac{\left( 64 {\alpha_0}^2 + 80 \alpha_0 \right){A_0}^2 \e^{8 \alpha_0 \left( \phi - N_0 \right)}}{\left( 1 + A_0 \e^{4 \alpha_0 \left( \phi - N_0 \right)} \right)^3}
 - \frac{\left(32 \alpha_0 + 40 \right) {A_0}^3 \e^{12 \alpha_0 \left( \phi - N_0 \right)}}{\left( 1 + A_0 \e^{4 \alpha_0 \left( \phi - N_0 \right)} \right)^3} \right\} \right] \nonumber \\
%%%%%%%%%%%%
=&\, \frac{2{H_0}^2 \left( 9 + A_0 \e^{4\alpha_0 \left( \phi - N_0 \right)}\right)}{\left(1 + A_0 \e^{4\alpha_0 \left( \phi - N_0 \right)}\right)^{\frac{1}{\alpha_0} +1}}
+ 24 \alpha {H_0}^4 \left[ \frac{4 \left( 6 + A_0 \e^{4\alpha_0 \left( \phi - N_0 \right)}\right)}{\left(1 + A_0 \e^{4\alpha_0 \left( \phi - N_0 \right)}\right)^{\frac{2}{\alpha_0} +2}}
+ \frac{1}{\left(1 + A_0 \e^{4\alpha_0 \left( \phi - N_0 \right)}\right)^{\frac{2}{\alpha_0}+ 3}} \right. \nonumber \\
&\, \left. \times \left\{ \left( - 40 - 32{\alpha_0}^2 \right) A_0 \e^{4 \alpha_0 \left( \phi - N_0 \right)}
+ \left( - 28 + 112 \alpha_0 + 32{\alpha_0}^2 \right) {A_0}^2 \e^{8 \alpha_0 \left( \phi - N_0 \right)}
 - 28 {A_0}^3 \e^{12 \alpha_0 \left( \phi - N_0 \right)} \right\} \right] \nonumber \\
=&\, \frac{2{H_0}^2 \left( 9 + A_0 \e^{4\alpha_0 \left( \phi - N_0 \right)}\right)}{\left(1 + A_0 \e^{4\alpha_0 \left( \phi - N_0 \right)}\right)^{\frac{1}{\alpha_0} +1}}
+ \frac{24 \alpha {H_0}^4}{\left(1 + A_0 \e^{4\alpha_0 \left( \phi - N_0 \right)}\right)^{\frac{2}{\alpha_0}+ 3}}
\left\{ 24 + \left( 28 - 40 - 32{\alpha_0}^2 \right) A_0 \e^{4 \alpha_0 \left( \phi - N_0 \right)} \right. \nonumber \\
&\, \left. + \left( 4 - 28 + 112 \alpha_0 + 32{\alpha_0}^2 \right) {A_0}^2 \e^{8 \alpha_0 \left( \phi - N_0 \right)}
 - 28 {A_0}^3 \e^{12 \alpha_0 \left( \phi - N_0 \right)} \right\} \nonumber \\
=&\, \frac{2{H_0}^2 \left( 9 + A_0 \e^{4\alpha_0 \left( \phi - N_0 \right)}\right)}{\left(1 + A_0 \e^{4\alpha_0 \left( \phi - N_0 \right)}\right)^{\frac{1}{\alpha_0} +1}}
+ \frac{96 \alpha {H_0}^4}{\left(1 + A_0 \e^{4\alpha_0 \left( \phi - N_0 \right)}\right)^{\frac{2}{\alpha_0}+ 3}}
\left\{ 6 + \left( - 3 - 8 {\alpha_0}^2 \right) A_0 \e^{4 \alpha_0 \left( \phi - N_0 \right)} \right. \nonumber \\
&\, \left. + \left( - 6 + 28 \alpha_0 + 8 {\alpha_0}^2 \right) {A_0}^2 \e^{8 \alpha_0 \left( \phi - N_0 \right)}
 - 7 {A_0}^3 \e^{12 \alpha_0 \left( \phi - N_0 \right)} \right\} \, .
\end{align}
We now derive \eqref{circsgmph}.
Because the scalar curvature $R$ is given by,
\begin{align}
\label{Ragain}
R=\frac{12{H_0}^2}{\left(1 + A_0 \e^{4\alpha_0 \left( \phi - N_0 \right)}\right)^{\frac{1}{\alpha_0}+1}}\, ,
\end{align}
with $\phi=N$, we also find,
\begin{align}
\label{esgm}
\e^{-\sigma} = f'(R) = 1 + 2\alpha R = 1 + \frac{24\alpha{H_0}^2}{\left(1 + A_0 \e^{4\alpha_0 \left( \phi - N_0 \right)}\right)^{\frac{1}{\alpha_0}+1}}\, .
\end{align}
Then the potential $V(\sigma)$ in \eqref{JGRG23sclr} has the following form,
\begin{align}
\label{Vsgma}
V(\sigma) =&\, \e^\sigma \frac{\e^{-\sigma} - 1}{2\alpha}
 - \e^{2\sigma} \left\{ \frac{\e^{-\sigma} - 1 }{2\alpha} + \alpha \frac{\left(\e^{-\sigma} - 1\right)^2}{4\alpha^2} \right\} \nonumber \\
=&\, \frac{1 - \e^\sigma}{2\alpha}
 - \e^{2\sigma} \left\{ \frac{\e^{-\sigma} - 1}{2\alpha} + \frac{\e^{-2\sigma} - 2 \e^{-\sigma} + 1}{4\alpha} \right\} \nonumber \\
=&\, \frac{1}{4\alpha} \left( 1 - 2 \e^\sigma + \e^{2\sigma} \right) \nonumber \\
=&\, \frac{\left( 1 - \e^\sigma \right)^2}{4\alpha} \nonumber \\
=&\, \frac{\left( \e^{-\sigma} - 1 \right)^2}{4\alpha \e^{-2\sigma}} \nonumber \\
=&\, \frac{144\alpha{H_0}^4}{\left(1 + A_0 \e^{4\alpha_0 \left( \phi - N_0 \right)}\right)^{\frac{2}{\alpha_0}+2}
\left( 1 + \frac{24\alpha{H_0}^2}{\left(1 + A_0 \e^{4\alpha_0 \left( \phi - N_0 \right)}\right)^{\frac{1}{\alpha_0}+1}} \right)^2}
\, .
\end{align}
Therefore, we obtain,
\begin{align}
\label{Vsgmadr}
V'(\sigma) =&\, - \frac{\left( 1 - \e^\sigma \right) \e^\sigma}{2\alpha} \nonumber \\
=&\, - \frac{\e^{-\sigma} - 1}{2\alpha \e^{-2\sigma}} \nonumber \\
=&\, - \frac{12{H_0}^2}{\left(1 + A_0 \e^{4\alpha_0 \left( \phi - N_0 \right)}\right)^{\frac{1}{\alpha_0}+1}
\left( 1 + \frac{24\alpha{H_0}^2}{\left(1 + A_0 \e^{4\alpha_0 \left( \phi - N_0 \right)}\right)^{\frac{1}{\alpha_0}+1}} \right)^2}
\, , \nonumber \\
V''(\sigma) =&\, - \frac{\e^\sigma - 2 \e^{2\sigma}}{2\alpha} \nonumber \\
=&\, - \frac{\e^{-\sigma} - 2}{2\alpha \e^{-2\sigma}} \nonumber \\
=&\, - \frac{ -1 + \frac{24\alpha{H_0}^2}{\left(1 + A_0 \e^{4\alpha_0 \left( \phi - N_0 \right)}\right)^{\frac{1}{\alpha_0}+1}}}
{2\alpha \left( 1 + \frac{24\alpha{H_0}^2}{\left(1 + A_0 \e^{4\alpha_0 \left( \phi - N_0 \right)}\right)^{\frac{1}{\alpha_0}+1}} \right)^2}
\, .
\end{align}
Then we find,
\begin{align}
\label{circsgmphAp}
\overset{\circ}{\sigma} =&\, \frac{\frac{96\alpha \alpha_0 A_0 {H_0}^2\left(\frac{1}{\alpha_0}+1\right)\e^{4\alpha_0 \left( \phi - N_0 \right)}}{\left(1 + A_0 \e^{4\alpha_0 \left( \phi - N_0 \right)}\right)^{\frac{1}{\alpha_0}+2}}}{1 + \frac{24\alpha{H_0}^2}{\left(1 + A_0 \e^{4\alpha_0 \left( \phi - N_0 \right)}\right)^{\frac{1}{\alpha_0}+1}}} H \e^{-\sigma} \nonumber \\
=&\, \frac{\frac{96\alpha \alpha_0 A_0 {H_0}^2\left(\frac{1}{\alpha_0}+1\right)\e^{4\alpha_0 \left( \phi - N_0 \right)}}{\left(1 + A_0 \e^{4\alpha_0 \left( \phi - N_0 \right)}\right)^{\frac{1}{\alpha_0}+2}}}{1 + \frac{24\alpha{H_0}^2}{\left(1 + A_0 \e^{4\alpha_0 \left( \phi - N_0 \right)}\right)^{\frac{1}{\alpha_0}+1}}}
\frac{H_0}{\left(1 + A_0 \e^{4\alpha_0 \left( N - N_0 \right)}\right)^\frac{1}{2\alpha_0}}
\left( 1 + \frac{24\alpha{H_0}^2}{\left(1 + A_0 \e^{4\alpha_0 \left( \phi - N_0 \right)}\right)^{\frac{1}{\alpha_0}+1}} \right) \nonumber \\
=&\, \frac{96\alpha \alpha_0 A_0 {H_0}^3\left(\frac{1}{\alpha_0}+1\right)\e^{4\alpha_0 \left( \phi - N_0 \right)}}{\left(1 + A_0 \e^{4\alpha_0 \left( \phi - N_0 \right)}\right)^{\frac{3}{2\alpha_0}+2}}
\, , \nonumber \\
\overset{\circ}{\phi} =&\, H \e^{-\sigma} \nonumber \\
=&\, \frac{H_0\left( 1 + \frac{24\alpha{H_0}^2}{\left(1 + A_0 \e^{4\alpha_0 \left( \phi - N_0 \right)}\right)^{\frac{1}{\alpha_0}+1}} \right)}
{\left(1 + A_0 \e^{4\alpha_0 \left( N - N_0 \right)}\right)^\frac{1}{2\alpha_0}}\, .
\end{align}
Therefore, $v^2$ in \eqref{vsqrt} is given by,
\begin{align}
\label{vsqrtAp}
v^2 =&\, \frac{27648\alpha^2 {\alpha_0}^2 {A_0}^2 {H_0}^6\left(\frac{1}{\alpha_0}+1\right)^2 \e^{8\alpha_0 \left( \phi - N_0 \right)}}
{\left(1 + A_0 \e^{4\alpha_0 \left( \phi - N_0 \right)}\right)^{\frac{6}{2\alpha_0}+4}} \nonumber \\
&\, + \left( 1 + \frac{24\alpha{H_0}^2}{\left(1 + A_0 \e^{4\alpha_0 \left( \phi - N_0 \right)}\right)^{\frac{1}{\alpha_0}+1}} \right)^{-1} \nonumber \\
&\, \times \left[ \frac{8 A_0 \e^{4 \alpha_0 \left( \phi - N_0 \right)}}{2\kappa^2 \left( 1 + A_0 \e^{4 \alpha_0 \left( \phi - N_0 \right)} \right)}
+ \frac{12\alpha {H_0}^2}{\kappa^2 \left(1 + A_0 \e^{4\alpha_0 \left( \phi - N_0 \right)}\right)^{\frac{1}{\alpha_0}+3}}
\left\{ 8A_0
+ \left(- 8 + 32 A_0 + 32{\alpha_0}^2 \right) A_0 \e^{4 \alpha_0 \left( \phi - N_0 \right)} \right. \right. \nonumber \\
&\, \left. \left. + \left(- 56 - 24\alpha_0 - 32{\alpha_0}^2 \right) {A_0}^2 \e^{8 \alpha_0 \left( \phi - N_0 \right)}
+ 64 \alpha_0 {A_0}^3 \e^{12 \alpha_0 \left( \phi - N_0 \right)} \right\} \right] \nonumber \\
&\, \times \frac{{H_0}^2\left( 1 + \frac{24\alpha{H_0}^2}{\left(1 + A_0 \e^{4\alpha_0 \left( \phi - N_0 \right)}\right)^{\frac{1}{\alpha_0}+1}} \right)^2}
{\left(1 + A_0 \e^{4\alpha_0 \left( N - N_0 \right)}\right)^\frac{1}{\alpha_0}} \nonumber \\
=&\, \frac{27648\alpha^2 {\alpha_0}^2 {A_0}^2 {H_0}^6\left(\frac{1}{\alpha_0}+1\right)^2 \e^{8\alpha_0 \left( \phi - N_0 \right)}}
{\left(1 + A_0 \e^{4\alpha_0 \left( \phi - N_0 \right)}\right)^{\frac{6}{2\alpha_0}+4}} \nonumber \\
&\, +\left[ \frac{8 A_0 \e^{4 \alpha_0 \left( \phi - N_0 \right)}}{2\kappa^2 \left( 1 + A_0 \e^{4 \alpha_0 \left( \phi - N_0 \right)} \right)}
+ \frac{12\alpha {H_0}^2}{\kappa^2 \left(1 + A_0 \e^{4\alpha_0 \left( \phi - N_0 \right)}\right)^{\frac{1}{\alpha_0}+3}}
\left\{ 8A_0
+ \left(- 8 + 32 A_0 + 32{\alpha_0}^2 \right) A_0 \e^{4 \alpha_0 \left( \phi - N_0 \right)} \right. \right. \nonumber \\
&\, \left. \left. + \left(- 56 - 24\alpha_0 - 32{\alpha_0}^2 \right) {A_0}^2 \e^{8 \alpha_0 \left( \phi - N_0 \right)}
+ 64 \alpha_0 {A_0}^3 \e^{12 \alpha_0 \left( \phi - N_0 \right)} \right\} \right]
\frac{{H_0}^2\left( 1 + \frac{24\alpha{H_0}^2}{\left(1 + A_0 \e^{4\alpha_0 \left( \phi - N_0 \right)}\right)^{\frac{1}{\alpha_0}+1}} \right)}
{\left(1 + A_0 \e^{4\alpha_0 \left( N - N_0 \right)}\right)^\frac{1}{\alpha_0}} \, .
\end{align}

We now derive \eqref{Ninf}.
For the preparation of the detailed checks, we calculate the following quantities,
\begin{align}
\label{sumModelI0dash}
\omega' \left( \phi \right)
=&\, \frac{32 A_0 \alpha_0 \e^{4 \alpha_0 \left( \phi - N_0 \right)}}{\left( 1 + A_0 \e^{4 \alpha_0 \left( \phi - N_0 \right)} \right)^2}
 - \frac{96\alpha \left( 1 + 3\alpha_0 \right) A_0 {H_0}^2\e^{4\alpha_0 \left( \phi - N_0 \right)}}
{\left(1 + A_0 \e^{4\alpha_0 \left( \phi - N_0 \right)}\right)^{\frac{1}{\alpha_0}+4}} \left\{ 8A_0
+ \left(- 8 + 32 A_0 + 32{\alpha_0}^2 \right) A_0 \e^{4 \alpha_0 \left( \phi - N_0 \right)} \right. \nonumber \\
&\, \left. + \left(- 56 - 24\alpha_0 - 32{\alpha_0}^2 \right) {A_0}^2 \e^{8 \alpha_0 \left( \phi - N_0 \right)}
+ 64 \alpha_0 {A_0}^3 \e^{12 \alpha_0 \left( \phi - N_0 \right)}
\right\} \nonumber \\
&\, + \frac{24\alpha {H_0}^2}{\left(1 + A_0 \e^{4\alpha_0 \left( \phi - N_0 \right)}\right)^{\frac{1}{\alpha_0}+3}}
\left\{ 32 \left(- 1 + 4 A_0 + 4{\alpha_0}^2 \right) \alpha_0 A_0 \e^{4 \alpha_0 \left( \phi - N_0 \right)} \right. \nonumber \\
&\, \left. + 64 \left(- 7 - 3\alpha_0 - 4{\alpha_0}^2 \right) \alpha_0 {A_0}^2 \e^{8 \alpha_0 \left( \phi - N_0 \right)}
+ 768 {\alpha_0}^2 {A_0}^3 \e^{12 \alpha_0 \left( \phi - N_0 \right)}
\right\} \nonumber \\
=&\, \frac{32 A_0 \alpha_0 \e^{4 \alpha_0 \left( \phi - N_0 \right)}}{\left( 1 + A_0 \e^{4 \alpha_0 \left( \phi - N_0 \right)} \right)^2}
+ \frac{768\alpha \left( 1 + 3\alpha_0 \right) A_0 {H_0}^2\e^{4\alpha_0 \left( \phi - N_0 \right)}}{\left(1 + A_0 \e^{4\alpha_0 \left( \phi - N_0 \right)}\right)^{\frac{1}{\alpha_0}+4}}
\left\{ - A_0 + \left( 1 - 4 A_0 - 4{\alpha_0}^2 \right) A_0 \e^{4 \alpha_0 \left( \phi - N_0 \right)} \right. \nonumber \\
&\, \left. + \left( 7 + 3\alpha_0 + 4 {\alpha_0}^2 \right) {A_0}^2 \e^{8 \alpha_0 \left( \phi - N_0 \right)}
+ 8 \alpha_0 {A_0}^3 \e^{12 \alpha_0 \left( \phi - N_0 \right)}
\right\} \nonumber \\
&\, + \frac{768\alpha A_0 {H_0}^2 \e^{4 \alpha_0 \left( \phi - N_0 \right)}}{\left(1 + A_0 \e^{4\alpha_0 \left( \phi - N_0 \right)}\right)^{\frac{1}{\alpha_0}+3}}
\left\{ \left(- 1 + 4 A_0 + 4{\alpha_0}^2 \right) \alpha_0 \right. \nonumber \\
&\, \left. + 2 \left(- 7 - 3\alpha_0 - 4{\alpha_0}^2 \right) \alpha_0 A_0 \e^{4 \alpha_0 \left( \phi - N_0 \right)}
+ 24 {\alpha_0}^2 {A_0}^2 \e^{8 \alpha_0 \left( \phi - N_0 \right)}
\right\} \nonumber \\
=&\, \frac{32 A_0 \alpha_0 \e^{4 \alpha_0 \left( \phi - N_0 \right)}}{\left( 1 + A_0 \e^{4 \alpha_0 \left( \phi - N_0 \right)} \right)^2}
+ \frac{768\alpha A_0 {H_0}^2\e^{4\alpha_0 \left( \phi - N_0 \right)}}{\left(1 + A_0 \e^{4\alpha_0 \left( \phi - N_0 \right)}\right)^{\frac{1}{\alpha_0}+4}}
\left\{ - A_0 \left( 1 + 3\alpha_0 \right) + \left( 1 - 4 A_0 - 4{\alpha_0}^2 \right) \left( 1 + 3\alpha_0 \right) A_0 \e^{4 \alpha_0 \left( \phi - N_0 \right)} \right. \nonumber \\
&\, \left. + \left( 7 + 3\alpha_0 + 4 {\alpha_0}^2 \right) \left( 1 + 3\alpha_0 \right) {A_0}^2 \e^{8 \alpha_0 \left( \phi - N_0 \right)}
+ 8 \alpha_0 \left( 1 + 3\alpha_0 \right) {A_0}^3 \e^{12 \alpha_0 \left( \phi - N_0 \right)}
\right\} \nonumber \\
&\, + \frac{768\alpha A_0 {H_0}^2 \e^{4 \alpha_0 \left( \phi - N_0 \right)}}{\left(1 + A_0 \e^{4\alpha_0 \left( \phi - N_0 \right)}\right)^{\frac{1}{\alpha_0}+4}}
\left\{ \left(- 1 + 4 A_0 + 4{\alpha_0}^2 \right) \alpha_0 \right. \nonumber \\
&\, + \left(- 14 - 6\alpha_0 - 8{\alpha_0}^2 - 1 + 4 A_0 + 4{\alpha_0}^2 \right) \alpha_0 A_0 \e^{4 \alpha_0 \left( \phi - N_0 \right)}
+ \left( - 14 - 6\alpha_0 - 8{\alpha_0}^2 + 24 \alpha_0 \right) \alpha_0 {A_0}^2 \e^{8 \alpha_0 \left( \phi - N_0 \right)} \nonumber \\
&\, \left. + 24 {\alpha_0}^2 {A_0}^3 \e^{12 \alpha_0 \left( \phi - N_0 \right)}
\right\} \nonumber \\
=&\, \frac{32 A_0 \alpha_0 \e^{4 \alpha_0 \left( \phi - N_0 \right)}}{\left( 1 + A_0 \e^{4 \alpha_0 \left( \phi - N_0 \right)} \right)^2}
+ \frac{768\alpha A_0 {H_0}^2\e^{4\alpha_0 \left( \phi - N_0 \right)}}{\left(1 + A_0 \e^{4\alpha_0 \left( \phi - N_0 \right)}\right)^{\frac{1}{\alpha_0}+4}}
\left\{ - A_0 - 3\alpha_0 A_0 - \alpha_0 + 4 A_0 + 4{\alpha_0}^3 \right. \nonumber \\
&\, + \left( 1 - 4 A_0 - 4{\alpha_0}^2 + 3 \alpha_0 - 12 \alpha_0 A_0 - 12 {\alpha_0}^3
 - 14 \alpha_0 - 6 {\alpha_0}^2 - 8{\alpha_0}^3 - \alpha_0 + 4 \alpha_0 A_0 + 4{\alpha_0}^3
\right) A_0 \e^{4 \alpha_0 \left( \phi - N_0 \right)} \nonumber \\
&\, + \left( 7 + 3\alpha_0 + 4 {\alpha_0}^2 + 21 \alpha_0 + 9 {\alpha_0}^2 + 12 {\alpha_0}^3
 - 14 \alpha_0 - 6 {\alpha_0}^2 - 8 {\alpha_0}^3 + 24 {\alpha_0}^2
\right) {A_0}^2 \e^{8 \alpha_0 \left( \phi - N_0 \right)} \nonumber \\
&\, \left. + \left( 8 \alpha_0 + 24 {\alpha_0}^2 + 24 {\alpha_0}^2 \right) {A_0}^3 \e^{12 \alpha_0 \left( \phi - N_0 \right)}
\right\} \nonumber \\
=&\, \frac{32 A_0 \alpha_0 \e^{4 \alpha_0 \left( \phi - N_0 \right)}}{\left( 1 + A_0 \e^{4 \alpha_0 \left( \phi - N_0 \right)} \right)^2}
+ \frac{768\alpha A_0 {H_0}^2\e^{4\alpha_0 \left( \phi - N_0 \right)}}{\left(1 + A_0 \e^{4\alpha_0 \left( \phi - N_0 \right)}\right)^{\frac{1}{\alpha_0}+4}}
\left\{ - \alpha_0 + 3 A_0 - 3\alpha_0 A_0 + 4{\alpha_0}^3 \right. \nonumber \\
&\, + \left( 1 - 12 \alpha_0 - 4 A_0 - 10 {\alpha_0}^2 - 8 \alpha_0 A_0 - 16 {\alpha_0}^3 \right) A_0 \e^{4 \alpha_0 \left( \phi - N_0 \right)} \nonumber \\
&\, \left. + \left( 7 + 10 \alpha_0 + 31 {\alpha_0}^2 + 4 {\alpha_0}^3 \right) {A_0}^2 \e^{8 \alpha_0 \left( \phi - N_0 \right)}
+ \left( 8 \alpha_0 + 48 {\alpha_0}^2 \right) {A_0}^3 \e^{12 \alpha_0 \left( \phi - N_0 \right)}
\right\} \nonumber \\
%%%%%%%
\omega''(\phi) =&\, \frac{128 A_0 {\alpha_0}^2 \e^{4 \alpha_0 \left( \phi - N_0 \right)}\left( 1 - A_0 \e^{4 \alpha_0 \left( \phi - N_0 \right)} \right)}
{\left( 1 + A_0 \e^{4 \alpha_0 \left( \phi - N_0 \right)} \right)^3} \nonumber \\
&\, + \frac{3072 \alpha A_0 \alpha_0 {H_0}^2\e^{4\alpha_0 \left( \phi - N_0 \right)}\left\{ \left(1 + A_0 \e^{4\alpha_0 \left( \phi - N_0 \right)}\right)
 - \left( \frac{1}{\alpha_0}+4 \right) A_0 \e^{4\alpha_0 \left( \phi - N_0 \right)} \right\}
}{\left(1 + A_0 \e^{4\alpha_0 \left( \phi - N_0 \right)}\right)^{\frac{1}{\alpha_0}+5}}
\left\{ - \alpha_0 + 3 A_0 - 3\alpha_0 A_0 + 4{\alpha_0}^3 \right. \nonumber \\
&\, + \left( 1 - 12 \alpha_0 - 4 A_0 - 10 {\alpha_0}^2 - 8 \alpha_0 A_0 - 16 {\alpha_0}^3 \right) A_0 \e^{4 \alpha_0 \left( \phi - N_0 \right)} \nonumber \\
&\, \left. + \left( 7 + 10 \alpha_0 + 31 {\alpha_0}^2 + 4 {\alpha_0}^3 \right) {A_0}^2 \e^{8 \alpha_0 \left( \phi - N_0 \right)}
+ \left( 8 \alpha_0 + 48 {\alpha_0}^2 \right) {A_0}^3 \e^{12 \alpha_0 \left( \phi - N_0 \right)}
\right\} \nonumber \\
&\, + \frac{3072\alpha A_0 {H_0}^2\e^{4\alpha_0 \left( \phi - N_0 \right)}}{\left(1 + A_0 \e^{4\alpha_0 \left( \phi - N_0 \right)}\right)^{\frac{1}{\alpha_0}+4}}
\left\{ \alpha_0 \left( 1 - 12 \alpha_0 - 4 A_0 - 10 {\alpha_0}^2 - 8 \alpha_0 A_0 - 16 {\alpha_0}^3 \right) A_0 \e^{4 \alpha_0 \left( \phi - N_0 \right)} \right. \nonumber \\
&\, \left. + 2\alpha_0 \left( 7 + 10 \alpha_0 + 31 {\alpha_0}^2 + 4 {\alpha_0}^3 \right) {A_0}^2 \e^{8 \alpha_0 \left( \phi - N_0 \right)}
+ 3 \alpha_0 \left( 8 \alpha_0 + 48 {\alpha_0}^2 \right) {A_0}^3 \e^{12 \alpha_0 \left( \phi - N_0 \right)}
\right\} \nonumber \\
=&\, \frac{128 A_0 {\alpha_0}^2 \e^{4 \alpha_0 \left( \phi - N_0 \right)}\left( 1 - A_0 \e^{4 \alpha_0 \left( \phi - N_0 \right)} \right)}
{\left( 1 + A_0 \e^{4 \alpha_0 \left( \phi - N_0 \right)} \right)^3} \nonumber \\
%%%%%%%%%%%%%%%%%
%%%%%%%%%%
&\, + \frac{3072\alpha A_0 {H_0}^2\e^{4\alpha_0 \left( \phi - N_0 \right)}}{\left(1 + A_0 \e^{4\alpha_0 \left( \phi - N_0 \right)}\right)^{\frac{1}{\alpha_0}+4}}
\left\{ - {\alpha_0}^2 + 3 A_0 \alpha_0 - 3{\alpha_0}^2 A_0 + 4{\alpha_0}^4 \right. \nonumber \\
&\, + \left( \alpha_0 - 3 A_0 + 3\alpha_0 A_0 - 4{\alpha_0}^3
+ 3 {\alpha_0}^2 - 9 A_0 \alpha_0 + 9 {\alpha_0}^2 A_0 - 12 {\alpha_0}^4
+ \alpha_0 - 12 {\alpha_0}^2 - 4 A_0 \alpha_0 - 10 {\alpha_0}^3 \right. \nonumber \\
&\, \left. - 8 {\alpha_0}^2 A_0 - 16 {\alpha_0}^4
+ \alpha_0 - 12 {\alpha_0}^2 - 4 A_0 \alpha_0 - 10 {\alpha_0}^3 - 8 {\alpha_0}^2 A_0 - 16 {\alpha_0}^4 \right) A_0 \e^{4 \alpha_0 \left( \phi - N_0 \right)} \nonumber \\
&\, + \left(
7 \alpha_0 + 10 {\alpha_0}^2 + 31 {\alpha_0}^3 + 4 {\alpha_0}^4
 - 1 + 12 \alpha_0 + 4 A_0 + 10 {\alpha_0}^2 + 8 \alpha_0 A_0 + 16 {\alpha_0}^3
 - 3 \alpha_0 + 36 {\alpha_0}^2 + 12 A_0 \alpha_0 \right. \nonumber \\
&\, \left. + 30 {\alpha_0}^3 + 24 {\alpha_0}^2 A_0 - 16 {\alpha_0}^4
+ 14 \alpha_0 + 20 {\alpha_0}^2 + 62 {\alpha_0}^3 + 8 {\alpha_0}^4 \right) {A_0}^2 \e^{8 \alpha_0 \left( \phi - N_0 \right)} \nonumber \\
&\, + \left( 8 {\alpha_0}^2 + 48 {\alpha_0}^3
- 7 - 10 \alpha_0 - 31 {\alpha_0}^2 - 4 {\alpha_0}^3
- 21 \alpha_0 + 30 {\alpha_0}^2 + 93 {\alpha_0}^3 + 12 {\alpha_0}^4
+ 24 {\alpha_0}^2 + 144 {\alpha_0}^3 \right) {A_0}^3 \e^{12 \alpha_0 \left( \phi - N_0 \right)} \nonumber \\
&\, \left. + \left( - 8 \alpha_0 - 48 {\alpha_0}^2 - 24 \alpha_0 - 144 {\alpha_0}^2 \right) {A_0}^4 \e^{16 \alpha_0 \left( \phi - N_0 \right)}
\right\} \nonumber \\
%%%%%%%%%%%%
%%%%%%%%%%%%
=&\, \frac{128 A_0 {\alpha_0}^2 \e^{4 \alpha_0 \left( \phi - N_0 \right)}\left( 1 - A_0 \e^{4 \alpha_0 \left( \phi - N_0 \right)} \right)}
{\left( 1 + A_0 \e^{4 \alpha_0 \left( \phi - N_0 \right)} \right)^3}
+ \frac{3072\alpha A_0 {H_0}^2\e^{4\alpha_0 \left( \phi - N_0 \right)}}{\left(1 + A_0 \e^{4\alpha_0 \left( \phi - N_0 \right)}\right)^{\frac{1}{\alpha_0}+4}}
\left\{ - {\alpha_0}^2 + 3 A_0 \alpha_0 - 3{\alpha_0}^2 A_0 + 4{\alpha_0}^4 \right. \nonumber \\
&\, + \left( 3 \alpha_0 - 21 {\alpha_0}^2 - 24 {\alpha_0}^3 - 44 {\alpha_0}^4 - 3 A_0 - 14 A_0 \alpha_0 - 7 A_0 {\alpha_0}^2 \right) A_0 \e^{4 \alpha_0 \left( \phi - N_0 \right)} \nonumber \\
&\, + \left( -1 + 30 \alpha_0 + 76 {\alpha_0}^2 + 139 {\alpha_0}^3 - 4 {\alpha_0}^4 + 4 A_0 + 20 A_0 \alpha_0 + 24 A_0 {\alpha_0}^2 \right) {A_0}^2 \e^{8 \alpha_0 \left( \phi - N_0 \right)} \nonumber \\
&\, \left. + \left( - 7 - 31 \alpha_0 + 31 {\alpha_0}^2 + 281 {\alpha_0}^3 + 12 {\alpha_0}^4 \right) {A_0}^3 \e^{12 \alpha_0 \left( \phi - N_0 \right)}
+ \left( - 32 \alpha_0 - 192 {\alpha_0}^2 \right) {A_0}^4 \e^{16 \alpha_0 \left( \phi - N_0 \right)}
\right\} \nonumber \\
U'\left( \phi \right)
=&\, \frac{2{H_0}^2 \left( 4 \alpha_0 A_0 \e^{4\alpha_0 \left( \phi - N_0 \right)}\left(1 + A_0 \e^{4\alpha_0 \left( \phi - N_0 \right)}\right)
- \left( \frac{1}{\alpha_0} +1 \right) \left(9 + A_0 \e^{4\alpha_0 \left( \phi - N_0 \right)}\right) 4 \alpha_0 A_0 \e^{4\alpha_0 \left( \phi - N_0 \right)} \right)}
{\left(1 + A_0 \e^{4\alpha_0 \left( \phi - N_0 \right)}\right)^{\frac{1}{\alpha_0} +2}}
\nonumber \\
&\, + \frac{96 \alpha {H_0}^4}{\left(1 + A_0 \e^{4\alpha_0 \left( \phi - N_0 \right)}\right)^{\frac{2}{\alpha_0}+ 4}}
\left[ - 4 \left( 2 + 3\alpha_0 \right) A_0 \e^{4 \alpha_0 \left( \phi - N_0 \right)} \left\{
6 + \left( - 3 - 8 {\alpha_0}^2 \right) A_0 \e^{4 \alpha_0 \left( \phi - N_0 \right)} \right. \right. \nonumber \\
&\, \left. + \left( - 6 + 28 \alpha_0 + 8 {\alpha_0}^2 \right) {A_0}^2 \e^{8 \alpha_0 \left( \phi - N_0 \right)}
 - 7 {A_0}^3 \e^{12 \alpha_0 \left( \phi - N_0 \right)} \right\} \nonumber \\
&\, + \left(1 + A_0 \e^{4\alpha_0 \left( \phi - N_0 \right)}\right) \left\{
4 \left( - 3 - 8 {\alpha_0}^2 \right) \alpha_0 A_0 \e^{4 \alpha_0 \left( \phi - N_0 \right)} \right. \nonumber \\
&\, \left. \left. + 8 \left( - 6 + 28 \alpha_0 + 8 {\alpha_0}^2 \right) \alpha_0 {A_0}^2 \e^{8 \alpha_0 \left( \phi - N_0 \right)}
 - 84 \alpha_0 {A_0}^3 \e^{12 \alpha_0 \left( \phi - N_0 \right)} \right\} \right] \nonumber \\
=&\, - \frac{8 {H_0}^2 \left( 9 + 8 \alpha_0 \right) A_0 \e^{4\alpha_0 \left( \phi - N_0 \right)}}
{\left(1 + A_0 \e^{4\alpha_0 \left( \phi - N_0 \right)}\right)^{\frac{1}{\alpha_0} +2}}
\nonumber \\
&\, + \frac{384 \alpha {H_0}^4 A_0 \e^{4 \alpha_0 \left( \phi - N_0 \right)} }{\left(1 + A_0 \e^{4\alpha_0 \left( \phi - N_0 \right)}\right)^{\frac{2}{\alpha_0}+ 4}}
\left[ - \left( 2 + 3\alpha_0 \right) \left\{
6 + \left( - 3 - 8 {\alpha_0}^2 \right) A_0 \e^{4 \alpha_0 \left( \phi - N_0 \right)} \right. \right. \nonumber \\
&\, \left. + \left( - 6 + 28 \alpha_0 + 8 {\alpha_0}^2 \right) {A_0}^2 \e^{8 \alpha_0 \left( \phi - N_0 \right)}
 - 7 {A_0}^3 \e^{12 \alpha_0 \left( \phi - N_0 \right)} \right\} \nonumber \\
&\, \left. + \left(1 + A_0 \e^{4\alpha_0 \left( \phi - N_0 \right)}\right) \left\{
\left( - 3 - 8 {\alpha_0}^2 \right) \alpha_0 + 2 \left( - 6 + 28 \alpha_0 + 8 {\alpha_0}^2 \right) \alpha_0 A_0 \e^{4 \alpha_0 \left( \phi - N_0 \right)}
 - 21 \alpha_0 {A_0}^2 \e^{8 \alpha_0 \left( \phi - N_0 \right)} \right\} \right] \nonumber \\
%%%%%%%%%%%%%%
=&\, - \frac{8 {H_0}^2 \left( 9 + 8 \alpha_0 \right) A_0 \e^{4\alpha_0 \left( \phi - N_0 \right)}}
{\left(1 + A_0 \e^{4\alpha_0 \left( \phi - N_0 \right)}\right)^{\frac{1}{\alpha_0} +2}}
\nonumber \\
&\, + \frac{384 \alpha {H_0}^4 A_0 \e^{4 \alpha_0 \left( \phi - N_0 \right)} }{\left(1 + A_0 \e^{4\alpha_0 \left( \phi - N_0 \right)}\right)^{\frac{2}{\alpha_0}+ 4}}
\left\{ - 12 - 18 \alpha_0 - 3 \alpha_0 - 8 {\alpha_0}^3 \right. \nonumber \\
&\, + \left( 6 + 9 \alpha_0 + 16{\alpha_0}^2 + 24 {\alpha_0}^3
 - 12 \alpha_0 + 56 {\alpha_0}^2 + 16 {\alpha_0}^3
 - 3 \alpha_0 - 8 {\alpha_0}^3 \right) A_0 \e^{4 \alpha_0 \left( \phi - N_0 \right)} \nonumber \\
&\, + \left( 12 - 56 \alpha_0 - 16 {\alpha_0}^2 + 18\alpha_0 - 84{\alpha_0}^2 - 24 {\alpha_0}^3
 - 21 \alpha_0 - 12 \alpha_0 + 56 {\alpha_0}^2 + 16 {\alpha_0}^3 \right)
{A_0}^2 \e^{8 \alpha_0 \left( \phi - N_0 \right)} \nonumber \\
&\, \left. + \left( 14 + 21 \alpha_0 - 21 \alpha_0 \right)
{A_0}^3 \e^{12 \alpha_0 \left( \phi - N_0 \right)} \right\} \nonumber \\
%%%%%%%%%%
=&\, - \frac{8 {H_0}^2 \left( 9 + 8 \alpha_0 \right) A_0 \e^{4\alpha_0 \left( \phi - N_0 \right)}}
{\left(1 + A_0 \e^{4\alpha_0 \left( \phi - N_0 \right)}\right)^{\frac{1}{\alpha_0} +2}}
+ \frac{384 \alpha {H_0}^4 A_0 \e^{4 \alpha_0 \left( \phi - N_0 \right)} }{\left(1 + A_0 \e^{4\alpha_0 \left( \phi - N_0 \right)}\right)^{\frac{2}{\alpha_0}+ 4}}
\left\{ - 12 - 21 \alpha_0 - 8 {\alpha_0}^3 \right. \nonumber \\
&\, + \left( 6 - 6 \alpha_0 + 72{\alpha_0}^2 + 32 {\alpha_0}^3 \right) A_0 \e^{4 \alpha_0 \left( \phi - N_0 \right)}
+ \left( 12 - 71 \alpha_0 - 44 {\alpha_0}^2
 - 8 {\alpha_0}^3 \right)
{A_0}^2 \e^{8 \alpha_0 \left( \phi - N_0 \right)} \nonumber \\
&\, \left. + 14 {A_0}^3 \e^{12 \alpha_0 \left( \phi - N_0 \right)} \right\} \, , \nonumber \\
U''\left( \phi \right)
=&\, - \frac{32 {H_0}^2 \alpha_0 \left( 9 + 8 \alpha_0 \right) A_0 \e^{4\alpha_0 \left( \phi - N_0 \right)}\left(1 + A_0 \e^{4\alpha_0 \left( \phi - N_0 \right)}
 - \left( \frac{1}{\alpha_0} + 2 \right) A_0 \e^{4\alpha_0 \left( \phi - N_0 \right)}\right)}
{\left(1 + A_0 \e^{4\alpha_0 \left( \phi - N_0 \right)}\right)^{\frac{1}{\alpha_0} +3}} \nonumber \\
&\, + \frac{1536 \alpha {H_0}^4 \alpha_0 A_0 \e^{4 \alpha_0 \left( \phi - N_0 \right)} \left(1 + A_0 \e^{4\alpha_0 \left( \phi - N_0 \right)}
 - \left( \frac{2}{\alpha_0} + 4 \right) A_0 \e^{4\alpha_0 \left( \phi - N_0 \right)} \right)
}{\left(1 + A_0 \e^{4\alpha_0 \left( \phi - N_0 \right)}\right)^{\frac{2}{\alpha_0}+ 5}}
\left\{ - 12 - 21 \alpha_0 - 8 {\alpha_0}^3 \right. \nonumber \\
&\, + \left( 6 - 6 \alpha_0 + 72{\alpha_0}^2 + 32 {\alpha_0}^3 \right) A_0 \e^{4 \alpha_0 \left( \phi - N_0 \right)}
+ \left( 12 - 71 \alpha_0 - 44 {\alpha_0}^2
 - 8 {\alpha_0}^3 \right)
{A_0}^2 \e^{8 \alpha_0 \left( \phi - N_0 \right)} \nonumber \\
&\, \left. + 14 {A_0}^3 \e^{12 \alpha_0 \left( \phi - N_0 \right)} \right\} \nonumber \\
%%%%%%%%%%%%
&\, + \frac{1536 \alpha {H_0}^4 \alpha_0 A_0 \e^{4 \alpha_0 \left( \phi - N_0 \right)}}{\left(1 + A_0 \e^{4\alpha_0 \left( \phi - N_0 \right)}\right)^{\frac{2}{\alpha_0}+ 5}}
\left(1 + A_0 \e^{4\alpha_0 \left( \phi - N_0 \right)}\right)
\left\{ \left( 6 - 6 \alpha_0 + 72{\alpha_0}^2 + 32 {\alpha_0}^3 \right) A_0 \e^{4 \alpha_0 \left( \phi - N_0 \right)} \right. \nonumber \\
&\, \left. + 2 \left( 12 - 71 \alpha_0 - 44 {\alpha_0}^2
 - 8 {\alpha_0}^3 \right) {A_0}^2 \e^{8 \alpha_0 \left( \phi - N_0 \right)}
+ 42 {A_0}^3 \e^{12 \alpha_0 \left( \phi - N_0 \right)} \right\} \nonumber \\
\, , \nonumber \\
%%%%%%%%%%%%%%%%%%
=&\, - \frac{32 {H_0}^2 \alpha_0 \left( 9 + 8 \alpha_0 \right) A_0 \e^{4\alpha_0 \left( \phi - N_0 \right)}\left(1 + A_0 \e^{4\alpha_0 \left( \phi - N_0 \right)}
 - \left( \frac{1}{\alpha_0} + 2 \right) A_0 \e^{4\alpha_0 \left( \phi - N_0 \right)}\right)}
{\left(1 + A_0 \e^{4\alpha_0 \left( \phi - N_0 \right)}\right)^{\frac{1}{\alpha_0} +3}} \nonumber \\
&\, + \frac{1536 \alpha {H_0}^4 \alpha_0 A_0 \e^{4 \alpha_0 \left( \phi - N_0 \right)}
}{\left(1 + A_0 \e^{4\alpha_0 \left( \phi - N_0 \right)}\right)^{\frac{2}{\alpha_0}+ 5}}
\left\{ - 12 - 21 \alpha_0 - 8 {\alpha_0}^3 \right. \nonumber \\
&\, + \left( \frac{24}{\alpha_0} + 42 + 8 {\alpha_0}^2 + 36 + 63\alpha_0 + 24 {\alpha_0}^3
+ 6 - 6 \alpha_0 + 72{\alpha_0}^2 + 32 {\alpha_0}^3 \right) A_0 \e^{4 \alpha_0 \left( \phi - N_0 \right)} \nonumber \\
&\, + \left( - \frac{12}{\alpha_0} + 12 - 144 \alpha_0 - 64 {\alpha_0}^2
 - 18 + 18 \alpha_0 - 216 {\alpha_0}^2 - 96 {\alpha_0}^3
+ 12 - 71 \alpha_0 - 44 {\alpha_0}^2 - 8 {\alpha_0}^3 \right)
{A_0}^2 \e^{8 \alpha_0 \left( \phi - N_0 \right)} \nonumber \\
&\, + \left( - \frac{24}{\alpha_0} + 142 + 88 \alpha_0 + 16 {\alpha_0}^2
 - 36 + 213 \alpha_0 + 132 {\alpha_0}^2 + 24 {\alpha_0}^3
+ 14 \right) {A_0}^3 \e^{12 \alpha_0 \left( \phi - N_0 \right)} \nonumber \\
&\, \left. + \left( - \frac{24}{\alpha_0} - 42 \right) {A_0}^4 \e^{16 \alpha_0 \left( \phi - N_0 \right)}
\right\} \nonumber \\
%%%%%%%%%%%%
&\, + \frac{1536 \alpha {H_0}^4 \alpha_0 A_0 \e^{4 \alpha_0 \left( \phi - N_0 \right)}}{\left(1 + A_0 \e^{4\alpha_0 \left( \phi - N_0 \right)}\right)^{\frac{2}{\alpha_0}+ 5}}
\left\{ \left( 6 - 6 \alpha_0 + 72{\alpha_0}^2 + 32 {\alpha_0}^3 \right) A_0 \e^{4 \alpha_0 \left( \phi - N_0 \right)} \right. \nonumber \\
&\, + \left( 6 - 6 \alpha_0 + 72{\alpha_0}^2 + 32 {\alpha_0}^3 + 24 - 142 \alpha_0 - 88 {\alpha_0}^2 - 16 {\alpha_0}^3 \right) {A_0}^2 \e^{8 \alpha_0 \left( \phi - N_0 \right)} \nonumber \\
&\, \left. + \left(24 - 142 \alpha_0 - 88 {\alpha_0}^2 - 16 {\alpha_0}^3
+ 42 \right){A_0}^3 \e^{12 \alpha_0 \left( \phi - N_0 \right)}
+ 42 {A_0}^4 \e^{16 \alpha_0 \left( \phi - N_0 \right)}
\right\} \nonumber \\
=&\, - \frac{32 {H_0}^2 \alpha_0 \left( 9 + 8 \alpha_0 \right) A_0 \e^{4\alpha_0 \left( \phi - N_0 \right)}\left(1 + A_0 \e^{4\alpha_0 \left( \phi - N_0 \right)}
 - \left( \frac{1}{\alpha_0} + 2 \right) A_0 \e^{4\alpha_0 \left( \phi - N_0 \right)}\right)}
{\left(1 + A_0 \e^{4\alpha_0 \left( \phi - N_0 \right)}\right)^{\frac{1}{\alpha_0} +3}} \nonumber \\
&\, + \frac{1536 \alpha {H_0}^4 \alpha_0 A_0 \e^{4 \alpha_0 \left( \phi - N_0 \right)}
}{\left(1 + A_0 \e^{4\alpha_0 \left( \phi - N_0 \right)}\right)^{\frac{2}{\alpha_0}+ 5}}
\left\{ - 12 - 21 \alpha_0 - 8 {\alpha_0}^3
+ \left( \frac{24}{\alpha_0} + 90 + 51 \alpha_0
+ 152 {\alpha_0}^2 + 88 {\alpha_0}^3 \right) A_0 \e^{4 \alpha_0 \left( \phi - N_0 \right)} \right. \nonumber \\
&\, + \left( - \frac{12}{\alpha_0} + 36 - 345 \alpha_0 - 340 {\alpha_0}^2 - 88 {\alpha_0}^3 \right) {A_0}^2 \e^{8 \alpha_0 \left( \phi - N_0 \right)} \nonumber \\
&\, \left. + \left( - \frac{24}{\alpha_0} + 186 + 159 \alpha_0 + 60 {\alpha_0}^2 + 8 {\alpha_0}^3 \right) {A_0}^3 \e^{12 \alpha_0 \left( \phi - N_0 \right)}
 - \frac{24}{\alpha_0} {A_0}^4 \e^{16 \alpha_0 \left( \phi - N_0 \right)}
\right\} \, .
\end{align}
In order to obtain the explicit forms of \eqref{det} and \eqref{ratio}, we need to calculate the following quantities,
\begin{align}
\label{esgm0}
\e^{-\sigma} =&\, 1 + \frac{24\alpha{H_0}^2}{\left(1 + A_0 \e^{4\alpha_0 \left( \phi - N_0 \right)}\right)^{\frac{1}{\alpha_0}+1}}\, , \nonumber \\
- \frac{d\e^{-\sigma}}{dN}=&\, - \frac{d}{dN} \left( \frac{1}{\e^\sigma} \right) = \e^{-2\sigma}\frac{d\e^\sigma}{dN}
= \frac{96\alpha{H_0}^2 A_0 \left( 1 + \alpha_0 \right) \e^{4\alpha_0 \left( \phi - N_0 \right)}}{\left(1 + A_0 \e^{4\alpha_0 \left( \phi - N_0 \right)}\right)^{\frac{1}{\alpha_0}+2}}\, , \\
\label{mdl2_0}
H^2 =&\, \frac{{H_0}^2}{\left(1 + A_0 \e^{4\alpha_0 \left( N - N_0 \right)}\right)^\frac{1}{\alpha_0}}\, , \nonumber \\
H H' =&\, \frac{1}{2} \frac{dH^2}{dN} = \frac{2{H_0}^2 A_0 \e^{4\alpha_0 \left( N - N_0 \right)}}{\left(1 + A_0 \e^{4\alpha_0 \left( N - N_0 \right)}\right)^{\frac{1}{\alpha_0}+ 1}}\, ,
\end{align}
and
\begin{align}
\label{GammaIJKL}
\Gamma^\sigma_{\phi\phi,\sigma} =&\, \Gamma^\sigma_{\phi\phi} \nonumber \\
=&\, \frac{\e^\sigma}{6} \left[ \frac{8 A_0 \e^{4 \alpha_0 \left( \phi - N_0 \right)}}{1 + A_0 \e^{4 \alpha_0 \left( \phi - N_0 \right)}}
+ \frac{24\alpha {H_0}^2}{\left(1 + A_0 \e^{4\alpha_0 \left( \phi - N_0 \right)}\right)^{\frac{1}{\alpha_0}+3}}
\left\{ 8A_0
+ \left(- 8 + 32 A_0 + 32{\alpha_0}^2 \right) A_0 \e^{4 \alpha_0 \left( \phi - N_0 \right)} \right. \right. \nonumber \\
&\, \left. \left. + \left(- 56 - 24\alpha_0 - 32{\alpha_0}^2 \right) {A_0}^2 \e^{8 \alpha_0 \left( \phi - N_0 \right)}
+ 64 \alpha_0 {A_0}^3 \e^{12 \alpha_0 \left( \phi - N_0 \right)} \right\} \right] \, , \nonumber \\
%%%%%%%%%
 =&\, - \frac{1}{6} \left\{ 1 + \frac{24\alpha{H_0}^2}{\left(1 + A_0 \e^{4\alpha_0 \left( N - N_0 \right)}\right)^{\frac{1}{\alpha_0}+1}} \right\}^{-1}
\left[ \frac{8 A_0 \e^{4 \alpha_0 \left( N - N_0 \right)}}{1 + A_0 \e^{4 \alpha_0 \left( N - N_0 \right)}} \right. \nonumber \\
&\, + \frac{24\alpha {H_0}^2}{\left(1 + A_0 \e^{4\alpha_0 \left( N - N_0 \right)}\right)^{\frac{1}{\alpha_0}+3}}
\left\{ 8A_0 + \left(- 8 + 32 A_0 + 32{\alpha_0}^2 \right) A_0 \e^{4 \alpha_0 \left( N - N_0 \right)} \right. \nonumber \\
&\, \left. \left. + \left(- 56 - 24\alpha_0 - 32{\alpha_0}^2 \right) {A_0}^2 \e^{8 \alpha_0 \left( N - N_0 \right)}
+ 64 \alpha_0 {A_0}^3 \e^{12 \alpha_0 \left( N - N_0 \right)} \right\} \right] \, , \nonumber \\
%%%%%%
%%%
\Gamma^\phi_{\phi\phi} =&\, \frac{\omega'(\phi)}{2\omega(\phi)} \nonumber \\
\Gamma^\phi_{\phi\phi, \phi} =&\, \frac{\omega''(\phi)}{2\omega(\phi)} - \frac{\omega'(\phi)^2}{2\omega(\phi)^2} \nonumber \\
=&\, \frac{1}{2} \left[ \frac{128 A_0 {\alpha_0}^2 \e^{4 \alpha_0 \left( \phi - N_0 \right)}\left( 1 - A_0 \e^{4 \alpha_0 \left( \phi - N_0 \right)} \right)}
{\left( 1 + A_0 \e^{4 \alpha_0 \left( \phi - N_0 \right)} \right)^3}
+ \frac{3072\alpha A_0 {H_0}^2\e^{4\alpha_0 \left( \phi - N_0 \right)}}{\kappa^2 \left(1 + A_0 \e^{4\alpha_0 \left( \phi - N_0 \right)}\right)^{\frac{1}{\alpha_0}+4}}
\left\{ - {\alpha_0}^2 + 3 A_0 \alpha_0 - 3{\alpha_0}^2 A_0 + 4{\alpha_0}^4 \right. \right. \nonumber \\
&\, + \left( 3 \alpha_0 - 21 {\alpha_0}^2 - 24 {\alpha_0}^3 - 44 {\alpha_0}^4 - 3 A_0 - 14 A_0 \alpha_0 - 7 A_0 {\alpha_0}^2 \right) A_0 \e^{4 \alpha_0 \left( \phi - N_0 \right)} \nonumber \\
&\, + \left( -1 + 30 \alpha_0 + 76 {\alpha_0}^2 + 139 {\alpha_0}^3 - 4 {\alpha_0}^4 + 4 A_0 + 20 A_0 \alpha_0 + 24 A_0 {\alpha_0}^2 \right) {A_0}^2 \e^{8 \alpha_0 \left( \phi - N_0 \right)} \nonumber \\
&\, \left. \left. + \left( - 7 - 31 \alpha_0 + 31 {\alpha_0}^2 + 281 {\alpha_0}^3 + 12 {\alpha_0}^4 \right) {A_0}^3 \e^{12 \alpha_0 \left( \phi - N_0 \right)}
+ \left( - 32 \alpha_0 - 192 {\alpha_0}^2 \right) {A_0}^4 \e^{16 \alpha_0 \left( \phi - N_0 \right)}
\right\} \right] \nonumber \\
&\, \times \left[
\frac{8 A_0 \e^{4 \alpha_0 \left( \phi - N_0 \right)}}{1 + A_0 \e^{4 \alpha_0 \left( \phi - N_0 \right)}}
+ \frac{24\alpha {H_0}^2}{\left(1 + A_0 \e^{4\alpha_0 \left( \phi - N_0 \right)}\right)^{\frac{1}{\alpha_0}+3}}
\left\{ 8A_0 + \left(- 8 + 32 A_0 + 32{\alpha_0}^2 \right) A_0 \e^{4 \alpha_0 \left( \phi - N_0 \right)} \right. \right. \nonumber \\
&\, \left. \left. + \left(- 56 - 24\alpha_0 - 32{\alpha_0}^2 \right) {A_0}^2 \e^{8 \alpha_0 \left( \phi - N_0 \right)}
+ 64 \alpha_0 {A_0}^3 \e^{12 \alpha_0 \left( \phi - N_0 \right)} \right\}
\right]^{-1} \nonumber \\
&\, - \frac{1}{2} \left[ \frac{32 A_0 \alpha_0 \e^{4 \alpha_0 \left( \phi - N_0 \right)}}{\left( 1 + A_0 \e^{4 \alpha_0 \left( \phi - N_0 \right)} \right)^2}
+ \frac{768\alpha A_0 {H_0}^2\e^{4\alpha_0 \left( \phi - N_0 \right)}}{\left(1 + A_0 \e^{4\alpha_0 \left( \phi - N_0 \right)}\right)^{\frac{1}{\alpha_0}+4}}
\left\{ - \alpha_0 + 3 A_0 - 3\alpha_0 A_0 + 4{\alpha_0}^3 \right. \right. \nonumber \\
&\, + \left( 1 - 12 \alpha_0 - 4 A_0 - 10 {\alpha_0}^2 - 8 \alpha_0 A_0 - 16 {\alpha_0}^3 \right) A_0 \e^{4 \alpha_0 \left( \phi - N_0 \right)} \nonumber \\
&\, \left. \left. + \left( 7 + 10 \alpha_0 + 31 {\alpha_0}^2 + 4 {\alpha_0}^3 \right) {A_0}^2 \e^{8 \alpha_0 \left( \phi - N_0 \right)}
+ \left( 8 \alpha_0 + 48 {\alpha_0}^2 \right) {A_0}^3 \e^{12 \alpha_0 \left( \phi - N_0 \right)} \right\} \right]^2 \nonumber \\
&\, \times \left[
\frac{8 A_0 \e^{4 \alpha_0 \left( \phi - N_0 \right)}}{1 + A_0 \e^{4 \alpha_0 \left( \phi - N_0 \right)}}
+ \frac{24\alpha {H_0}^2}{\left(1 + A_0 \e^{4\alpha_0 \left( \phi - N_0 \right)}\right)^{\frac{1}{\alpha_0}+3}}
\left\{ 8A_0 + \left(- 8 + 32 A_0 + 32{\alpha_0}^2 \right) A_0 \e^{4 \alpha_0 \left( \phi - N_0 \right)} \right. \right. \nonumber \\
&\, \left. \left. + \left(- 56 - 24\alpha_0 - 32{\alpha_0}^2 \right) {A_0}^2 \e^{8 \alpha_0 \left( \phi - N_0 \right)}
+ 64 \alpha_0 {A_0}^3 \e^{12 \alpha_0 \left( \phi - N_0 \right)} \right\}
\right]^{-2} \nonumber \\
=&\, \frac{1}{2} \left[ \frac{128 A_0 {\alpha_0}^2 \e^{4 \alpha_0 \left( N - N_0 \right)}\left( 1 - A_0 \e^{4 \alpha_0 \left( N - N_0 \right)} \right)}
{\left( 1 + A_0 \e^{4 \alpha_0 \left( N - N_0 \right)} \right)^3}
+ \frac{3072\alpha A_0 {H_0}^2\e^{4\alpha_0 \left( N - N_0 \right)}}{\kappa^2 \left(1 + A_0 \e^{4\alpha_0 \left( N - N_0 \right)}\right)^{\frac{1}{\alpha_0}+4}}
\left\{ - {\alpha_0}^2 + 3 A_0 \alpha_0 - 3{\alpha_0}^2 A_0 + 4{\alpha_0}^4 \right. \right. \nonumber \\
&\, + \left( 3 \alpha_0 - 21 {\alpha_0}^2 - 24 {\alpha_0}^3 - 44 {\alpha_0}^4 - 3 A_0 - 14 A_0 \alpha_0 - 7 A_0 {\alpha_0}^2 \right) A_0 \e^{4 \alpha_0 \left( N - N_0 \right)} \nonumber \\
&\, + \left( -1 + 30 \alpha_0 + 76 {\alpha_0}^2 + 139 {\alpha_0}^3 - 4 {\alpha_0}^4 + 4 A_0 + 20 A_0 \alpha_0 + 24 A_0 {\alpha_0}^2 \right) {A_0}^2 \e^{8 \alpha_0 \left( N - N_0 \right)} \nonumber \\
&\, \left. \left. + \left( - 7 - 31 \alpha_0 + 31 {\alpha_0}^2 + 281 {\alpha_0}^3 + 12 {\alpha_0}^4 \right) {A_0}^3 \e^{12 \alpha_0 \left( N - N_0 \right)}
+ \left( - 32 \alpha_0 - 192 {\alpha_0}^2 \right) {A_0}^4 \e^{16 \alpha_0 \left( N - N_0 \right)}
\right\} \right] \nonumber \\
&\, \times \left[
\frac{8 A_0 \e^{4 \alpha_0 \left( N - N_0 \right)}}{1 + A_0 \e^{4 \alpha_0 \left( N - N_0 \right)}}
+ \frac{24\alpha {H_0}^2}{\left(1 + A_0 \e^{4\alpha_0 \left( N - N_0 \right)}\right)^{\frac{1}{\alpha_0}+3}}
\left\{ 8A_0 + \left(- 8 + 32 A_0 + 32{\alpha_0}^2 \right) A_0 \e^{4 \alpha_0 \left( N - N_0 \right)} \right. \right. \nonumber \\
&\, \left. \left. + \left(- 56 - 24\alpha_0 - 32{\alpha_0}^2 \right) {A_0}^2 \e^{8 \alpha_0 \left( N - N_0 \right)}
+ 64 \alpha_0 {A_0}^3 \e^{12 \alpha_0 \left( N - N_0 \right)} \right\}
\right]^{-1} \nonumber \\
&\, - \frac{1}{2} \left[ \frac{32 A_0 \alpha_0 \e^{4 \alpha_0 \left( N - N_0 \right)}}{\left( 1 + A_0 \e^{4 \alpha_0 \left( N - N_0 \right)} \right)^2}
+ \frac{768\alpha A_0 {H_0}^2\e^{4\alpha_0 \left( N - N_0 \right)}}{\left(1 + A_0 \e^{4\alpha_0 \left( N - N_0 \right)}\right)^{\frac{1}{\alpha_0}+4}}
\left\{ - \alpha_0 + 3 A_0 - 3\alpha_0 A_0 + 4{\alpha_0}^3 \right. \right. \nonumber \\
&\, + \left( 1 - 12 \alpha_0 - 4 A_0 - 10 {\alpha_0}^2 - 8 \alpha_0 A_0 - 16 {\alpha_0}^3 \right) A_0 \e^{4 \alpha_0 \left( N - N_0 \right)} \nonumber \\
&\, \left. \left. + \left( 7 + 10 \alpha_0 + 31 {\alpha_0}^2 + 4 {\alpha_0}^3 \right) {A_0}^2 \e^{8 \alpha_0 \left( N - N_0 \right)}
+ \left( 8 \alpha_0 + 48 {\alpha_0}^2 \right) {A_0}^3 \e^{12 \alpha_0 \left( N - N_0 \right)} \right\} \right]^2 \nonumber \\
&\, \times \left[
\frac{8 A_0 \e^{4 \alpha_0 \left( N - N_0 \right)}}{1 + A_0 \e^{4 \alpha_0 \left( N - N_0 \right)}}
+ \frac{24\alpha {H_0}^2}{\left(1 + A_0 \e^{4\alpha_0 \left( N - N_0 \right)}\right)^{\frac{1}{\alpha_0}+3}}
\left\{ 8A_0 + \left(- 8 + 32 A_0 + 32{\alpha_0}^2 \right) A_0 \e^{4 \alpha_0 \left( N - N_0 \right)} \right. \right. \nonumber \\
&\, \left. \left. + \left(- 56 - 24\alpha_0 - 32{\alpha_0}^2 \right) {A_0}^2 \e^{8 \alpha_0 \left( N - N_0 \right)}
+ 64 \alpha_0 {A_0}^3 \e^{12 \alpha_0 \left( N - N_0 \right)} \right\}
\right]^{-2} \, , \nonumber \\
&\, \mbox{other components}=0 \, .
\end{align}
When $N\to -\infty$, the behaviors of the relevant quantities are given by,
\begin{align}
\label{NinfAp}
H^2 \to &\, {H_0}^2 \, , \quad
H H' \to 2{H_0}^2 A_0 \e^{4\alpha_0 \left( N - N_0 \right)}\, , \quad
\e^{-\sigma} \to 1 + 24\alpha {H_0}^2\, , \quad
 - \frac{d\e^{-\sigma}}{dN}
\to 96\alpha{H_0}^2 A_0 \left( 1 + \alpha_0 \right) \e^{4\alpha_0 \left( N - N_0 \right)}
\, , \nonumber \\
G_{\sigma\sigma} =&\, 3\, , \quad G_{\sigma\phi}=G_{\phi\sigma}=0\, , \nonumber \\
G_{\phi\phi}=&\, \left\{ 1 + 24\alpha{H_0}^2 \right\}^{-1}
\left[ 18{H_0}^2 + 576 \alpha {H_0}^4 \right]
= \frac{18{H_0}^2 \left( 1 + 32 \alpha {H_0}^2\right)}{1 + 24\alpha{H_0}^2}
\, , \nonumber \\
\Gamma^\sigma_{\sigma\sigma}=&\, \Gamma^\sigma_{\sigma\phi} = \Gamma^\sigma_{\phi\sigma}=\Gamma^\phi_{\sigma\sigma}= 0 \, , \quad
\Gamma^\phi_{\sigma\phi} = \Gamma^\phi_{\phi\sigma} = \frac{1}{2} \, , \nonumber \\
\Gamma^\sigma_{\phi\phi,\sigma} =&\, \Gamma^\sigma_{\phi\phi} \nonumber \\
\to &\, - \frac{1}{6} \left\{ 1 + 24\alpha{H_0}^2 \right\}^{-1} 192 \alpha A_0 {H_0}^2
= - \frac{32 \alpha A_0 {H_0}^2}{1 + 24\alpha{H_0}^2}
\, , \nonumber \\
\Gamma^\sigma_{\phi\phi,\phi} \to &\, - \frac{1}{6} \left\{ 1 + 24\alpha{H_0}^2 \right\}^{-1}
\left\{ 32 A_0 \alpha_0 \e^{4 \alpha_0 \left( N - N_0 \right)}
+ 768 \alpha A_0 {H_0}^2 \left( - \alpha_0 + 3 A_0 - 3\alpha_0 A_0 + 4{\alpha_0}^3 \right)\e^{4\alpha_0 \left( N - N_0 \right)}
\right\} \nonumber \\
=&\, - \frac{A_0 \e^{4\alpha_0 \left( N - N_0 \right)}}{3 \left( 1 + 24\alpha{H_0}^2 \right)}
\left\{ 16 \alpha_0 + 384\alpha {H_0}^2 \left( - \alpha_0 + 3 A_0 - 3\alpha_0 A_0 + 4{\alpha_0}^3 \right) \right\} \nonumber \\
%%%
\Gamma^\phi_{\phi\phi} \to &\,
\frac{1}{2} \frac{\left( 32 A_0 \alpha_0 - 768\alpha A_0 \alpha_0 {H_0}^2\right)\e^{4 \alpha_0 \left( \phi - N_0 \right)}}{192\alpha{H_0}^2 A_0}
= \frac{\left( \alpha_0 - 24 \alpha A_0 \alpha_0 {H_0}^2\right)\e^{4 \alpha_0 \left( \phi - N_0 \right)}}{12\alpha{H_0}^2} \, , \nonumber \\
\Gamma^\phi_{\phi\phi, \phi}
\to &\, \frac{1}{2} \frac{ \left\{ 128 A_0 {\alpha_0}^2 + 3072\alpha A_0 {H_0}^2 \left( - {\alpha_0}^2 + 3 A_0 \alpha_0 - 3{\alpha_0}^2 A_0 + 4{\alpha_0}^4 \right) \right\} \e^{4 \alpha_0 \left( \phi - N_0 \right)}}
 {192\alpha{H_0}^2 A_0} \nonumber \\
 = &\, \frac{ \left\{ {\alpha_0}^2 + 24\alpha {H_0}^2 \left( - {\alpha_0}^2 + 3 A_0 \alpha_0 - 3{\alpha_0}^2 A_0 + 4{\alpha_0}^4 \right) \right\} \e^{4 \alpha_0 \left( \phi - N_0 \right)}}
 {6 \alpha{H_0}^2}
\, , \nonumber \\
&\, \mbox{other components}=0 \, .
\end{align}
Then Eq.~\eqref{Ninf} can be obtained.

Similarly, \eqref{dsprsnlt} and \eqref{ratiolt} are obtained as follows:
At the end of the inflation defined by \eqref{einf}, that is, when $A_0 \e^{4 \alpha_0 \left( N - N_0 \right)}=1$, we obtain the following quantities,
\begin{align}
\omega(\phi)=&\, 4 + 3\cdot 2^\frac{1}{\alpha_0} \alpha {H_0}^2
\left\{ 8A_0 + \left(- 8 + 32 A_0 + 32{\alpha_0}^2 \right) \right. \nonumber \\
&\, \left. + \left(- 56 - 24\alpha_0 - 32{\alpha_0}^2 \right) + 64 \alpha_0 \right\} \nonumber \\
=&\, 4 + 3\cdot 2^{-\frac{1}{\alpha_0}} \alpha {H_0}^2 \left( - 64 + 40 A_0 + 40 \alpha_0 \right) \nonumber \\
=&\, 4 + 24\cdot 2^{-\frac{1}{\alpha_0}} \alpha {H_0}^2 \left( - 8 + 5 \alpha_0 + 5 A_0 \right) \, ,
\end{align}
\begin{align}
\omega'(\phi) =&\, 8 + 48\cdot 2^{-\frac{1}{\alpha_0}} \alpha {H_0}^2
\left\{ - \alpha_0 + 3 A_0 - 3\alpha_0 A_0 + 4{\alpha_0}^3 \right. \nonumber \\
&\, + \left( 1 - 12 \alpha_0 - 4 A_0 - 10 {\alpha_0}^2 - 8 \alpha_0 A_0 - 16 {\alpha_0}^3 \right) \nonumber \\
&\, \left. + \left( 7 + 10 \alpha_0 + 31 {\alpha_0}^2 + 4 {\alpha_0}^3 \right)
+ \left( 8 \alpha_0 + 48 {\alpha_0}^2 \right) \right\} \nonumber \\
=&\, 8 + 48\cdot 2^{-\frac{1}{\alpha_0}} \alpha {H_0}^2
\left( 8 + 5 \alpha_0 + 69 {\alpha_0}^2 - 8 {\alpha_0}^3 - A_0 - 11 \alpha_0 A_0 \right) \, ,
\end{align}
\begin{align}
\omega''(\phi)
=&\, 16 {\alpha_0}^2
+ 192 \cdot 2^{-\frac{1}{\alpha_0}} \alpha {H_0}^2
\left\{ - {\alpha_0}^2 + 3 A_0 \alpha_0 - 3{\alpha_0}^2 A_0 + 4{\alpha_0}^4 \right. \nonumber \\
&\, + \left( 3 \alpha_0 - 21 {\alpha_0}^2 - 24 {\alpha_0}^3 - 44 {\alpha_0}^4 - 3 A_0 - 14 A_0 \alpha_0 - 7 A_0 {\alpha_0}^2 \right) \nonumber \\
&\, + \left( -1 + 30 \alpha_0 + 76 {\alpha_0}^2 + 139 {\alpha_0}^3 - 4 {\alpha_0}^4 + 4 A_0 + 20 A_0 \alpha_0 + 24 A_0 {\alpha_0}^2 \right) \nonumber \\
&\, \left. + \left( - 7 - 31 \alpha_0 + 31 {\alpha_0}^2 + 281 {\alpha_0}^3 + 12 {\alpha_0}^4 \right)
+ \left( - 32 \alpha_0 - 192 {\alpha_0}^2 \right) \right\} \nonumber \\
=&\, 16 {\alpha_0}^2 + 192 \cdot 2^{-\frac{1}{\alpha_0}} \alpha {H_0}^2 \left( - 8 - 30 \alpha_0 - 107 {\alpha_0}^2 + 396 {\alpha_0}^3 - 32 {\alpha_0}^4 - 3 A_0 + 9 \alpha_0 A_0 + 14 {\alpha_0}^2 A_0 \right) \, .
\end{align}
\begin{align}
\e^{-\sigma} =&\, 1 + 12 \cdot 2^{-\frac{1}{\alpha_0}} \alpha {H_0}^2\, , \quad
- \frac{d\e^{-\sigma}}{dN} = 24 \cdot 2^{-\frac{1}{\alpha_0}} \alpha {H_0}^2 \left( 1 + \alpha_0 \right) \, , \nonumber \\
H^2 =&\, 2^{-\frac{1}{\alpha_0}} {H_0}^2\, , \quad
H H' = 2^{-\frac{1}{\alpha_0}} {H_0}^2 \, ,
\end{align}
\begin{align}
G_{\sigma\sigma} =&\, 3\, , \quad G_{\sigma\phi}=G_{\phi\sigma}=0\, , \nonumber \\
G_{\phi\phi}=&\, \e^\sigma \omega(\phi) \nonumber \\
G_{\phi\phi} =&\, G_{\phi\phi,\sigma} \nonumber \\
=&\, \left\{ 1 + 12 \cdot 2^{-\frac{1}{\alpha_0}} \alpha {H_0}^2 \right\}^{-1}
\left[ 10 \cdot 2^{-\frac{1}{\alpha_0}} {H_0}^2
+ 12 \cdot 2^{-\frac{1}{\alpha_0}} \alpha {H_0}^4
\left\{ 6 + \left( - 3 - 8 {\alpha_0}^2 \right)
+ \left( - 6 + 28 \alpha_0 + 8 {\alpha_0}^2 \right)
 - 7 \right\} \right] \nonumber \\
=&\, \frac{10 \cdot 2^{-\frac{1}{\alpha_0}} {H_0}^2 + 12 \cdot 2^{-\frac{1}{\alpha_0}} \alpha {H_0}^4 \left( - 10 + 28 \alpha_0 \right)}
{1 + 12 \cdot 2^{-\frac{1}{\alpha_0}} \alpha {H_0}^2} \, , \nonumber \\
G_{\phi\phi,\phi} =&\, \e^\sigma \omega'(\phi) \nonumber \\
=&\, \frac{8 + 48\cdot 2^{-\frac{1}{\alpha_0}} \alpha {H_0}^2 \left( 8 + 5 \alpha_0 + 69 {\alpha_0}^2 - 8 {\alpha_0}^3 - A_0 - 11 \alpha_0 A_0 \right)}
{1 + 12 \cdot 2^{-\frac{1}{\alpha_0}} \alpha {H_0}^2}\, ,
\end{align}
\begin{align}
\Gamma^\sigma_{\sigma\sigma}=&\, \Gamma^\sigma_{\sigma\phi} = \Gamma^\sigma_{\phi\sigma}=\Gamma^\phi_{\sigma\sigma}= 0 \, , \quad
\Gamma^\phi_{\sigma\phi} = \Gamma^\phi_{\phi\sigma} = \frac{1}{2} \, , \nonumber \\
\Gamma^\sigma_{\phi\phi,\sigma} =&\, \Gamma^\sigma_{\phi\phi} \nonumber \\
 =&\, - \frac{1}{6 \left( 1 + 12 \cdot 2^{-\frac{1}{\alpha_0}} \alpha{H_0}^2 \right)}
\left[ 4
+ 3 \cdot 2^{-\frac{1}{\alpha_0}} \alpha {H_0}^2
\left\{ 8A_0 + \left(- 8 + 32 A_0 + 32{\alpha_0}^2 \right)
+ \left(- 56 - 24\alpha_0 - 32{\alpha_0}^2 \right)
+ 64 \alpha_0 \right\} \right] \nonumber \\
 =&\, - \frac{2 + 3 \cdot 2^{-\frac{1}{\alpha_0}} \alpha {H_0}^2
\left(-32 + 20 \alpha_0 + 20 A_0 \right)}{3 \left( 1 + 12 \cdot 2^{-\frac{1}{\alpha_0}} \alpha{H_0}^2 \right)}
\nonumber \\
%%%%%%
%%%
\Gamma^\phi_{\phi\phi}
=&\, \frac{1}{2} \frac{8 + 48\cdot 2^{-\frac{1}{\alpha_0}} \alpha {H_0}^2
\left( 8 + 5 \alpha_0 + 69 {\alpha_0}^2 - 8 {\alpha_0}^3 - A_0 - 11 \alpha_0 A_0 \right) }
{4 + 24\cdot 2^{-\frac{1}{\alpha_0}} \alpha {H_0}^2 \left( - 8 + 5 \alpha_0 + 5 A_0 \right)}
\nonumber \\
=&\, \frac{1 + 6\cdot 2^{-\frac{1}{\alpha_0}} \alpha {H_0}^2
\left( 8 + 5 \alpha_0 + 69 {\alpha_0}^2 - 8 {\alpha_0}^3 - A_0 - 11 \alpha_0 A_0 \right) }
{1 + 6\cdot 2^{-\frac{1}{\alpha_0}} \alpha {H_0}^2 \left( - 8 + 5 \alpha_0 + 5 A_0 \right)} \, , \nonumber \\
\Gamma^\phi_{\phi\phi, \phi} =&\, \frac{\omega''(\phi)}{2\omega(\phi)} - \frac{\omega'(\phi)^2}{2\omega(\phi)^2} \nonumber \\
=&\, \frac{1}{2} \frac{16 {\alpha_0}^2 + 192 \cdot 2^{-\frac{1}{\alpha_0}} \alpha {H_0}^2 \left( - 8 - 30 \alpha_0 - 107 {\alpha_0}^2 + 396 {\alpha_0}^3 - 32 {\alpha_0}^4 - 3 A_0
+ 9 \alpha_0 A_0 + 14 {\alpha_0}^2 A_0 \right) }
{4 + 24\cdot 2^{-\frac{1}{\alpha_0}} \alpha {H_0}^2 \left( - 8 + 5 \alpha_0 + 5 A_0 \right) }
\nonumber \\
&\, - \frac{1}{2} \frac{ \left\{ 8 + 48\cdot 2^{-\frac{1}{\alpha_0}} \alpha {H_0}^2
\left( 8 + 5 \alpha_0 + 69 {\alpha_0}^2 - 8 {\alpha_0}^3 - A_0 - 11 \alpha_0 A_0 \right) \right\}^2}
{\left\{ 4 + 24\cdot 2^{-\frac{1}{\alpha_0}} \alpha {H_0}^2 \left( - 8 + 5 \alpha_0 + 5 A_0 \right) \right\}^2} \nonumber \\
=&\, \frac{2 {\alpha_0}^2 + 24 \cdot 2^{-\frac{1}{\alpha_0}} \alpha {H_0}^2 \left( - 8 - 30 \alpha_0 - 107 {\alpha_0}^2 + 396 {\alpha_0}^3 - 32 {\alpha_0}^4 - 3 A_0
+ 9 \alpha_0 A_0 + 14 {\alpha_0}^2 A_0 \right) }
{1 + 6 \cdot 2^{-\frac{1}{\alpha_0}} \alpha {H_0}^2 \left( - 8 + 5 \alpha_0 + 5 A_0 \right) }
\nonumber \\
&\, - \frac{ 2 \left\{ 1 + 6 \cdot 2^{-\frac{1}{\alpha_0}} \alpha {H_0}^2
\left( 8 + 5 \alpha_0 + 69 {\alpha_0}^2 - 8 {\alpha_0}^3 - A_0 - 11 \alpha_0 A_0 \right) \right\}^2}
{\left\{ 1 + 6 \cdot 2^{-\frac{1}{\alpha_0}} \alpha {H_0}^2 \left( - 8 + 5 \alpha_0 + 5 A_0 \right) \right\}^2}
\, , \nonumber \\
&\, \mbox{other components}=0 \, .
\end{align}
\begin{align}
U\left( \phi \right)
=&\, 10 \cdot 2^{-\frac{1}{\alpha_0}} {H_0}^2
+ 12 \cdot 2^{-\frac{2}{\alpha_0}} \alpha {H_0}^4
\left\{ 6 + \left( - 3 - 8 {\alpha_0}^2 \right)
+ \left( - 6 + 28 \alpha_0 + 8 {\alpha_0}^2 \right) - 7 \right\} \nonumber \\
=&\, 10 \cdot 2^{-\frac{1}{\alpha_0}} {H_0}^2
+ 12 \cdot 2^{-\frac{2}{\alpha_0}} \alpha {H_0}^4 \left( - 10 + 28 \alpha_0 \right)\, .
\end{align}
\begin{align}
U'\left( \phi \right)
=&\, - 2 \cdot 2^{-\frac{1}{\alpha_0}} {H_0}^2 \left( 9 + 8 \alpha_0 \right)
+ 24 \cdot 2^{-\frac{2}{\alpha_0}} {H_0}^4
\left\{ - 12 - 21 \alpha_0 - 8 {\alpha_0}^3 \right. \nonumber \\
&\, \left. + \left( 6 - 6 \alpha_0 + 72{\alpha_0}^2 + 32 {\alpha_0}^3 \right)
+ \left( 12 - 71 \alpha_0 - 44 {\alpha_0}^2 - 8 {\alpha_0}^3 \right) + 14 \right\} \nonumber \\
=&\, - 2 \cdot 2^{-\frac{1}{\alpha_0}} {H_0}^2 \left( 9 + 8 \alpha_0 \right)
+ 24 \cdot 2^{-\frac{2}{\alpha_0}} {H_0}^4 \left( 20 - 98 \alpha_0 + 26 {\alpha_0}^2 + 16{\alpha_0}^3 \right) \, , \nonumber \\
U''\left( \phi \right)
=&\, 4 \cdot 2^{-\frac{1}{\alpha_0}} {H_0}^2 \left( 9 + 8 \alpha_0 \right) \nonumber \\
&\, + 48 \cdot 2^{-\frac{2}{\alpha_0}} \alpha {H_0}^4 \alpha_0
\left\{ - 12 - 21 \alpha_0 - 8 {\alpha_0}^3
+ \left( \frac{24}{\alpha_0} + 90 + 51 \alpha_0
+ 152 {\alpha_0}^2 + 88 {\alpha_0}^3 \right) \right. \nonumber \\
&\, + \left( - \frac{12}{\alpha_0} + 36 - 345 \alpha_0 - 340 {\alpha_0}^2 - 88 {\alpha_0}^3 \right) \nonumber \\
&\, \left. + \left( - \frac{24}{\alpha_0} + 186 + 159 \alpha_0 + 60 {\alpha_0}^2 + 8 {\alpha_0}^3 \right)
 - \frac{24}{\alpha_0} \right\} \nonumber \\
=&\, 4 \cdot 2^{-\frac{1}{\alpha_0}} {H_0}^2 \left( 9 + 8 \alpha_0 \right)
+ 48 \cdot 2^{-\frac{2}{\alpha_0}} \alpha {H_0}^4 \left( - 36 + 291 \alpha_0 + 146 {\alpha_0}^2 - 128 {\alpha_0}^3 \right) \, .
\end{align}
%%%%%%%%%%%%%%
\begin{align}
V(\sigma) =&\,
\frac{36 \cdot 2^{-\frac{2}{\alpha_0}} \alpha{H_0}^4}{\left(1 + 12 \cdot 2^{-\frac{1}{\alpha_0}} \alpha{H_0}^2 \right)^2} \, ,
\end{align}
\begin{align}
V'(\sigma) =&\,
 - \frac{6 \cdot 2^{-\frac{1}{\alpha_0}} {H_0}^2}{\left(1 + 12 \cdot 2^{-\frac{1}{\alpha_0}} \alpha{H_0}^2 \right)^2} \, , \nonumber \\
V''(\sigma)
=&\, - \frac{\e^{-\sigma} - 2}{2\alpha \e^{-2\sigma}}
= - \frac{ -1 + 12 \cdot 2^{-\frac{1}{\alpha_0}} \alpha{H_0}^2}
{2\alpha \left( 1 + 12 \cdot 2^{-\frac{1}{\alpha_0}} \alpha{H_0}^2 \right)^2} \, .
\end{align}
Then, by using the following expressions,
\begin{align}
\label{Ex1}
&\, H^2 \e^{-\sigma} \Omega(N)^2 - i \left( \frac{d\e^\sigma}{dN} H^2 \e^{-2\sigma} + 3 H^2 \e^{-\sigma} + HH' \e^{-\sigma} \right) \Omega(N) - \e^{-\sigma} a^{-2} k^2 \nonumber \\
&\, - \sum_{J,K=\sigma,\phi} \Gamma^1_{JK,1} \e^{-\sigma} H^2 \frac{d\Phi^K}{dN} \frac{d\Phi^J}{dN}
+ \sum_{J=\sigma,\phi} \Gamma^1_{1J} \left( i\Omega(N) \e^{-\sigma} H^2 \frac{d \Phi^J}{dN}
 - G^{1J}_{,1} U_{\mathrm{eff}, J} - G^{1J} U_{\mathrm{eff}, J1} \right) \nonumber \\
=&\, 2^{-\frac{1}{\alpha_0}} {H_0}^2 \left( 1 + 12 \cdot 2^{-\frac{1}{\alpha_0}} \alpha {H_0}^2 \right) \Omega(N)^2
 - i \left( 2^{-\frac{1}{\alpha_0}} {H_0}^2 24 \cdot 2^{-\frac{1}{\alpha_0}} \alpha {H_0}^2 \left( 1 + \alpha_0 \right) \right. \nonumber \\
&\, \left. + 3 \cdot 2^{-\frac{1}{\alpha_0}} {H_0}^2 \left( 1 + 12 \cdot 2^{-\frac{1}{\alpha_0}} \alpha {H_0}^2 \right)
+ 2^{-\frac{1}{\alpha_0}} {H_0}^2 \left( 1 + 12 \cdot 2^{-\frac{1}{\alpha_0}} \alpha {H_0}^2 \right) \right) \Omega(N) - \e^{-\sigma} a^{-2} k^2 \nonumber \\
&\, + \frac{2 + 3 \cdot 2^{-\frac{1}{\alpha_0}} \alpha {H_0}^2 \left(-32 + 20 \alpha_0 + 20 A_0 \right)}{3 \left( 1 + 12 \cdot 2^{-\frac{1}{\alpha_0}} \alpha{H_0}^2 \right)}
2^{-\frac{1}{\alpha_0}} {H_0}^2 \left( 1 + 12 \cdot 2^{-\frac{1}{\alpha_0}} \alpha {H_0}^2 \right) \nonumber \\
=&\, 2^{-\frac{1}{\alpha_0}} {H_0}^2 \left( 1 + 12 \cdot 2^{-\frac{1}{\alpha_0}} \alpha {H_0}^2 \right) \Omega(N)^2
 - i \left\{ 24 \cdot 2^{-\frac{2}{\alpha_0}} \alpha {H_0}^4 \left( 1 + \alpha_0 \right) \right. \nonumber \\
&\, \left. + 4 \cdot 2^{-\frac{1}{\alpha_0}} {H_0}^2 \left( 1 + 12 \cdot 2^{-\frac{1}{\alpha_0}} \alpha {H_0}^2 \right) \right\} \Omega(N) - \e^{-\sigma} a^{-2} k^2 \nonumber \\
&\, + \frac{2^{-\frac{1}{\alpha_0}} {H_0}^2 }{3} \left\{ 2 + 3 \cdot 2^{-\frac{1}{\alpha_0}} \alpha {H_0}^2 \left(-32 + 20 \alpha_0 + 20 A_0 \right)\right\} \, ,
\end{align}
\begin{align}
\label{Ex2}
&\, H^2 \e^{-\sigma} \Omega(N)^2 - i \left( \frac{d\e^\sigma}{dN} H^2 \e^{-2\sigma} + 3 H^2 \e^{-\sigma} + HH' \e^{-\sigma} \right) \Omega(N) - \e^{-\sigma} a^{-2} k^2 \nonumber \\
&\, - \sum_{J,K=\sigma,\phi} \Gamma^2_{JK,2} \e^{-\sigma} H^2 \frac{d\Phi^K}{dN} \frac{d\Phi^J}{dN}
+ \sum_{J=\sigma,\phi} \Gamma^2_{2J} \left( i\Omega(N) \e^{-\sigma} H^2 \frac{d \Phi^J}{dN}
 - G^{2J}_{,2} U_{\mathrm{eff}, J} - G^{2J} U_{\mathrm{eff}, J2} \right) \nonumber \\
=&\, 2^{-\frac{1}{\alpha_0}} {H_0}^2 \left( 1 + 12 \cdot 2^{-\frac{1}{\alpha_0}} \alpha {H_0}^2 \right) \Omega(N)^2
 - i \left\{ 24 \cdot 2^{-\frac{2}{\alpha_0}} \alpha {H_0}^4 \left( 1 + \alpha_0 \right) \right. \nonumber \\
&\, \left. + 4 \cdot 2^{-\frac{1}{\alpha_0}} {H_0}^2 \left( 1 + 12 \cdot 2^{-\frac{1}{\alpha_0}} \alpha {H_0}^2 \right) \right\} \Omega(N) - \e^{-\sigma} a^{-2} k^2 \nonumber \\
&\, - \left[ \frac{2 {\alpha_0}^2 + 24 \cdot 2^{-\frac{1}{\alpha_0}} \alpha {H_0}^2 \left( - 8 - 30 \alpha_0 - 107 {\alpha_0}^2 + 396 {\alpha_0}^3 - 32 {\alpha_0}^4 - 3 A_0
+ 9 \alpha_0 A_0 + 14 {\alpha_0}^2 A_0 \right) }
{1 + 6 \cdot 2^{-\frac{1}{\alpha_0}} \alpha {H_0}^2 \left( - 8 + 5 \alpha_0 + 5 A_0 \right) }
\right. \nonumber \\
&\, \left. - \frac{ 2 \left\{ 1 + 6 \cdot 2^{-\frac{1}{\alpha_0}} \alpha {H_0}^2
\left( 8 + 5 \alpha_0 + 69 {\alpha_0}^2 - 8 {\alpha_0}^3 - A_0 - 11 \alpha_0 A_0 \right) \right\}^2}
{\left\{ 1 + 6 \cdot 2^{-\frac{1}{\alpha_0}} \alpha {H_0}^2 \left( - 8 + 5 \alpha_0 + 5 A_0 \right) \right\}^2} \right] 2^{-\frac{1}{\alpha_0}} {H_0}^2 \nonumber \\
&\, + \frac{1}{2} \left[ i\Omega(N) 2^{-\frac{1}{\alpha_0}} \alpha {H_0}^2 24 \cdot 2^{-\frac{1}{\alpha_0}} \alpha {H_0}^2 \left( 1 + \alpha_0 \right) \right. \nonumber \\
&\, + \frac{\left(1 + 12 \cdot 2^{-\frac{1}{\alpha_0}} \alpha {H_0}^2\right)^2}{\left\{10 \cdot 2^{-\frac{1}{\alpha_0}} {H_0}^2 + 12 \cdot 2^{-\frac{1}{\alpha_0}} \alpha {H_0}^4 \left( - 10 + 28 \alpha_0 \right)\right\}^2}
\frac{8 + 48\cdot 2^{-\frac{1}{\alpha_0}} \alpha {H_0}^2 \left( 8 + 5 \alpha_0 + 69 {\alpha_0}^2 - 8 {\alpha_0}^3 - A_0 - 11 \alpha_0 A_0 \right)}
{1 + 12 \cdot 2^{-\frac{1}{\alpha_0}} \alpha {H_0}^2} \nonumber \\
&\, \times \left\{ - 2 \cdot 2^{-\frac{1}{\alpha_0}} {H_0}^2 \left( 9 + 8 \alpha_0 \right)
+ 24 \cdot 2^{-\frac{2}{\alpha_0}} {H_0}^4 \left( 20 - 98 \alpha_0 + 26 {\alpha_0}^2 + 16{\alpha_0}^3 \right) \right\}
\frac{1}{1 + 12 \cdot 2^{-\frac{1}{\alpha_0}} \alpha {H_0}^2} \nonumber \\
&\, - \frac{1 + 12 \cdot 2^{-\frac{1}{\alpha_0}} \alpha {H_0}^2}{10 \cdot 2^{-\frac{1}{\alpha_0}} {H_0}^2 + 12 \cdot 2^{-\frac{1}{\alpha_0}} \alpha {H_0}^4 \left( - 10 + 28 \alpha_0 \right)} \nonumber \\
&\, \left. \times \left\{ 4 \cdot 2^{-\frac{1}{\alpha_0}} {H_0}^2 \left( 9 + 8 \alpha_0 \right)
+ 48 \cdot 2^{-\frac{2}{\alpha_0}} \alpha {H_0}^4 \left( - 36 + 291 \alpha_0 + 146 {\alpha_0}^2 - 128 {\alpha_0}^3 \right) \right\}
\frac{1}{1 + 12 \cdot 2^{-\frac{1}{\alpha_0}} \alpha {H_0}^2} \right] \nonumber \\
=&\, 2^{-\frac{1}{\alpha_0}} {H_0}^2 \left( 1 + 12 \cdot 2^{-\frac{1}{\alpha_0}} \alpha {H_0}^2 \right) \Omega(N)^2
 - i \left\{ 24 \cdot 2^{-\frac{2}{\alpha_0}} \alpha {H_0}^4 \left( 1 + \alpha_0 \right) \right. \nonumber \\
&\, \left. + 4 \cdot 2^{-\frac{1}{\alpha_0}} {H_0}^2 \left( 1 + 12 \cdot 2^{-\frac{1}{\alpha_0}} \alpha {H_0}^2 \right) \right\} \Omega(N) - \e^{-\sigma} a^{-2} k^2 \nonumber \\
&\, - \left[ \frac{2 {\alpha_0}^2 + 24 \cdot 2^{-\frac{1}{\alpha_0}} \alpha {H_0}^2 \left( - 8 - 30 \alpha_0 - 107 {\alpha_0}^2 + 396 {\alpha_0}^3 - 32 {\alpha_0}^4 - 3 A_0
+ 9 \alpha_0 A_0 + 14 {\alpha_0}^2 A_0 \right) }
{1 + 6 \cdot 2^{-\frac{1}{\alpha_0}} \alpha {H_0}^2 \left( - 8 + 5 \alpha_0 + 5 A_0 \right) }
\right. \nonumber \\
&\, \left. - \frac{ 2 \left\{ 1 + 6 \cdot 2^{-\frac{1}{\alpha_0}} \alpha {H_0}^2
\left( 8 + 5 \alpha_0 + 69 {\alpha_0}^2 - 8 {\alpha_0}^3 - A_0 - 11 \alpha_0 A_0 \right) \right\}^2}
{\left\{ 1 + 6 \cdot 2^{-\frac{1}{\alpha_0}} \alpha {H_0}^2 \left( - 8 + 5 \alpha_0 + 5 A_0 \right) \right\}^2} \right] 2^{-\frac{1}{\alpha_0}} {H_0}^2 \nonumber \\
&\, + \frac{1}{2} \left[ i\Omega(N) 2^{-\frac{1}{\alpha_0}} \alpha {H_0}^2 24 \cdot 2^{-\frac{1}{\alpha_0}} \alpha {H_0}^2 \left( 1 + \alpha_0 \right) \right. \nonumber \\
&\, + \frac{8 + 48\cdot 2^{-\frac{1}{\alpha_0}} \alpha {H_0}^2 \left( 8 + 5 \alpha_0 + 69 {\alpha_0}^2 - 8 {\alpha_0}^3 - A_0 - 11 \alpha_0 A_0 \right)}{\left\{10 \cdot 2^{-\frac{1}{\alpha_0}} {H_0}^2 + 12 \cdot 2^{-\frac{1}{\alpha_0}} \alpha {H_0}^4 \left( - 10 + 28 \alpha_0 \right)\right\}^2} \nonumber \\
&\, \times \left\{ - 2 \cdot 2^{-\frac{1}{\alpha_0}} {H_0}^2 \left( 9 + 8 \alpha_0 \right)
+ 24 \cdot 2^{-\frac{2}{\alpha_0}} {H_0}^4 \left( 20 - 98 \alpha_0 + 26 {\alpha_0}^2 + 16{\alpha_0}^3 \right) \right\} \nonumber \\
&\, \left. - \frac{4 \cdot 2^{-\frac{1}{\alpha_0}} {H_0}^2 \left( 9 + 8 \alpha_0 \right)
+ 48 \cdot 2^{-\frac{2}{\alpha_0}} \alpha {H_0}^4 \left( - 36 + 291 \alpha_0 + 146 {\alpha_0}^2 - 128 {\alpha_0}^3 \right)}{10 \cdot 2^{-\frac{1}{\alpha_0}} {H_0}^2 + 12 \cdot 2^{-\frac{1}{\alpha_0}} \alpha {H_0}^4 \left( - 10 + 28 \alpha_0 \right)} \right] \, ,
\end{align}
\begin{align}
\label{Ex3}
&\, - \sum_{J,K=\sigma,\phi} \Gamma^1_{JK,2} \e^{-\sigma} H^2 \frac{d\Phi^K}{dN} \frac{d\Phi^J}{dN}
+ \sum_{J=\sigma,\phi} \Gamma^1_{2J} \left( i\Omega(N) \e^{-\sigma} H^2 \frac{d \Phi^J}{dN}
 - G^{1J}_{,2} U_{\mathrm{eff}, J} - G^{1J} U_{\mathrm{eff}, J2} \right) \nonumber \\
=&\, \frac{2 + 3 \cdot 2^{-\frac{1}{\alpha_0}} \alpha {H_0}^2 \left(-32 + 20 \alpha_0 + 20 A_0 \right)}{3 \left( 1 + 12 \cdot 2^{-\frac{1}{\alpha_0}} \alpha{H_0}^2 \right)}
2^{-\frac{1}{\alpha_0}} {H_0}^2 \left( 1 + 12 \cdot 2^{-\frac{1}{\alpha_0}} \alpha {H_0}^2 \right) \nonumber \\
&\, - \frac{2 + 3 \cdot 2^{-\frac{1}{\alpha_0}} \alpha {H_0}^2 \left(-32 + 20 \alpha_0 + 20 A_0 \right)}{3 \left( 1 + 12 \cdot 2^{-\frac{1}{\alpha_0}} \alpha{H_0}^2 \right)}
\left\{ 2^{-\frac{1}{\alpha_0}} {H_0}^2 \left( 1 + 12 \cdot 2^{-\frac{1}{\alpha_0}} \alpha {H_0}^2 \right) i\Omega(N) \right. \nonumber \\
&\, \left. - \frac{1}{3} 2 \left( 1 + 12 \cdot 2^{-\frac{1}{\alpha_0}} \alpha {H_0}^2 \right)^{-2}
\left( - 2 \cdot 2^{-\frac{1}{\alpha_0}} {H_0}^2 \left( 9 + 8 \alpha_0 \right)
+ 24 \cdot 2^{-\frac{2}{\alpha_0}} {H_0}^4 \left( 20 - 98 \alpha_0 + 26 {\alpha_0}^2 + 16{\alpha_0}^3 \right) \right) \right\}
\nonumber \\
=&\, \frac{2^{-\frac{1}{\alpha_0}} {H_0}^2 \left\{ 2 + 3 \cdot 2^{-\frac{1}{\alpha_0}} \alpha {H_0}^2 \left(-32 + 20 \alpha_0 + 20 A_0 \right)\right\} }{3} \nonumber \\
&\, - \frac{2 + 3 \cdot 2^{-\frac{1}{\alpha_0}} \alpha {H_0}^2 \left(-32 + 20 \alpha_0 + 20 A_0 \right)}{3 \left( 1 + 12 \cdot 2^{-\frac{1}{\alpha_0}} \alpha{H_0}^2 \right)}
\left\{ 2^{-\frac{1}{\alpha_0}} {H_0}^2 \left( 1 + 12 \cdot 2^{-\frac{1}{\alpha_0}} \alpha {H_0}^2 \right) i\Omega(N) \right. \nonumber \\
&\, \left. \frac{ - 4 \cdot 2^{-\frac{1}{\alpha_0}} {H_0}^2 \left( 9 + 8 \alpha_0 \right) + 48 \cdot 2^{-\frac{2}{\alpha_0}} {H_0}^4 \left( 20 - 98 \alpha_0 + 26 {\alpha_0}^2 + 16{\alpha_0}^3 \right)}
{3\left( 1 + 12 \cdot 2^{-\frac{1}{\alpha_0}} \alpha {H_0}^2 \right)^2} \right\} \, ,
\end{align}
\begin{align}
\label{Ex4}
&\, - \sum_{J,K=\sigma,\phi} \Gamma^2_{JK,1} \e^{-\sigma} H^2 \frac{d\Phi^K}{dN} \frac{d\Phi^J}{dN}
+ \sum_{J=\sigma,\phi} \Gamma^2_{1J} \left( i\Omega(N) \e^{-\sigma} H^2 \frac{d \Phi^J}{dN}
 - G^{2J}_{,1} U_{\mathrm{eff}, J} - G^{2J} U_{\mathrm{eff}, J1} \right)
\nonumber \\
= &\, \frac{1}{2} \left\{ i\Omega(N) 2^{-\frac{1}{\alpha_0}} {H_0}^2 \left( 1 + 12 \cdot 2^{-\frac{1}{\alpha_0}} \alpha {H_0}^2 \right) \right. \nonumber \\
&\, + \frac{1 + 12 \cdot 2^{-\frac{1}{\alpha_0}} \alpha {H_0}^2}{10 \cdot 2^{-\frac{1}{\alpha_0}} {H_0}^2 + 12 \cdot 2^{-\frac{1}{\alpha_0}} \alpha {H_0}^4 \left( - 10 + 28 \alpha_0 \right)}
\left( 1 + 12 \cdot 2^{-\frac{1}{\alpha_0}} \alpha {H_0}^2 \right)^{-1} \nonumber \\
&\, \times \left( - 2 \cdot 2^{-\frac{1}{\alpha_0}} {H_0}^2 \left( 9 + 8 \alpha_0 \right)
+ 24 \cdot 2^{-\frac{2}{\alpha_0}} {H_0}^4 \left( 20 - 98 \alpha_0 + 26 {\alpha_0}^2 + 16{\alpha_0}^3 \right) \right) \nonumber \\
&\, - \frac{1 + 12 \cdot 2^{-\frac{1}{\alpha_0}} \alpha {H_0}^2}{10 \cdot 2^{-\frac{1}{\alpha_0}} {H_0}^2 + 12 \cdot 2^{-\frac{1}{\alpha_0}} \alpha {H_0}^4 \left( - 10 + 28 \alpha_0 \right)}
\left( 1 + 12 \cdot 2^{-\frac{1}{\alpha_0}} \alpha {H_0}^2 \right)^{-1} \nonumber \\
&\, \left. \times \left( - 2 \cdot 2^{-\frac{1}{\alpha_0}} {H_0}^2 \left( 9 + 8 \alpha_0 \right)
+ 24 \cdot 2^{-\frac{2}{\alpha_0}} {H_0}^4 \left( 20 - 98 \alpha_0 + 26 {\alpha_0}^2 + 16{\alpha_0}^3 \right) \right)
\right\} \nonumber \\
= &\, i\Omega(N) 2^{-1 -\frac{1}{\alpha_0}} {H_0}^2 \left( 1 + 12 \cdot 2^{-\frac{1}{\alpha_0}} \alpha {H_0}^2 \right) \, .
\end{align}
we obtain the dispersion relation \eqref{det} and the ratio of $\delta \Phi^1 = \delta \sigma$ and $\delta \Phi^2 = \delta \phi$ in \eqref{ratio} as in \eqref{dsprsnlt} and \eqref{ratiolt}.

\bibliographystyle{apsrev4-1}
\bibliography{References4}

\end{document}